\pdfoutput=1
\documentclass{article}
\PassOptionsToPackage{numbers, compress}{natbib}

\usepackage[main,final]{neurips_2026}

\usepackage[utf8]{inputenc}
\usepackage[T1]{fontenc}
\usepackage{microtype}
\usepackage{xspace}

\usepackage{caption}

\usepackage{amsmath}
\usepackage{amssymb}
\usepackage{amsthm}
\usepackage{amsfonts}
\usepackage{mathtools}
\usepackage{mathrsfs}
\usepackage{bm}
\usepackage{bbm}

\usepackage{booktabs}
\usepackage{multirow}
\usepackage{graphicx}
\usepackage{float}
\usepackage[table,dvipsnames]{xcolor}
\usepackage{enumitem}
\usepackage{wrapfig}
\usepackage{siunitx}

\usepackage{pgfplots}
\pgfplotsset{compat=1.18}
\usepackage[ruled,vlined]{algorithm2e}

\usepackage{url}
\usepackage{hyperref}
\hypersetup{
  colorlinks=true,
  linkcolor=blue,
  citecolor=blue,
  urlcolor=blue,
  breaklinks=true,
}

\definecolor{colorbest}{HTML}{2A9D8F}
\definecolor{colorsecond}{HTML}{264653}

\newcommand{\best}[1]{{\color{colorbest}\bm{#1}}}
\newcommand{\second}[1]{{\color{colorsecond}\underline{#1}}}
\newcommand{\tied}[1]{#1^{\dagger}}

\newcommand{\ZINC}{$\mathrm{ZINC}_{\mathrm{subset}}^{*}$\xspace}
\newcommand{\initZINC}{$\mathrm{ZINC}_{\mathrm{subset}}$}

\newcommand{\C}{$\boldsymbol{\mathcal{C}}$}
\newcommand{\Cone}{$\boldsymbol{\mathcal{C}}^{(1)}_{i}$}
\newcommand{\Ctwo}{$\boldsymbol{\mathcal{C}}^{(2)}_{ij}$}

\usepackage[normalem]{ulem}

\usepackage{orcidlink}

\title{
Neutral-atom quantum features as complementary structural encodings for graph learning
}

\author{%
  Mathieu Garrigues~\thanks{Equal contribution.} ~\thanks{All authors are with Pasqal, 24 avenue Émile Baudot, 91120 Palaiseau, France.}\,~\orcidlink{0009-0008-8630-5177}\\
  \texttt{mathieu.garrigues@pasqal.com}
  \And
  Igor O.~Sokolov~\footnotemark[1] ~\footnotemark[2]\,~\orcidlink{0000-0002-0022-5686}  \\
  \texttt{igor.sokolov@pasqal.com}
  \And
  Matthias Hecker~\footnotemark[2]\,~\orcidlink{0000-0002-3021-1074} \\
  \texttt{matthias.hecker@pasqal.com}
  \And
  Lorenzo Moro~\footnotemark[2]\,~\orcidlink{0000-0001-9818-8854}  \\
  \texttt{lorenzo.moro@pasqal.com}
  \And
  Shaheen Acheche~\footnotemark[2]\,~\orcidlink{0009-0004-9987-8223}  \\
  \texttt{shaheen.acheche@pasqal.com}
}

\date{\today}

\begin{document}

\maketitle

\begin{abstract}
Structural descriptors provide graph neural networks (GNNs) with positional and topological inductive biases.
However, widely used classical encodings based on random walks or Laplacian spectra emphasize specific structural statistics of a graph, leaving room for alternative descriptors.
Neutral-atom quantum processing units (QPUs) can realize graph-dependent Rydberg Hamiltonians, allowing graph topology to be probed through dynamical correlation measurements.
In this work, we use these measurements as \emph{quantum graph features} (QGFs), providing a complementary, hardware-derived family of structural encoding
that --- as we demonstrate empirically --- is not accurately described by classical features derived from Laplacian and adjacency matrices, even when supplemented with an extended set of random-walk, geometric descriptors and classical physical models.
By appropriately selecting the quench parameters of the Hamiltonian, we access two characteristic correlation regimes with distinct structural signatures: antiferromagnetic-like and ferromagnetic-like.
On an intrinsic information benchmark of 15 topological and spectral probing tasks, quantum and hybrid quantum-classical features match or surpass classical descriptors on local topological properties.

When supplied to message-passing, attention, and graph-transformer backbones, hybrid quantum-classical features yield small improvements that are within cross-fold noise on most cells but sign-consistent across architectures.
When comparing noiseless, shot-sampled emulation-derived quantum features with an end-to-end neutral-atom QPU implementation, the downstream hybrid results with compressed hardware features match their emulated counterparts within $0.01$ F$_1$ on all six backbones.
On a synthetic family of ladder graphs, QGFs distinguish two specific classes, outperforming the strongest classical structural descriptors by a margin exceeding cross-fold variability.
\end{abstract}

\section{Introduction}

Graph neural networks (GNNs) and graph transformers are the default
tools for learning on relational data such as molecules, materials,
and biological networks~\citep{kipf2017semi,gilmer2017neural,velickovic2018graph,rampasek2022GPS}.
Their performance depends strongly on how graph structure is exposed
to the model.
Standard choices include shortest-path distances,
Laplacian eigenvectors, heat kernels, and random-walk profiles such as
relative random walk probabilities (RRWP) or random walk structural encodings (RWSE) or even application-specific like ring-membership, which are concatenated to
node or edge features or injected into attention
mechanisms~\citep{ma2023graph,dwivedi2022benchmarking,kreuzer2021san, menegaux2023self}.
These classical descriptors provide useful inductive biases and are efficiently
computable, nonetheless, they encode specific structural statistics only.
They inherit known limitations: for example, even expressive relative encodings such as RRWP fail on certain adversarial graph families, including non-isomorphic strongly regular graphs~\citep{ma2023graph,zhang2023gdwl,thabet2024quantum,kasture2025quantum}.
This motivates the search for complementary graph descriptors.
Physics-inspired architectures, which cast message passing as diffusion~\citep{chamberlain2021grand}, magnetic-Laplacian propagation~\citep{zhang2021magnet}, or coupled oscillators~\citep{rusch2022graphcon}, show that dynamics on graphs provide useful inductive biases; here, the dynamics is instead executed on a physical quantum system.
Programmable neutral-atom quantum processors offer a concrete route
towards such graph descriptors.
In such a device, atoms can be arranged in user-defined registers.
Atomic energy levels are used as qubit states which are time-evolved under a particular many-body Hamiltonian~\citep{henry2021quantum,Albrecht2023}.

Embedding a graph into such a register imprints its topology onto the quantum state, which can be extracted via observable measurement.
Prior work has exploited this principle to define quantum evolution kernels and
quantum positional encodings for graph learning
~\citep{henry2021quantum,Albrecht2023,kyriienko2022qek,thabet2024quantum}.
In particular, \citet{thabet2024quantum} showed that quantum
correlation-based encodings can be strictly more expressive than RRWP
on selected graph families, while recent kernel studies have used
neutral-atom observables to construct graph kernels competitive with
classical baselines~\citep{Djellabi_2026}.
These results suggest that QPU-derived observables may capture information not contained in standard random-walk
and spectral encodings.

\begin{figure}[t]
\centering
\includegraphics[width=\textwidth]{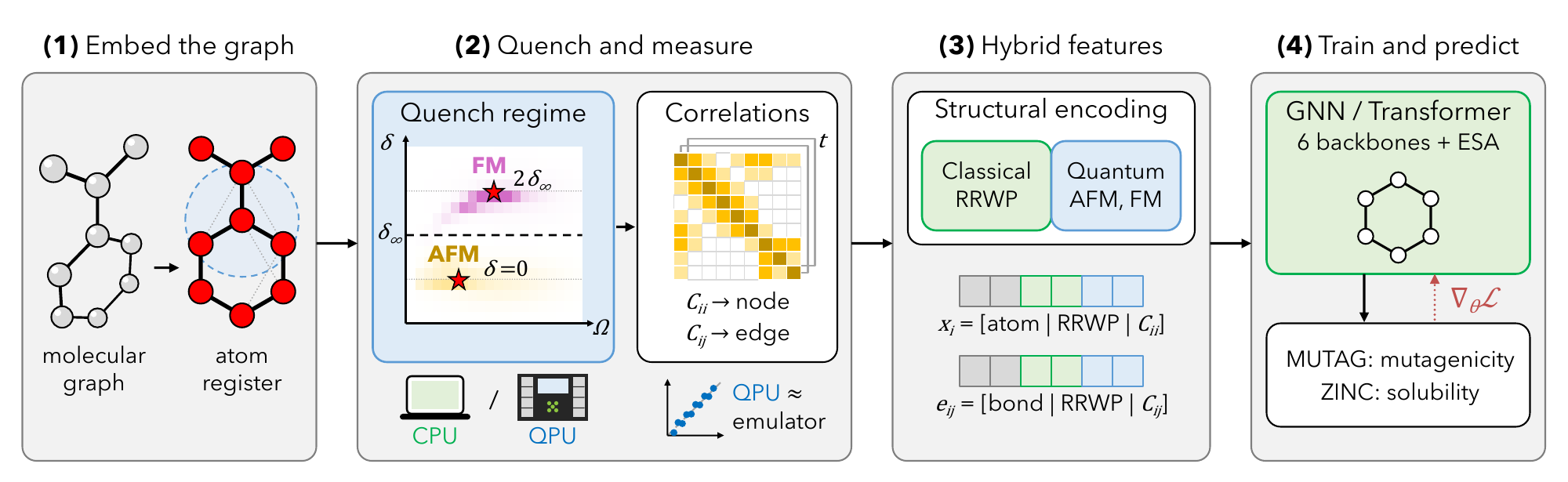}
\caption{ \textbf{Pipeline.}
(1)~Embed the graph in an atom register;
(2)~Run a quench in the antiferromagnetic (AFM) or ferromagnetic (FM) regime and measure the correlations $C_{i,j}$;
(3)~Diagonal/off-diagonal elements $C_{i,j}$ serve as node/edge features and are combined with classical RRWP;
(4)~Train GNNs and ESA on the combined encodings to predict molecular properties.}
\vspace{-0.7cm}
\label{fig:master}
\end{figure}

In this work, using quantum correlations as \emph{quantum graph features} (QGFs) we address three open questions.
First, what intrinsic topological and
spectral information do QPU-derived correlations carry, independently
of any downstream neural architecture?
Second, do they consistently improve GNNs across various architectures, rather than a single favorable backbone?
Third, do they remain useful when produced by actual hardware involving noise sources from shot noise, over decoherence, to state-preparation-and-measurement (SPAM) errors?
We focus on MUTAG and ZINC~\citep{MUTAG_1991,Sterling2015}, two standard molecular graph
benchmarks spanning classification and regression, whose small
molecular graphs largely admit two-dimensional unit-disk embeddings
compatible with neutral-atom registers.
In our datasets this covers $100\%$ of MUTAG and $95\%$ of \initZINC.

\textbf{Our contributions.}
First,
using two physically motivated operating regimes of the graph-encoded Rydberg Hamiltonian, antiferromagnetic (AFM) and ferromagnetic (FM) [Sec.~\ref{subsec:Pulse-shaping}], we introduce an intrinsic benchmark of $15$ topological and spectral prediction tasks evaluated with permutation-invariant probes, showing that hybrid quantum--classical descriptors match or surpass purely classical baselines based on RRWP, RWSE and Laplacian eigenvectors, with AFM and FM correlations emphasizing different structural aspects of the graph [Sec.~\ref{subsec:intrinsic-method}].
Second, regression analyses indicate that QGFs are not accurately recoverable from the classical descriptors considered here [Sec.~\ref{subsec:predictability_of_quantum_features}]
, even after adding ring, distance, register-geometry and restricted physical descriptors.
Third, we show that concatenating QGFs with molecular features yields small but consistent gains across six message-passing, attention-based, and graph-transformer backbones, including GCN~\citep{kipf2017semi}, GINE~\citep{hu2020strategies}, GeneralConv~\citep{li2020deepergcn}, GAT~\citep{velickovic2018graph}, Transformer~\citep{vaswani2017attention,shi2021masked}, and our customized graph transformer (GT) [Sec.~\ref{subsec:architectures}]; we further validate the trend on ESA~\citep{buterez2025end}, a graph architecture outside the standard transformer family [Sec.~\ref{subsec:esa}].
Fourth, we execute the full feature-extraction pipeline on a neutral-atom QPU for all $188$ MUTAG graphs at four measurement time steps.
Despite the unavoidable noise in the hardware, the QPU-derived compressed quantum features match their emulated counterparts within $0.01$ F$_1$ on all six backbones, and the full time-resolved features within $0.014$ F$_1$ on five of six [Sec.~\ref{subsec:architectures}].
Lastly, we introduce geometry-augmented baselines, atom-placement perturbations [App.~\ref{app:placement}] and a synthetic family of ladder graphs where QGF bring a performance boost over RRWP outside of cross-fold noise [Sec.~\ref{sec:controls}].
Together, these results establish QPU-derived correlations as hardware-validated structural graph descriptors that complement, rather than replace, classical positional encodings.
The complete workflow is presented in Fig.~\ref{fig:master}.

\vspace{-0.1cm}
\section{Quantum graph feature extraction \label{sec:quantum-feature-extraction}}
\vspace{-0.1cm}

To obtain QGFs, we encode each graph into a graph-dependent Rydberg Hamiltonian,
let a fixed initial product state evolve under this Hamiltonian for a chosen time,
after which we measure the desired observable: the occupation-based correlation.

\subsection{Rydberg Hamiltonian}\label{subsec:Rydberg-Hamiltonian}

A neutral-atom register of $N$ qubits encoded in a ground and a Rydberg state evolves under the
analog Rydberg Hamiltonian~\citep{Gallagher_1994, Browaeys_2020, Saffman_2016, Labuhn_2016, Lukin_2017}
\begin{align}
\hat{\mathcal{H}}(\mathbb{J},\Omega,\delta)
=\sum_{i<j} J_{ij}\,\hat{n}_{i}\hat{n}_{j}
+\hbar\sum_{i}\Big[(\Omega/2)\,\hat{\sigma}_{i}^{x}-\delta\,\hat{n}_{i}\Big],
\label{eq:Rydberg_Ham}
\end{align}
with the Rabi frequency $\Omega$, the global detuning $\delta$, and the positive van-der-Waals coupling strength $J_{ij}=C_{6}/r_{ij}^{6}$ set by the inter-atomic distance $r_{ij}$, and the Rydberg coefficient $C_{6}$, see App.~\ref{app:Ising_Hamiltonian} for details.
We write $(\mathbb{J})_{ij}=J_{ij}$ for the coupling matrix.
By design, the QPU is initialized in the all-zero product state $\left|\boldsymbol{0}\right\rangle \equiv\left|0\right\rangle ^{\otimes N}$, i.e. $\left|\psi\left(0\right)\right\rangle =\left|\boldsymbol{0}\right\rangle$.
The implemented quantum-feature extraction scheme focuses on a specific observable: the occupation-based connected correlation matrix $C_{i,j}$ defined by
\begin{align}
C_{i,j}(t)&=\langle\hat{n}_{i}(t)\,\hat{n}_{j}(t)\rangle-\langle\hat{n}_{i}(t)\rangle\langle\hat{n}_{j}(t)\rangle & &\in[-1/4,\,1/4].
\label{eq:obs_C}
\end{align}
Due to symmetry, it contains $M=N(N+1)/2$ real-valued and independent entries.
Importantly, the operator $\hat{n}_{i}$ is diagonal in the computational basis such that $C_{i,j}$ can be estimated efficiently from projective readout.
Diagonal entries $C_{i,i}(t)$ effectively encode the per-site occupations $n_{i}(t)=\langle\hat{n}_{i}(t)\rangle$.
The QGFs are defined by the entries of the correlation matrices: diagonal entries provide node-level features, whereas off-diagonal entries provide pairwise edge-level features for the models studied here.
To obtain these features, we embed graphs into the QPU register, as described next.

\subsection{Graph embedding}\label{subsec:Graph_Embedding}

We focus on molecular graphs drawn from two established benchmark datasets: MUTAG (188 graphs, binary mutagenicity) and \initZINC (12{'}000 graphs, constrained solubility regression).
Each molecule is represented as an attributed graph $\mathcal{G}=(\mathcal{V},\mathcal{E},\mathcal{L}_\mathcal{V},\mathcal{L}_\mathcal{E})$ with nodes (atoms) $\mathcal{V}$, edges (chemical bonds) $\mathcal{E}$, and attributes $\mathcal{L}_\mathcal{V}$ and $\mathcal{L}_\mathcal{E}$ for nodes and edges, respectively~\citep{Djellabi_2026}.
To encode a graph into the Rydberg Hamiltonian of Eq.~(\ref{eq:Rydberg_Ham}), we assign each graph node to a qubit in the QPU register plane.
The atomic positions are chosen such that pairs of atoms corresponding to graph edges interact strongly, while non-edge pairs are separated far enough that their residual van der Waals interactions are weak.
The resulting graph-dependent coupling matrix is
\begin{equation}
\mathbb{J}_{ij}^{\mathcal{G}}=
\begin{cases}
J_{\mathrm{NN}} & (i,j)\in\mathcal{E},\\
J_{ij}\ll J_{\mathrm{NN}} & (i,j)\notin\mathcal{E},
\end{cases}
\label{eq:graph_dependent_J}
\end{equation}
where $J_{\mathrm{NN}}=C_{6}/r_{\mathrm{NN}}^{6}$ is the target nearest-neighbor coupling strength set by the edge distance $r_{\mathrm{NN}}$.
Unless stated otherwise, we fix $r_{\mathrm{NN}}=5.01\,\mathrm{\mu m}$ in the emulations, corresponding to $J_{\mathrm{NN}}/\hbar\approx54.75\,\mathrm{rad}/\mu\mathrm{s}$.

This embedding strategy is naturally related to unit-disk graphs~\citep{CLARK1990165}.
Recognizing whether an arbitrary graph admits a unit-disk representation is NP-hard~\citep{Kuhn_2004, Breu_1998}.
Molecular graphs, however, originate from physical constraints and thus, they often admit a geometric realization consistent with the scheme of Eq.~(\ref{eq:graph_dependent_J}); the primary limitation in current neutral-atom devices is that the programmable register is effectively two-dimensional.
We therefore restrict the datasets to molecules that can be embedded in two dimensions: all $188$ MUTAG graphs and $11{,}368$ of \initZINC ($95\%$; train/val/test $9{,}475/946/947$), denoted \ZINC.
The atomic layouts are computed using a force-based placement method following standard graph-drawing principles, \texttt{BLaDE}~\citep{Blade}.
Here, we use a purely structural embedding: neither node attributes $\mathcal{L}_{\mathcal{V}}$ nor edge attributes $\mathcal{L}_{\mathcal{E}}$ are encoded, e.g., molecular bond types are not reflected in the coupling strengths $J_{ij}$.
QGFs are therefore descriptors of a specified graph-to-register protocol: small positional noise leaves them only moderately perturbed (Spearman $\rho = 0.82$ at 2\% noise), whereas alternative valid layouts yield measurably different features (App.~\ref{app:placement}).

\begin{wrapfigure}[]{r}{0.5\textwidth}
\centering
\vspace{-0.5cm}
\includegraphics[width=0.5\columnwidth]{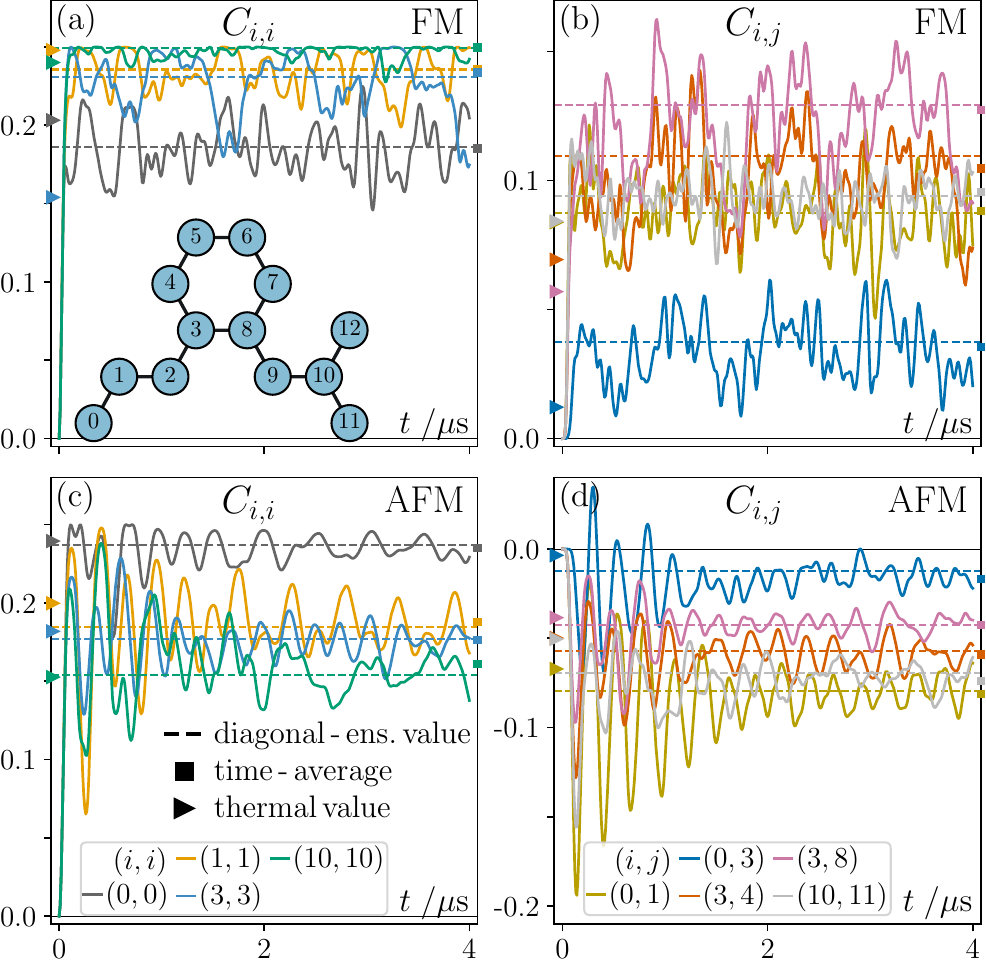}
\caption{
Time evolution of the correlation matrix $C_{i,j}$ (\ref{eq:obs_C}) for the indicated molecular graph from the \initZINC dataset,  under the two characteristic Hamiltonians of Eq.~(\ref{eq:H_AFM_FM}).
Panels (a)--(b) show the FM setting $\hat{\mathcal{H}}_{\mathrm{FM}}^{\mathcal{G}}$, while panels (c)--(d) show the AFM setting $\hat{\mathcal{H}}_{\mathrm{AFM}}^{\mathcal{G}}$.
The left panels display diagonal elements $C_{i,i}$, whereas the right panels show off-diagonal elements $C_{i,j}$ with $i\neq j$.
The dashed lines indicate the diagonal-ensemble values $C_{i,j}^{\mathrm{diag}}$,
square markers on the right indicate finite-time averages over $[0,4\,\mu\mathrm{s}]$, and triangular markers on the left indicate thermal expectation values $C_{i,j}^{\mathrm{th}}$ (see App.~\ref{app:quench_parameter_selection} for details).
}
\label{fig:time_evolution}
\vspace{-0.99cm}
\end{wrapfigure}

\subsection{Quench-parameter selection: AFM and FM regimes}
\label{subsec:Pulse-shaping}

Having embedded a graph into the Rydberg Hamiltonian $\hat{\mathcal{H}}^{\mathcal{G}}=\hat{\mathcal{H}}\left(\mathbb{J}^{\mathcal{G}},\Omega,\delta\right)$, we select two constant quench settings $(\Omega,\delta)$ that generate distinct correlation patterns.
The first setting produces antiferromagnetic (AFM)-like correlations, while the second produces ferromagnetic (FM)-like correlations.
In both cases, the parameters are chosen to yield pronounced connected density correlations $C_{i,j}$ [Eq.~(\ref{eq:obs_C})], which are used as quantum features in the downstream graph-learning task.
Our parameter-selection strategy is guided by thermalization arguments \citep{DAlessio_2016}, detailed in App. \ref{app:thermalization} and applied to the Rydberg Hamiltonian in App.~\ref{app:quench_parameter_selection}.
The two quench settings used in this work define the AFM-like and FM-like
Hamiltonians [cf. Eqs.~(\ref{eq:Rydberg_Ham}), (\ref{eq:graph_dependent_J})]
\begin{align}
&\hat{\mathcal{H}}_{\mathrm{AFM}}^{\mathcal{G}}  =\hat{\mathcal{H}}\left(\mathbb{J}^{\mathcal{G}},\,\Omega=0.3\,J_{\mathrm{NN}}/\hbar,\,\delta=0\right), \nonumber\\
&\hat{\mathcal{H}}_{\mathrm{FM}}^{\mathcal{G}} =\hat{\mathcal{H}}\left(\mathbb{J}^{\mathcal{G}},\,\Omega=0.5\,J_{\mathrm{NN}}/\hbar,\,\delta=2\delta_\infty^{\mathcal{G}}\right),
\label{eq:H_AFM_FM}
\end{align}
where $\delta_{\infty}^{\mathcal{G}} = \sum_{i\neq j}J_{ij}^{\mathcal{G}} / (4N\hbar)$.
The value $\delta_\infty^{\mathcal{G}}$ corresponds to the detuning for which
the all-zero initial state lies at infinite effective temperature.
The choice $\delta=2\delta_\infty^{\mathcal{G}}$ cancels the spatially averaged
longitudinal field and therefore places the FM setting near a field-balanced
transverse-field Ising regime, in the vicinity of a finite-size thermal
crossover.
Figure~\ref{fig:time_evolution} confirms the expected qualitative behavior on a small molecular graph.
In the FM setting, neighboring atoms develop positive off-diagonal correlations, whereas in the AFM setting the corresponding correlations are negative.
This distinction follows from the sign of the effective interaction, set by $\delta$ relative to $\delta_\infty^{\mathcal{G}}$, rather than from finely tuned pulse values:
both working points, AFM and FM, are chosen without task-level tuning, inside a broad region of pronounced correlations (App.~\ref{app:quench_parameter_selection}).

\subsection{Quantum graph feature pipeline: emulation and hardware}\label{subsec:pipeline}

For both datasets we emulate the quench dynamics under $\hat{\mathcal{H}}_{\mathrm{AFM}}^{\mathcal{G}}$ and $\hat{\mathcal{H}}_{\mathrm{FM}}^{\mathcal{G}}$ [Eq.~(\ref{eq:H_AFM_FM})] with Pasqal's open-source emulators~\citep{quelle2025emulators}, interfaced with Pulser~\citep{silverio2022pulser}, which is also used for the hardware runs, ensuring consistency between emulated and hardware pulses.
For the emulated benchmark, we record the correlation elements $C_{i,j}(t)$ [Eq.~(\ref{eq:obs_C})] at $N_{t}=20$ equally spaced time steps $t_{\nu}\in[0.1,4.0]\,\mu\mathrm{s}$.
At each time step, the observables are sampled $N_{\mathrm{shots}}=1000$ times, which---based on a central-limit-theorem-derived budget (App.~\ref{annex:shot_number})---is sufficient to resolve all correlation elements at $95\%$ simultaneous confidence for all MUTAG graphs and for \ZINC graphs with $N \le 30$ (App.~\ref{annex:shot_number}).
The derived quantum descriptors are reused as fixed inputs across all downstream models.
For each graph and each quench-pulse setting, AFM and FM, we concatenate the sampled correlation elements to form distinct descriptor sets
\begin{align}
\boldsymbol{\mathcal{C}} &=\bigl[C_{i,j}(t_{\nu})\bigr]_{\nu=1}^{N_t}\!\in\! \mathbb{R}^{MN_t},
&\boldsymbol{\mathcal{C}}^{(1)}_{i} &=\bigl[C_{i,i}(t_{\nu})\bigr]_{\nu=1}^{N_t}\!\in\! \mathbb{R}^{N_t},
&
\boldsymbol{\mathcal{C}}^{(2)}_{ij}&=\bigl[C_{i,j\neq i}(t_{\nu})\bigr]_{\nu=1}^{N_t}\!\in\! \mathbb{R}^{N_t},
\label{eq:quantum_descriptors}
\end{align}
of indicated dimensions, which are interchangeably used in the ML protocols.
For example, if a model permits, \Cone can be used for node-feature injection while \Ctwo can be injected in edge features.
Depending on which quench-pulse setting the downstream models are trained on, their respective method is labeled by AFM or FM.
Additionally, assuming that much of the graph information is encoded in the time average [see App.~\ref{app:thermalization}], we also train the model on the four-dimensional  summary statistics $(\mathrm{mean},\mathrm{variance},\mathrm{minimal\ value},\mathrm{maximal\ value})$, indicated later via AFM$_{\mathrm{stat}}$ and FM$_{\mathrm{stat}}$, respectively.
A summary of all feature sets and their dimensionalities is given in App.~\ref{app:features} (Table~\ref{tab:feature-notation}).
Emulation details are given in App.~\ref{app:emu-budget}.
For the hardware run, the full pipeline is executed on a Pasqal Fresnel neutral-atom QPU~\citep{pasqal2025fresnel} for all 188 MUTAG graphs at four selected time steps due to budget and time restrictions.
The required QPU wall time for the chosen setting is $\sim\!7.5$ QPU-days.
The four selected time steps are determined based on a principal component analysis, see App.~\ref{app:pca-analysis}.
Despite SPAM, decoherence and shot noise, the per-graph Spearman concordance between hardware and emulator features is high, with medians $\rho_{s}=0.924$ for diagonal elements \Cone and $\rho_{s}=0.870$ for off-diagonal elements \Ctwo, see App.~\ref{app:emu-qpu-concordance}.
Models trained with hardware correlations are indicated via prefix HW-, for example HW-AFM.

\subsection{Classical baselines for graph structural features}
\label{subsec:classical_baselines}

We mainly use two classical baselines: (i) RRWP~\citep{ma2023graph}
involving transition-matrix powers up to order $k_\mathrm{max}=20$
to match $N_t=20$ and (ii) Laplacian~\citep{dwivedi2021generalization} with four leading eigenvectors to match dimension of the summary statistics, e.g. AFM$_{\mathrm{stat}}$.
We choose RRWP over other positional encoding methods such as RWPE~\citep{dwivedi2022graph}, SignNet~\citep{lim2023sign}, GRIT~\citep{ma2023graph}, because RRWP underpins GRIT on the one hand, and on the other hand, it matches or exceeds the alternatives on the studied molecular benchmarks.
RWSE with ring-membership features are used only for the ESA model, since this feature set gives the best performance for that particular architecture~\citep{buterez2025end}.
Finally, we use two control baselines: (i) Random (Gaussian pseudo-features of matched dimension, fixed per graph) controls for input-dimensionality gains, and (ii) \textit{No data} (atom/bond attributes only) gives each backbone's bare performance and is restricted to the downstream setting.
To control for information beyond random walks, including the register geometry seen by the quantum dynamics, we also use a strengthened baseline RRWP$+$Laplacian$+$Rings$+$Geometry, which adds ring-membership descriptors, the inter-atomic distances $r_{ij}$ and the couplings $J_{ij}$ (Sec.~\ref{sec:controls}).
All features are z-scored using training-split statistics.

\vspace{-0.2cm}
\section{Intrinsic Information Benchmark}
\label{sec:intrinsic-information-benchmark}

We first measure how much graph-theoretic information each descriptor family encodes
intrinsically [Sec.~\ref{subsec:intrinsic-method}], then test whether QGFs can be reconstructed from classical
descriptors [Sec.~\ref{subsec:predictability_of_quantum_features}].

\subsection{Predicting graph properties}
\label{subsec:intrinsic-method}

We design a lightweight probing benchmark to measure graph-theoretic information encoded in each descriptor while minimizing model-specific inductive biases.
It spans 15 tasks, formally defined in App.~\ref{app:benchmark-targets}: node-level tasks (degree, betweenness, eccentricity, bridge-node detection, smallest cycle through node), edge-level tasks (link prediction, edge betweenness, bridge-edge detection, local connectivity, smallest cycle through edge), and graph-level tasks (number of cycles, diameter, density, Wiener index, algebraic connectivity).

The probe is a DeepSet~\citep{zaheer2017deep} with a shared encoder $\phi$ and predictor $\rho$.
Node- and edge-level targets are processed independently through $\rho \circ \phi$.
For graph-level targets, node features are enriched with $\boldsymbol{\mathcal{C}}^{(1)}_{i}$ [Eq.~(\ref{eq:quantum_descriptors})] and the row-wise mean of $\boldsymbol{\mathcal{C}}^{(2)}_{ij}$ over $j$, encoded, sum-pooled, and passed to $\rho$.
The probe has no message passing, attention, or learned graph traversal; any recovered signal must already be present in the input encoding.
It is diagnostic rather than competitive, revealing which structural targets are naturally accessible from each feature family.
Training details are deferred to App.~\ref{app:benchmark-targets}.

Table~\ref{tab:pe-benchmark-main} reports results on \ZINC for ten feature sets: classical descriptors (RRWP, Laplacian, RRWP+Laplacian), quantum descriptors (FM, AFM, FM$_{\mathrm{stat}}$, AFM$_{\mathrm{stat}}$; see Sec.~\ref{subsec:pipeline}), and three RRWP-based hybrids.
A random-feature control, per-Hamiltonian mean variants, and MUTAG results are given in App.~\ref{app:full-benchmark} and App.~\ref{app:mutag}.
The random control is beaten by every structural feature by a large margin on non-trivial tasks, while link prediction and bridge-edge detection are close to saturation and retained only as sanity checks.

\textbf{Quantum correlations carry strong local information and complement RRWP};
RRWP alone is a strong classical baseline, consistent with its use as a positional encoding~\citep{ma2023graph}.
Quantum correlations provide a strong signal on selected local tasks: on degree prediction, AFM alone reaches $1.26$ ($\times 10^{-2}$ MAE), clearly ahead of RRWP+Laplacian ($6.98$) and far ahead of RRWP ($14.4$).
Combined with RRWP, this signal yields clear hybrid gains: degree prediction improves from $14.4$ to $1.18$ and $1.25$ ($\times 10^{-2}$ MAE, statistically tied) when AFM and then FM+AFM are added.
A monotone pattern holds for local connectivity ($10.0 \to 6.07 \to 5.14$), smallest cycle through a node ($0.73 \to 0.36 \to 0.28$), and Wiener index ($10.8 \to 8.81 \to 7.24$, $\times 10^{-2}$).
These gains span node-, edge-, and graph-level probes, showing that QGFs provide information compatible with, but not redundant with, random-walk transition probabilities.

\textbf{AFM and FM correlations probe different structural regimes.}
RRWP+AFM is stronger than RRWP+FM on local neighbourhood and cycle-related tasks, including degree, betweenness, and smallest cycle through a node or edge (e.g., node cycle: $0.36$ vs.~$0.48$; edge cycle: $7.62$ vs.~$10.0$, $\times 10^{-2}$).
Conversely, RRWP+FM is stronger on global-distance targets, reaching $1.48$ MAE on eccentricity against $1.77$ for RRWP+AFM, and close to RRWP+AFM on diameter.
This suggests, without formally establishing, that AFM-like dynamics are more sensitive to local frustration and cycle structure, whereas FM-like dynamics capture more coherent graph-scale correlation propagation.
RRWP+FM+AFM inherits both strengths and is best, tied with the best, or second-best on most non-saturating structural tasks.
The few remaining cases correspond to global or spectral targets better captured by RRWP+Laplacian, or by the simpler RRWP+AFM hybrid.
Statistic-compressed variants impact and hardware results for MUTAG are discussed in resp. App.~\ref{app:intrinsic-compression} and App.~\ref{app:intrinsic-hardware}.

\begin{table}[t]
\centering
\caption{%
Intrinsic information benchmark on \ZINC.
Means over 5-fold cross-validation are reported; standard deviations over folds are omitted for readability and provided in Table~\ref{tab:pe-benchmark} (Appendix~\ref{app:full-benchmark}).
Metrics are MAE ($\downarrow$) for regression tasks and accuracy ($\uparrow$) for classification tasks.
\mbox{$\best{\text{Best}}$} and \mbox{$\second{\text{2nd best}}$} indicate the best and runner-up feature sets among the columns shown; entries marked \mbox{$\tied{}$} are tied with the best.
R: RRWP ($d=20$); L: Laplacian ($d=4$); F: FM; A: AFM.
Quantum columns FM$_{\mathrm{stat}}$ and AFM$_{\mathrm{stat}}$ use the four-statistic compression ($d=4$), whereas FM and AFM use the full time series ($d=20$).
}
\label{tab:pe-benchmark-main}
\footnotesize
\setlength{\tabcolsep}{3pt}
\resizebox{0.9\columnwidth}{!}{%
\begin{tabular}{l ccc cccc ccc}
\toprule
 & \multicolumn{3}{c}{Classical} & \multicolumn{4}{c}{Quantum} & \multicolumn{3}{c}{Hybrid} \\
  \cmidrule(lr){2-4} \cmidrule(lr){5-8} \cmidrule(lr){9-11}
Task & R & L & R+L & F$_{\mathrm{stat}}$ & F & A$_{\mathrm{stat}}$ & A & R+F & R+A & R+F+A \\
\midrule
\multicolumn{11}{l}{\textit{Node Prediction}}\\ [-0.5ex]
Bridge Node$^\uparrow$ & $0.93$ & $0.76$ & $0.95$ & $0.74$ & $0.83$ & $0.85$ & $0.91$ & $0.97$ & $\tied{\best{0.98}}$ & $\tied{\best{0.98}}$  \\
\specialrule{0.4pt}{1pt}{1pt}
Degree$^\downarrow$ ($\times 10^{2}$) & $14.4$ & $46.7$ & $6.98$ & $30.5$ & $18.7$ & $9.07$ & $\tied{1.26}$ & $4.71$ & $\tied{1.18}$ & $\tied{1.25}$  \\
Betweenness$^\downarrow$ ($\times 10^{2}$) & $7.77$ & $12.5$ & $4.76$ & $9.10$ & $7.17$ & $10.1$ & $7.11$ & $4.33$ & $\second{3.90}$ & $\best{3.31}$  \\
Eccentricity$^\downarrow$ & $1.88$ & $1.51$ & $\best{1.08}$ & $1.90$ & $1.65$ & $2.02$ & $1.91$ & $1.48$ & $1.77$ & $\second{1.47}$  \\
Min Cycle (Node)$^\downarrow$ & $0.73$ & $1.68$ & $0.60$ & $1.80$ & $1.42$ & $1.40$ & $0.88$ & $0.48$ & $\second{0.36}$ & $\best{0.28}$  \\
\addlinespace[2pt]\midrule\addlinespace[1pt]
\multicolumn{11}{l}{\textit{Edge Prediction}}\\ [-0.5ex]
Link Pred.$^\uparrow$ & $\tied{1.00}$ & $0.94$ & $\tied{1.00}$ & $0.99$ & $0.99$ & $\tied{1.00}$ & $\tied{1.00}$ & $\tied{1.00}$ & $\tied{1.00}$ & $\tied{1.00}$ \\
Bridge Edge$^\uparrow$ & $\tied{0.99}$ & $0.98$ & $\tied{0.99}$ & $0.97$ & $0.98$ & $0.97$ & $0.99$ & $\tied{0.99}$ & $\tied{0.99}$ & $\tied{1.00}$  \\
\specialrule{0.4pt}{1pt}{1pt}
Edge Betw.$^\downarrow$ ($\times 10^{3}$) & $11.6$ & $20.6$ & $\second{6.09}$ & $13.0$ & $9.89$ & $9.76$ & $8.61$ & $6.69$ & $6.13$ & $\best{5.36}$  \\
Local Conn.$^\downarrow$ ($\times 10^{2}$) & $10.0$ & $13.9$ & $7.39$ & $13.9$ & $12.6$ & $15.2$ & $11.3$ & $6.83$ & $\second{6.07}$ & $\best{5.14}$  \\
Min Cycle (Edge)$^\downarrow$ ($\times 10^{2}$) & $13.3$ & $42.4$ & $10.2$ & $23.0$ & $19.4$ & $23.1$ & $13.2$ & $10.0$ & $\tied{\best{7.62}}$ & $\tied{\best{7.17}}$  \\
\addlinespace[2pt]\midrule\addlinespace[1pt]
\multicolumn{11}{l}{\textit{Graph Prediction}}\\ [-0.5ex]
Num Cycles$^\downarrow$ ($\times 10^{2}$) & $6.68$ & $27.4$ & $5.90$ & $18.4$ & $15.3$ & $16.4$ & $9.85$ & $\tied{4.83}$ & $\tied{4.48}$ & $\tied{4.67}$  \\
Diameter$^\downarrow$ & $0.49$ & $0.69$ & $0.47$ & $0.71$ & $0.65$ & $0.75$ & $0.60$ & $\second{0.36}$ & $0.37$ & $\best{0.32}$  \\
Density$^\downarrow$ ($\times 10^{3}$) & $\tied{1.03}$ & $2.48$ & $\tied{1.14}$ & $1.86$ & $1.70$ & $1.72$ & $1.86$ & $\tied{1.21}$ & $\tied{1.10}$ & $\tied{1.25}$  \\
Wiener Index$^\downarrow$ ($\times 10^{2}$) & $10.8$ & $19.2$ & $9.61$ & $16.4$ & $14.1$ & $19.5$ & $15.0$ & $\tied{\best{7.13}}$ & $8.81$ & $\tied{\best{7.24}}$  \\
Alg. Connect.$^\downarrow$ ($\times 10^{3}$) & $2.63$ & $5.59$ & $\tied{\best{2.22}}$ & $5.46$ & $5.21$ & $6.30$ & $5.32$ & $2.70$ & $\tied{\best{2.34}}$ & $2.38$  \\
\bottomrule
\end{tabular}
}
\vspace{-0.3cm}
\end{table}

\subsection{A ceiling on quantum feature reconstruction with classical descriptors}
\label{subsec:predictability_of_quantum_features}

\begin{wrapfigure}[]{r}{0.5\textwidth}
\centering
\vspace{-0.8cm}
\includegraphics[width=0.5\columnwidth]{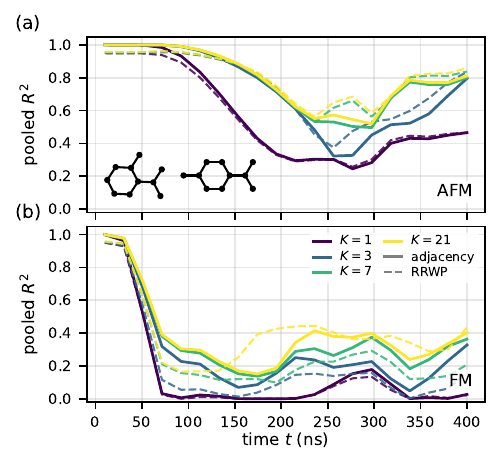}
\vspace{-0.6cm}
\caption{
Linear recovery of quantum features from adjacency powers and RRWP features including matrix powers up to order $K$, for the two indicated graphs, and the two quench-parameter settings (a) AFM and (b) FM [Eq.~(\ref{eq:H_AFM_FM})].
After an early-time period, quantum features are no longer fully recovered.
}
\label{fig:recovery-qf}
\vspace{-0.3cm}
\end{wrapfigure}

We empirically assess whether quantum features can be reconstructed from classical graph descriptors.
For this purpose we construct a regression target $y(t)$ that, at a given time step $t$, combines the off-diagonal correlation elements $C_{i,j}(t)$ from all graphs in the dataset via a specific pooling scheme, see App.~\ref{app:walk-ceiling-setup}.
We test whether $y(t)$ is reducible to two classical pair-feature families up to a chosen order $K$: (i) walk-powers, i.e., powers of the adjacency matrix $A$, and (ii) RRWP, i.e., powers of the degree-normalized transition matrix $D^{-1}A$.
For both physical regimes (AFM and FM), Figure~\ref{fig:recovery-qf}
illustrates on a toy dataset ($N_\mathrm{data}=2$)
that quantum features at early times are well approximated by linear combination of powers of RRWP or adjacency ($R^2 \approx1$), whereas at later times ($t \gtrsim 200\,\mathrm{ns}$ for AFM and $t \gtrsim 100\,\mathrm{ns}$ for FM) this is not the case anymore.

The results on the full MUTAG and \ZINC datasets are provided in App.~\ref{app:walk-ceiling-setup}, and summarized in the following.
Under a pooled linear probe, the explained variance of $y(t)$ saturates at the measurement times: $R^2$ reaches $\sim 0.83$ on MUTAG and $\sim 0.78$ on \ZINC at $K=21$, improving by at most $2 \%$ up to $K=50$ (see Table~\ref{tab:walk-ceiling} and Figs.~\ref{fig:linear-mutag},\ref{fig:linear-zinc}).
To rule out a linearity limitation, we fit kernel ridge (RBF) and random forest on the same features and score them by 5-fold held-out $R^2$; both close at most $7$--$8$ percent points of the gap for walk powers on MUTAG at $K=21,\,t=1536$\,ns (graph-level folds), leaving a residual of $\sim 9\%$ on AFM and $\sim 22\%$ on FM
that is not recovered by any nonlinear function of adjacency or RRWP (see Table~\ref{tab:nonlinear}).
Augmenting the design with the first $K$ Laplacian eigenpairs additionally lifts the FM result to $0.85$, identifying the FM beyond-walk content as partly spectral, while the AFM residual is nearly unchanged ($\le 1\%$) and remains beyond any walk- or spectrum-derived feature we tested (see Table~\ref{tab:laplacian}).
We further expand the basis with ring, heat-kernel, geodesic and effective-resistance descriptors, the register distances $r_{ij}$ and couplings $J_{ij}$, and restricted physical approximations of the same dynamics (classical-spin dTWA~\cite{wootters1987wigner, kunimi2021performance}, exact two-atom and mean-field correlations).
The unexplained variance then remains $8.3\%$ for AFM and $14.0\%$ for FM (App.~\ref{app:residual}).
We therefore describe it as information not captured by the tested classical, geometric and restricted physical descriptors, rather than as purely quantum information.

\section{Downstream evaluation across GNN backbones}
\label{sec:gnn-downstream}
\subsection{Architectures, feature injection and results}
\label{subsec:architectures}

To test robustness across downstream models, we evaluate six backbones: 5 from PyTorch Geometric~\citep{fey2019fast}---GCN~\citep{kipf2017semi}, GAT~\citep{velickovic2018graph}, GINE~\citep{hu2020strategies}, Transformer~\citep{shi2021masked}, GeneralConv~\citep{li2020deepergcn}---plus a custom dense graph transformer (GT).
Diagonal features $\boldsymbol{\mathcal{C}}^{(1)}_{i}$ [Eq.~(\ref{eq:quantum_descriptors})] are concatenated to the node embedding (or, in GT, projected and added residually); off-diagonal features $\boldsymbol{\mathcal{C}}^{(2)}_{ij}$ enter through whichever pairwise interface each backbone exposes (edge weights for GCN, edge attributes for the four MPNN/attention layers, dense per-head attention biases for GT). Models details are given in App.~\ref{app:GNN}.
Architectural details of the GT, which consumes the dense tensor \Ctwo for all node pairs, are given in App.~\ref{app:model-details}.

To control for model capacity, we approximately match the number of trainable parameters across backbones separately for each dataset. We keep the training protocol fixed across all feature configurations and do not perform per-feature hyperparameter tuning, thereby prioritizing a clean comparison of features over peak predictive performance.
Hyperparameters are reported in App.~\ref{app:GNN}.
The choice of number of trainable parameters is justified by the model-size ablation in App.~\ref{app:gnn-ablation}.
It should be noted that increasing model size after a certain point does not increase performance for RRWP, motivating the addition of supplemental features as a source of improvement.

\begin{table*}
\vspace{-1.4cm}
\caption{%
Comparison of GNN architectures and feature sets on MUTAG.
Mean macro F$_1$ ($\uparrow$) $\pm$ standard deviation over 10 stratified cross-validation folds $\times$ 5 seeds is reported.
\mbox{$\best{\text{Best}}$} and \mbox{$\second{\text{2nd best}}$} indicate the best and runner-up feature sets in each column.
}
\label{tab:pe-comparison-mutag}
\centering
\resizebox{0.88\textwidth}{!}{
\begin{tabular}{l c c c c c c}
\toprule
\multirow{2}{*}{Method} & \multicolumn{6}{c}{MUTAG (F$_1$ macro $\uparrow$)} \\
\cmidrule(lr){2-7}
 & GT & GCN & GAT & GINE & Transformer & General \\
\midrule
\multicolumn{7}{l}{\textit{Baseline}} \\
  No Data & $0.690{_{\pm 0.104}}$ & $0.652{_{\pm 0.134}}$ & $0.734{_{\pm 0.103}}$ & $0.785{_{\pm 0.098}}$ & $0.781{_{\pm 0.084}}$ & $0.823{_{\pm 0.065}}$ \\
\midrule
\multicolumn{7}{l}{\textit{Classical}} \\
  Laplacian & $0.807{_{\pm 0.065}}$ & $0.800{_{\pm 0.099}}$ & $0.787{_{\pm 0.090}}$ & $0.818{_{\pm 0.106}}$ & $0.802{_{\pm 0.072}}$ & $0.821{_{\pm 0.079}}$  \\
  RRWP & $0.854{_{\pm 0.060}}$ & $0.849{_{\pm 0.081}}$ & $0.862{_{\pm 0.078}}$ & $0.861{_{\pm 0.071}}$ & $0.869{_{\pm 0.066}}$ & $0.868{_{\pm 0.080}}$ \\
  RRWP + Laplacian & $\second{0.868}{_{\pm 0.071}}$ & $0.851{_{\pm 0.073}}$ & $0.850{_{\pm 0.078}}$ & $0.843{_{\pm 0.073}}$ & $0.852{_{\pm 0.078}}$ & $0.854{_{\pm 0.073}}$ \\
\midrule
\multicolumn{7}{l}{\textit{Quantum}} \\
  AFM & $0.830{_{\pm 0.054}}$ & $0.817{_{\pm 0.098}}$ & $0.791{_{\pm 0.108}}$ & $0.851{_{\pm 0.093}}$ & $0.848{_{\pm 0.086}}$ & $0.849{_{\pm 0.090}}$ \\
  AFM$_{\text{stat}}$ & $0.806{_{\pm 0.047}}$ & $0.778{_{\pm 0.075}}$ & $0.782{_{\pm 0.102}}$ & $0.809{_{\pm 0.084}}$ & $0.811{_{\pm 0.081}}$ & $0.816{_{\pm 0.094}}$ \\
  FM & $0.851{_{\pm 0.059}}$ & $0.832{_{\pm 0.079}}$ & $0.821{_{\pm 0.082}}$ & $0.844{_{\pm 0.086}}$ & $0.860{_{\pm 0.070}}$ & $0.814{_{\pm 0.083}}$ \\
  FM$_{\text{stat}}$ & $0.841{_{\pm 0.069}}$ & $0.792{_{\pm 0.103}}$ & $0.828{_{\pm 0.077}}$ & $0.834{_{\pm 0.083}}$ & $0.829{_{\pm 0.089}}$ & $0.831{_{\pm 0.078}}$ \\
  AFM + FM & $0.835{_{\pm 0.077}}$ & $0.845{_{\pm 0.074}}$ & $0.836{_{\pm 0.092}}$ & $0.843{_{\pm 0.083}}$ & $0.853{_{\pm 0.075}}$ & $0.815{_{\pm 0.064}}$ \\
\midrule
\multicolumn{7}{l}{\textit{Hybrid}} \\
  RRWP + AFM & $0.855{_{\pm 0.078}}$ & $\best{0.878{_{\pm 0.078}}}$ & $0.862{_{\pm 0.070}}$ & $0.848{_{\pm 0.069}}$ & $0.868{_{\pm 0.079}}$ & $0.867{_{\pm 0.074}}$ \\
  RRWP + FM & $0.862{_{\pm 0.060}}$ & $0.856{_{\pm 0.083}}$ & $0.866{_{\pm 0.082}}$ & $0.847{_{\pm 0.080}}$ & $0.869{_{\pm 0.089}}$ & $0.848{_{\pm 0.065}}$ \\
  RRWP + AFM$_{\text{stat}}$ & $0.854{_{\pm 0.077}}$ & $0.852{_{\pm 0.075}}$ & $0.860{_{\pm 0.078}}$ & $0.846{_{\pm 0.064}}$ & $0.852{_{\pm 0.078}}$ & $0.873{_{\pm 0.077}}$ \\
  RRWP + FM$_{\text{stat}}$ & $0.851{_{\pm 0.070}}$ & $\second{0.867}{_{\pm 0.065}}$ & $\best{0.876{_{\pm 0.075}}}$ & $\second{0.866}{_{\pm 0.070}}$ & $0.870{_{\pm 0.072}}$ & $\second{0.873}{_{\pm 0.082}}$ \\
  RRWP + AFM + FM & $\best{0.874{_{\pm 0.081}}}$ & $0.850{_{\pm 0.090}}$ & $\second{0.872}{_{\pm 0.077}}$ & $\best{0.867{_{\pm 0.075}}}$ & $\best{0.878{_{\pm 0.080}}}$ & $0.859{_{\pm 0.079}}$ \\
  RRWP + AFM$_{\text{stat}}$ + FM$_{\text{stat}}$ & $0.862{_{\pm 0.053}}$ & $0.860{_{\pm 0.076}}$ & $0.868{_{\pm 0.074}}$ & $0.862{_{\pm 0.078}}$ & $\second{0.871}{_{\pm 0.071}}$ & $0.870{_{\pm 0.078}}$ \\
\midrule
\multicolumn{7}{l}{\textit{Quantum hardware}} \\
  HW-AFM & $0.784{_{\pm 0.098}}$ & $0.793{_{\pm 0.096}}$ & $0.796{_{\pm 0.052}}$ & $0.826{_{\pm 0.073}}$ & $0.805{_{\pm 0.075}}$ & $0.823{_{\pm 0.097}}$ \\
  HW-AFM$_{\text{stat}}$ & $0.778{_{\pm 0.086}}$ & $0.767{_{\pm 0.086}}$ & $0.801{_{\pm 0.059}}$ & $0.833{_{\pm 0.055}}$ & $0.810{_{\pm 0.053}}$ & $0.809{_{\pm 0.075}}$ \\

  RRWP + HW-AFM & $0.854{_{\pm 0.074}}$ & $0.837{_{\pm 0.076}}$ & $0.852{_{\pm 0.059}}$ & $0.841{_{\pm 0.059}}$ & $0.854{_{\pm 0.060}}$ & $0.859{_{\pm 0.079}}$ \\
  RRWP + HW-AFM$_{\text{stat}}$ & $0.849{_{\pm 0.072}}$ & $0.852{_{\pm 0.055}}$ & $0.853{_{\pm 0.062}}$ & $0.852{_{\pm 0.059}}$ & $0.861{_{\pm 0.060}}$ & $\best{0.877{_{\pm 0.066}}}$ \\
\bottomrule
\end{tabular}
}
\end{table*}

\begin{table*}
\caption{%
Comparison of GNN architectures and feature sets on ZINC$_{\text{subset}}^{*}$.
Mean MAE ($\downarrow$) $\pm$ standard deviation over 10 seeds (GT) or 5 seeds (other backbones) is reported.
\mbox{$\best{\text{Best}}$} and \mbox{$\second{\text{2nd best}}$} indicate the best and runner-up feature sets in each column, $^\dagger$ indicates a tie.}
\label{tab:pe-comparison-zinc}
\centering
\resizebox{0.88\textwidth}{!}{
\begin{tabular}{l c c c c c c}
\toprule
\multirow{2}{*}{Method} & \multicolumn{6}{c}{ZINC$^{*}_{\text{subset}}$ (MAE $\downarrow$)} \\
\cmidrule(lr){2-7}
 & GT & GCN & GAT & GINE & Transformer & General \\
\midrule
\multicolumn{7}{l}{\textit{Baseline}} \\
  No Data & $0.317{_{\pm 0.018}}$ & $0.591{_{\pm 0.066}}$ & $0.434{_{\pm 0.008}}$ & $0.236{_{\pm 0.041}}$ & $0.339{_{\pm 0.009}}$ & $0.331{_{\pm 0.006}}$ \\
\midrule
\multicolumn{7}{l}{\textit{Classical}} \\
  Laplacian & $0.232{_{\pm 0.020}}$ & $0.478{_{\pm 0.006}}$ & $0.434{_{\pm 0.005}}$ & $0.195{_{\pm 0.012}}$ & $0.243{_{\pm 0.034}}$ & $0.270{_{\pm 0.031}}$  \\
  RRWP & $0.113{_{\pm 0.003}}$ & $0.252{_{\pm 0.006}}$ & $0.223{_{\pm 0.002}}$ & $\tied{0.117}{_{\pm 0.007}}$ & $0.126{_{\pm 0.006}}$ & $0.148{_{\pm 0.004}}$ \\
  RRWP + Laplacian & $0.112{_{\pm 0.004}}$ & $0.278{_{\pm 0.017}}$ & $0.229{_{\pm 0.007}}$ & $\tied{0.117}{_{\pm 0.007}}$ & $\second{0.123}{_{\pm 0.008}}$ & $\best{0.145}{_{\pm 0.005}}$ \\
\midrule
\multicolumn{7}{l}{\textit{Quantum}} \\
  AFM & $0.163{_{\pm 0.004}}$ & $0.339{_{\pm 0.040}}$ & $0.288{_{\pm 0.010}}$ & $0.188{_{\pm 0.002}}$ & $0.191{_{\pm 0.007}}$ & $0.214{_{\pm 0.006}}$ \\
  AFM$_{\text{stat}}$ & $0.168{_{\pm 0.008}}$ & $0.419{_{\pm 0.071}}$ & $0.360{_{\pm 0.024}}$ & $0.192{_{\pm 0.007}}$ & $0.195{_{\pm 0.006}}$ & $0.218{_{\pm 0.009}}$ \\
  FM & $0.183{_{\pm 0.010}}$ & $0.479{_{\pm 0.017}}$ & $0.430{_{\pm 0.005}}$ & $0.248{_{\pm 0.040}}$ & $0.283{_{\pm 0.036}}$ & $0.325{_{\pm 0.005}}$ \\
  FM$_{\text{stat}}$ & $0.183{_{\pm 0.010}}$ & $0.465{_{\pm 0.016}}$ & $0.413{_{\pm 0.004}}$ & $0.208{_{\pm 0.012}}$ & $0.292{_{\pm 0.013}}$ & $0.309{_{\pm 0.005}}$ \\
  AFM + FM & $0.154{_{\pm 0.006}}$ & $0.314{_{\pm 0.020}}$ & $0.304{_{\pm 0.009}}$ & $0.189{_{\pm 0.003}}$ & $0.195{_{\pm 0.012}}$ & $0.215{_{\pm 0.012}}$ \\
\midrule
\multicolumn{7}{l}{\textit{Hybrid}} \\
  RRWP + AFM & $\tied{0.111}{_{\pm 0.004}}$ & $0.255{_{\pm 0.011}}$ & $0.215{_{\pm 0.008}}$ & $0.127{_{\pm 0.008}}$ & $0.128{_{\pm 0.006}}$ & $0.157{_{\pm 0.007}}$ \\
  RRWP + FM & $0.120{_{\pm 0.004}}$ & $0.285{_{\pm 0.016}}$ & $0.242{_{\pm 0.008}}$ & $0.130{_{\pm 0.004}}$ & $0.146{_{\pm 0.006}}$ & $0.166{_{\pm 0.005}}$ \\
  RRWP + AFM$_{\text{stat}}$ & $\best{0.108{_{\pm 0.007}}}$ & $\best{0.244{_{\pm 0.004}}}$ & $\second{0.210}{_{\pm 0.011}}$ & $0.124{_{\pm 0.009}}$ & $0.125{_{\pm 0.006}}$ & $0.150{_{\pm 0.006}}$ \\
  RRWP + FM$_{\text{stat}}$ & $\tied{0.111}{_{\pm 0.004}}$ & $0.257{_{\pm 0.010}}$ & $0.228{_{\pm 0.004}}$ & $0.121{_{\pm 0.009}}$ & $\best{0.120{_{\pm 0.010}}}$ & $\best{0.145{_{\pm 0.006}}}$ \\
  RRWP + AFM + FM & $0.114{_{\pm 0.005}}$ & $0.262{_{\pm 0.011}}$ & $0.235{_{\pm 0.006}}$ & $0.135{_{\pm 0.011}}$ & $0.147{_{\pm 0.004}}$ & $0.165{_{\pm 0.010}}$ \\
  RRWP + AFM$_{\text{stat}}$ + FM$_{\text{stat}}$ & $\tied{0.111}{_{\pm 0.005}}$ & $\second{0.247}{_{\pm 0.008}}$ & $\best{0.209{_{\pm 0.004}}}$ & $\best{0.116{_{\pm 0.009}}}$ & $0.128{_{\pm 0.002}}$ & $0.149{_{\pm 0.008}}$ \\
\bottomrule
\end{tabular}
}
\vspace{-0.6cm}
\end{table*}

\textbf{Results.}
Tables~\ref{tab:pe-comparison-mutag} and~\ref{tab:pe-comparison-zinc} report resp. 19 and 15 feature sets on MUTAG and ZINC$_{\text{subset}}^{*}$ across six backbones.
We emphasize that \ZINC is a non-standard subset, so our results should not be directly compared with published results on \initZINC.
The feature sets fall into three blocks: classical structural encodings (Random, Laplacian, RRWP, RRWP$+$Laplacian) [Sec.~\ref{subsec:classical_baselines}], standalone quantum features (AFM,FM, AFM$_{\mathrm{stat}}$ and FM$_{\mathrm{stat}}$) [Sec.~\ref{subsec:pipeline}], and hybrids RRWP$+$\{quantum pool\}.
As demonstrated in the tables, standalone quantum features do not substitute for RRWP in this setup. Yet, the column-best entry is always attained by a quantum--RRWP hybrid, with the GeneralConv column on ZINC$_{\text{subset}}^{*}$ tied between RRWP$+$Laplacian and RRWP$+$FM$_{\mathrm{stat}}$ at $0.145$ MAE.
RRWP$+$AFM$_{\mathrm{stat}}+$FM$_{\mathrm{stat}}$ stays within one standard deviation of RRWP+AFM+FM and attains the column best in two cells (GAT and GINE on \ZINC).

Because per-cell recipe selection is a free parameter, we test this observation with fixed champions selected by best average performance across the sweep: RRWP for the classical block and RRWP$+$AFM$_{\mathrm{stat}}+$FM$_{\mathrm{stat}}$ for the hybrid block.
The hybrid champion ties or improves on the classical champion in ten of twelve (architecture, dataset) cells; the two losses (Transformer and GeneralConv on ZINC) sit within $0.5\sigma$ of equality.
A one-sided sign test~\citep{demsar2006statistical} gives $p = 0.019$, supporting a sign-consistent but small-amplitude signal: most per-cell gaps lie inside the cross-fold noise band, with only GAT on ZINC reaching a $3\sigma$ effect ($0.209$ vs $0.223$ MAE).
Bayesian decompositions in App.~\ref{app:stat-analysis}~\citep{benavoli2017time} support the same calibration: the effect is directionally asymmetric in favor of the hybrid champion, but mostly compatible with practical equivalence once cell- or architecture-level uncertainty is accounted for.
Altogether, no single feature set dominates uniformly, but pooling with both quench settings combined, RRWP$+$AFM$+$FM and RRWP$+$AFM$_{\mathrm{stat}}+$FM$_{\mathrm{stat}}$, ranks in the top two in eight of the twelve cells, mirroring downstream the AFM/FM complementarity observed in the information benchmark.

\textbf{Hardware.} We repeat the MUTAG benchmark with AFM features obtained by executing the corresponding pulse on a neutral-atom device; the FM pulse and the ZINC$_{\text{subset}}^{*}$ sweep are deferred to future work. Rows HW-AFM and HW-AFM$_{\mathrm{stat}}$ of Table~\ref{tab:pe-comparison-mutag} show that the standalone hardware features remain close to their emulated counterparts despite SPAM errors, decoherence, and finite shots;
readout errors are corrected in post-processing (App.~\ref{annex:shot_number}).
HW-AFM differs from AFM by at most $0.046$ F$_1$ across the six backbones, while HW-AFM$_{\mathrm{stat}}$ stays within $0.028$ F$_1$ and even exceeds the emulator on GAT and GINE. The downstream hybrid picture is strongest for the compressed features. RRWP$+$HW-AFM is within $0.014$ F$_1$ of RRWP$+$AFM on five of six backbones, with GCN the only larger deviation ($0.837$ vs $0.878$). In contrast, RRWP$+$HW-AFM$_{\mathrm{stat}}$ matches RRWP$+$AFM$_{\mathrm{stat}}$ within $0.01$ F$_1$ on all six backbones and reaches the MUTAG column best on GeneralConv ($0.877$); per-backbone differences are listed in Table~\ref{tab:hw-deltas} (App.~\ref{app:hw-deltas}).
 The feature-level ranking of the compressed hybrid benchmark therefore transfers to a real Rydberg device under realistic finite-shot noise and residual coherent errors.

\vspace{-0.3cm}
\subsection{Classical--quantum feature fusion}
\label{subsec:esa}

\begin{wraptable}{r}{0.49\textwidth}

\vspace{-0.4cm}

\caption{%
ESA model performance on MUTAG with concatenation or gated fusion strategy.
RWSE$^{*}$ denotes RWSE augmented with ring-membership features.
Mean binary F$_1$ ($\uparrow$) $\pm$ standard deviation over 10 stratified cross-validation folds $\times$ 5 seeds is reported.
\mbox{$\best{\text{Best}}$} and \mbox{$\second{\text{2nd best}}$} indicate the best and runner-up methods overall.
}
\label{tab:mutag-esa}
\centering
\small
\setlength{\tabcolsep}{2pt}
\resizebox{\linewidth}{!}{
\begin{tabular}{l c c}
\toprule
\multirow{2}{*}{Method} & \multicolumn{2}{c}{MUTAG (F$_1$ binary $\uparrow$)} \\
\cmidrule(lr){2-3}
 & Concat. & Gated \\
\midrule
\multicolumn{3}{l}{\textit{Baseline}} \\
  $\text{RWSE}^{*}$                                & $0.905{_{\pm 0.045}}$ & --- \\
\midrule
\multicolumn{3}{l}{\textit{Hybrid (classical ${+}$ quantum emulation)}} \\
  $\text{RWSE}^{*}{+}\text{AFM}_{\text{stat}}$      & $0.899{_{\pm 0.044}}$ & $0.915{_{\pm 0.042}}$ \\
  $\text{RWSE}^{*}{+}\text{FM}_{\text{stat}}$      & $0.888{_{\pm 0.076}}$ & $0.918{_{\pm 0.046}}$ \\
  $\text{RWSE}^{*}{+}\text{AFM}$                    & $0.893{_{\pm 0.049}}$ & $0.915{_{\pm 0.048}}$ \\
  $\text{RWSE}^{*}{+}\text{FM}$                    & $0.913{_{\pm 0.042}}$ & $0.911{_{\pm 0.043}}$ \\
  $\text{RWSE}^{*}{+}\text{AFM}{+}\text{FM}$        & $0.903{_{\pm 0.050}}$ & $\best{0.922{_{\pm 0.042}}}$ \\
\midrule
\multicolumn{3}{l}{\textit{Hybrid (classical ${+}$ quantum hardware)}} \\
  $\text{RWSE}^{*}{+}\text{HW-AFM}$                 & $0.907{_{\pm 0.047}}$ & $\second{0.920}{_{\pm 0.044}}$ \\
  $\text{RWSE}^{*}{+}\text{HW-AFM}_{\text{stat}}$   & $0.899{_{\pm 0.046}}$ & $0.909{_{\pm 0.046}}$ \\
\bottomrule
\end{tabular}
}

\vspace{-0.4cm}

\end{wraptable}

The preceding benchmark shows that compressed quantum statistics are often on par with, or better than, the full time-resolved correlation vectors. This suggests that full QGFs are informative but not always easy to exploit downstream. Although the intrinsic benchmark (Sec.~\ref{sec:intrinsic-information-benchmark}) shows strong hybrid information content, fixed task-agnostic time samples may include redundant or weakly informative entries, making it difficult for standard GNN concatenation to isolate the useful quantum signal and combine it effectively with RRWP. We therefore replace unconditional concatenation by gated fusion, allowing the model to regulate how much information is passed from each feature stream~\citep{srivastava2015highway,arevalo2017gated}.

We use Edge Set Attention (ESA) as a stringent test case: it is a strong molecular graph architecture based on attention over edge-level representations, and is therefore both competitive with modern graph-learning baselines and naturally suited to incorporating pairwise structural information~\citep{buterez2025end}. Starting from the original ESA setup with RWSE$^{*}$ features, we augment the model with complete QGF feature sets, either by direct concatenation or by gated fusion.
Table~\ref{tab:mutag-esa} shows that direct concatenation is not consistently beneficial, with most hybrids remaining close to the RWSE$^{*}$ baseline. Gated fusion is more effective: it improves nearly all hybrid variants over their concatenated counterparts and reaches the best score with RWSE$^{*}$+AFM+FM, at $0.922\pm0.042$ binary F$_1$.
The best hardware hybrid, RWSE$^{*}$+HW-AFM with gating, reaches a comparable $0.920\pm0.044$. The gains remain modest relative to the cross-validation variance. Still, the learned gates are informative: from an initial value of $0.88$, they stay close to initialization for compressed statistics ($0.887$ for AFM$_{\mathrm{stat}}$, $0.855$ for FM$_{\mathrm{stat}}$) and decrease more for FM-containing full features (0.816 for FM, 0.830 for AFM+FM), whereas full AFM stays close to initialization (0.862).
This is consistent with the model selectively attenuating the higher-dimensional QGFs rather than ignoring them.

\vspace{-0.2cm}
\section{\texorpdfstring{Quantum graph features beyond classical structural encodings}{Quantum graph features beyond classical structural encodings}}
\label{sec:controls}
\vspace{-0.1cm}

RRWP-based encodings cannot separate certain non-isomorphic graphs~\citep{thabet2024quantum,kasture2025quantum}.
We construct ladder graphs whose label depends only on the relative parity of two diagonal crossings---a global combinatorial property---and keep only graphs whose opposite-label partners share identical RRWP and ring features (394 graphs, $18$--$26$ atoms, exact unit-disk registers; App.~\ref{app:ladder}).
With a logistic-regression probe, AFM$+$FM reach $0.875\pm0.043$ accuracy (ROC-AUC $0.941$), versus $0.641\pm0.077$ (ROC-AUC $0.680$) for the combined
classical baseline (Tab.~\ref{tab:ladder_main}); the gap of $0.23$ is roughly three times the larger cross-fold standard deviation.
This is a result on a synthetic family constructed to defeat these encodings, not evidence of a generic practical advantage for graph learning.

\begin{table}[t]
\begin{minipage}[c]{0.47\textwidth}
\caption{%
Ladder graphs: pinned versus frustrated ground state (394 graphs, 98 RRWP-2-blind groups, $18$--$26$ atoms).
Classical-All $=$ RRWP$+$Laplacian$+$Ring$+$Geometry.
Logistic regression, 5 folds grouped by blind group $\times$ 3 repeats; mean $\pm$ standard deviation over the 15 test folds.
Label-permutation test (100 shuffles): $p=0.010$ (the minimum) for all sets except RRWP ($p=0.020$) and Geometry ($p=0.446$).
\mbox{$\best{\text{Best}}$} and \mbox{$\second{\text{2nd best}}$} per metric.
}
\label{tab:ladder_main}
\end{minipage}\hfill
\begin{minipage}[c]{0.49\textwidth}
\centering
\small
\setlength{\tabcolsep}{3pt}
\resizebox{\linewidth}{!}{
\begin{tabular}{l c c c}
\toprule
Features & Acc.\ $\uparrow$ & F$_1$ $\uparrow$ & AUC $\uparrow$ \\
\midrule
\multicolumn{4}{l}{\textit{Classical}} \\
  RRWP      & $0.557{_{\pm 0.048}}$ & $0.559{_{\pm 0.047}}$ & $0.612{_{\pm 0.065}}$ \\
  Geom      & $0.503{_{\pm 0.053}}$ & $0.495{_{\pm 0.080}}$ & $0.501{_{\pm 0.063}}$ \\
  Classical-All & $0.641{_{\pm 0.077}}$ & $0.643{_{\pm 0.091}}$ & $0.680{_{\pm 0.065}}$ \\
\midrule
\multicolumn{4}{l}{\textit{Quantum emulation}} \\
  FM                    & $0.622{_{\pm 0.063}}$ & $0.620{_{\pm 0.071}}$ & $0.670{_{\pm 0.065}}$ \\
  AFM                   & $0.847{_{\pm 0.045}}$ & $0.849{_{\pm 0.045}}$ & $\second{0.914}{_{\pm 0.026}}$ \\
  AFM${+}$FM            & $\second{0.875}{_{\pm 0.043}}$ & $\second{0.875}{_{\pm 0.044}}$ & $\best{0.941}{_{\pm 0.022}}$ \\
\midrule
\multicolumn{4}{l}{\textit{Hybrid (classical ${+}$ quantum emulation)}} \\
  Classical-All${+}$AFM${+}$FM & $\best{0.882}{_{\pm 0.043}}$ & $\best{0.884}{_{\pm 0.043}}$ & $\best{0.941}{_{\pm 0.023}}$ \\
\bottomrule
\end{tabular}
}
\end{minipage}
\end{table}
\vspace{-0.2cm}
\section{Conclusion}\label{sec:conclusion}
\vspace{-0.2cm}

In this work we use the occupation-based connected correlation matrix $C_{i,j}$ \mbox{[Eq.~(\ref{eq:obs_C})]} as \emph{quantum graph features} (QGFs), whereby the observable results from two specific quench-parameter settings of a graph-encoded Rydberg Hamiltonian: FM-like and AFM-like \mbox{[Eq.~(\ref{eq:H_AFM_FM})]}. Diagonal and off-diagonal elements $C_{i,j}$ serve as node-level and edge-level injections, respectively.
Across all presented analyses (intrinsic information benchmark [Sec.~\ref{sec:intrinsic-information-benchmark}], downstream GNNs [Sec.~\ref{sec:gnn-downstream}], including hardware-derived quantum features),
the main conclusion is consistent: QGFs do not replace classical structural encodings, but provide complementary information that is most useful when concatenated with RRWP.
In the intrinsic information benchmark of Sec.~\ref{sec:intrinsic-information-benchmark}, the two quench settings, AFM and FM, exhibit distinct local and global biases.
About $8\%$ (AFM) and $14\%$ (FM) of QGFs are not captured by the tested classical, geometric features, and on a ladder family QGFs separate two structural classes well beyond cross-fold noise [Sec.~\ref{sec:controls}].
In the downstream benchmark of Sec.~\ref{sec:gnn-downstream}, a fixed compact RRWP+QGF champion ties or improves over RRWP in ten of twelve cells (one-sided sign test $p=0.019$), with Bayesian analyses indicating a small effect size.

Crucially, we demonstrate an end-to-end QPU implementation: all $188$ MUTAG graphs are embedded with quantum dynamics executed on a neutral-atom device, and the measured features drive the full machine-learning pipeline.
The hybrid AFM$_\mathrm{stat}$ result matches emulation within $\Delta F_1 \sim 0.01$ across all six MUTAG backbones, demonstrating that QGFs are viable structural features for graph learning on molecules on present-day Rydberg hardware.
The current limitations are clear. The encoding is restricted to two-dimensional unit-disk embeddings: although our protocol retains about $95\%$ of \ZINC, it still limits graph coverage and introduces a geometry-dependent preprocessing step. Future neutral-atom processors with three-dimensional registers should reduce this bottleneck.
The features also depend on the chosen placement: they are stable under small positional noise but not invariant under re-embedding, and canonical or placement-invariant quantum encodings remain an open problem. A further, more practical limitation is cost: hardware extraction remains much slower than classical preprocessing, and the current compute and QPU sampling time of the full pipeline is too high for practical use (see App.~\ref{annex:shot_number}).
Much higher repetition rates will be required before such features can be generated at graph-learning scale.
Finally, the downstream gains are modest, and the feature-extraction protocol is not yet optimized for learning.
Fixed quenches, task-agnostic readout times, and low-order summaries may miss informative dynamics or introduce redundant samples.
Future work should improve quantum-feature construction and hybrid fusion through optimized readout schedules, learned dynamical summaries, and gating mechanisms, as suggested by the ESA experiment in Sec.~\ref{subsec:esa}.

\textbf{Code and data:} \href{https://github.com/pasqal-io/Quantum-Graph-Features}{github.com/pasqal-io/Quantum-Graph-Features}, \href{https://doi.org/10.5281/zenodo.23099924}{doi:10.5281/zenodo.23099924}.

\begin{ack}
This work was funded by Pasqal SAS. All authors are employees of Pasqal,
which develops and commercialises the neutral-atom quantum computer used in this work.
\end{ack}

\bibliographystyle{unsrtnat}

\bibliography{bibliography}

\appendix

\section{Equivalence between Rydberg and Ising Hamiltonian}
\label{app:Ising_Hamiltonian}

In this appendix, we provide supporting information on the relevant neutral-atom Hamiltonian introduced in Sec.~\ref{subsec:Rydberg-Hamiltonian}.
In analog neutral-atom quantum simulators, each atom realizes an effective spin-$1/2$ degree of freedom spanned by a stable ground state $\left|0\right\rangle$ and a highly excited Rydberg state $\left|1\right\rangle$.
In typical $^{87}\mathrm{Rb}$ implementations, these states may be chosen as $\left|0\right\rangle=\left|5S_{1/2},F=2,m_F=2\right\rangle$ and $\left|1\right\rangle=\left|60S_{1/2},m_J=1/2\right\rangle$.
The Rydberg excitation gives rise to strong interactions between atoms occupying the $\left|1\right\rangle$ state.
For the Rydberg levels and interatomic distances relevant here, these interactions are well approximated by a repulsive van der Waals potential, $V\left(\boldsymbol{r}\right)\sim\left|\boldsymbol{r}\right|^{-6}$, acting over micrometer-scale distances.
The resulting dynamics of a register of $N$ atoms are described by the Rydberg Hamiltonian (\ref{eq:Rydberg_Ham}), cf. Refs. \cite{Gallagher_1994, Browaeys_2020, Saffman_2016, Labuhn_2016, Lukin_2017}.
The Hamiltonian acts on a $2^{N}$-dimensional Hilbert space spanned by computational basis states $\left|\boldsymbol{b}\right\rangle =\left|b_{1}b_{2}\cdots b_{N}\right\rangle $ with $b_{i}\in\left\{ 0,1\right\} $.
The operators satisfy $\hat{n}_{i}\left|b_{i}\right\rangle =b_{i}\left|b_{i}\right\rangle $, $\hat{\sigma}_{i}^{x}\left|b_{i}\right\rangle =\left|1-b_{i}\right\rangle $ and $\hat{\sigma}_{i}^{z}\left|b_{i}\right\rangle =\left(-1\right)^{b_{i}}\left|b_{i}\right\rangle $, with $\hat{n}_{i}=\big(1-\hat{\sigma}_{i}^{z}\big)/2$.
Naturally, Eq.~(\ref{eq:Rydberg_Ham}) can be rewritten in a conventional Ising form,
\begin{equation}
\hat{\mathcal{H}}\left(\mathbb{J},\Omega,\delta\right)=\frac{1}{4}\sum_{i<j}\,J_{ij}\,\hat{\sigma}_{i}^{z}\hat{\sigma}_{j}^{z}+\hbar\sum_{i}\left[(\Omega/2)\,\hat{\sigma}_{i}^{x}+h_{i}^{z}\hat{\sigma}_{i}^{z}\right]+\bar{E},\label{eq:Ising_Ham}
\end{equation}
where the local longitudinal field $h_{i}^{z}$ is given by $h_i^z=(\delta/2)-\bar J_i/(4\hbar)$ with $\bar{J}_{i}=\sum_{j=1}^{N}\bar{\delta}_{i,j}J_{ij}$ denoting the net interaction strength on atom $i$.
The constant energy shift $\bar{E}$, given by
\begin{align}
\bar{E} & =\mathrm{tr}\big\{\hat{\mathcal{H}}\big\}/2^{N}=(N\bar{J}/8)- (\hbar N\delta/2),\label{eq:E_bar}
\end{align}
coincides with the infinite-temperature average.
The average interaction strength is $\bar{J}=\sum_{i=1}^{N}\bar{J}_{i}/N$.
The positive van-der-Waals coupling strength $J_{ij}=C_{6}/r_{ij}^{6}$ is set by the inter-atomic distance $r_{ij}$ between atoms $i$ and $j$, and the Rydberg coefficient for the $n{=}60$ manifold used in this work evaluates to $C_{6}/\hbar=8.66\!\times\!10^{5}\, \mu\mathrm{m}^6\,  \mathrm{rad}/\mu \mathrm{s}  $ \cite{SIBALIC_2017}.
The transverse driving $\Omega$ (Rabi frequency) and global detuning $\delta$ are set by the laser fields; their values are discussed in Sec.~\ref{subsec:Pulse-shaping} and in App.~\ref{app:quench_parameter_selection}.
Since $\hat{\mathcal{H}}\left(\mathbb{J},\Omega,\delta\right)$ is time-independent during each experiment, the evolution corresponds to a quench, leading to the time-evolved state
$\left|\psi\left(t\right)\right\rangle   =\exp\left[-\mathsf{i}(t/\hbar)\,\hat{\mathcal{H}}\left(\mathbb{J},\Omega,\delta\right)\right]\,\left|\psi\left(0\right)\right\rangle $, leading to the desired correlation matrix
 \begin{align}
C_{i,j}\left(t\right) & = A^{nn}_{i,j}\left(t\right) -n_{i}\left(t\right)\,n_{j}\left(t\right),\label{eq:app_obs_C}
\end{align}
where
\begin{align}
A^{nn}_{i,j}\left(t\right) & =\left\langle \psi\left(t\right)\big|\,\hat{n}_{i}\,\hat{n}_{j}\,\big|\psi\left(t\right)\right\rangle, & n_{i}\left(t\right) & =\left\langle \psi\left(t\right)\big|\,\hat{n}_{i}\,\big|\psi\left(t\right)\right\rangle,\label{eq:app_obs_A_n}
\end{align}
see Eq.~(\ref{eq:obs_C}) where we used the Heisenberg operator $\hat{n}_{i}(t)$.
The respective value ranges are $C_{i,j} \in [-1/4, 1/4]$ and $n_i \in [0, 1]$.
Since $\hat{n}_{i}^{2}=\hat{n}_{i}$, it follows for the diagonal elements
$C_{i,i}\left(t\right)  =n_{i}\left(t\right)\left[1-n_{i}\left(t\right)\right]$, i.e. the diagonal element $C_{i,i}\left(t\right)  \in  \left[0,1/4\right]$ effectively encodes the site occupations $n_{i}\left(t\right)$.
Importantly, both $\hat{n}_{i}$ and $\hat{n}_{i}\hat{n}_{j}$ are
diagonal in the measurement basis, which allows $n_{i}\left(t\right)$
and $C_{i,j}\left(t\right)$
to be estimated efficiently by repeated projective readout of the
atomic register.

\section{Thermalization}
\label{app:thermalization}

In this appendix, we collect the thermalization arguments which are implicitly used in Sec.~\ref{subsec:Pulse-shaping}, and explicitly used in App.~\ref{app:quench_parameter_selection}.
We first recall the diagonal-ensemble description of long-time observables after a quench under a time-independent Hamiltonian.
We then explain how the eigenstate thermalization hypothesis (ETH) motivates a canonical thermal approximation to the diagonal ensemble.
Finally, we derive the simple criterion to determine the sign of the effective inverse temperature.
The scheme is then applied to the Rydberg
Hamiltonian (\ref{eq:Rydberg_Ham}) [or (\ref{eq:Ising_Ham})] in App.~\ref{app:quench_parameter_selection}.

Let $\hat A$ be a generic observable and let the system evolve under a
time-independent Hamiltonian $\hat{\mathcal H}$,
$|\psi(t)\rangle = e^{-\mathsf{i}\hat{\mathcal H}t/\hbar}
|\psi(0)\rangle$ .
We denote the eigenstates and eigenenergies of $\hat{\mathcal H}$ by
$\hat{\mathcal H}|E_\alpha\rangle = E_\alpha |E_\alpha\rangle$ ,
and expand the initial state as  $|\psi(0)\rangle=\sum_\alpha c_\alpha |E_\alpha\rangle$ with $c_\alpha=\langle E_\alpha|\psi(0)\rangle$.

The expectation value of $\hat A$ can then be written as
\begin{align}
A(t)
&=
\langle \psi(t)|\hat A|\psi(t)\rangle
&\equiv
A^{\mathrm{diag}}
+
A^{\mathrm{off\text{-}diag}}(t),
\label{eq:app_A_of_t}
\end{align}
where
\begin{align}
A^{\mathrm{diag}} &= \sum_\alpha |c_\alpha|^2 A_{\alpha\alpha} ,
&
A^{\mathrm{off\text{-}diag}}(t)
&=
\sum_{\alpha\neq\alpha^\prime} c_\alpha^*c_{\alpha^\prime}
e^{i\omega_{\alpha,\alpha^\prime}t} A_{\alpha\alpha^\prime} ,
\label{eq:app_A_diag}
\end{align}
with $A_{\alpha\alpha^\prime} = \langle E_\alpha |\hat{A}|E_{\alpha^\prime} \rangle$ and $\omega_{\alpha,\alpha^\prime} = (E_\alpha-E_{\alpha^\prime})/\hbar$.
The first term is the time-independent diagonal-ensemble value.
The second term contains oscillatory off-diagonal contributions.

For a finite closed quantum system, $A^{\mathrm{off\text{-}diag}}(t)$ does
not generally converge pointwise to zero: finite-size revivals and degeneracies
can occur. What is relevant for equilibration is instead dephasing.
Under
generic non-resonance assumptions, the off-diagonal contribution vanishes in
the infinite-time average.
Experimentally, we approximate the long-time value by the finite-time average
\begin{align}
\overline{A}(t_1,t_0)
&=
\frac{1}{t_1-t_0}
\int_{t_0}^{t_1}dt'\,A(t').
\label{eq:app_A_bar}
\end{align}
In the absence of long-lived nonergodic effects, one expects $\overline{A}(t_{1},t_{0})$ to approach $A^{\mathrm{diag}}$ as the averaging window becomes large.
In practice, this convergence can be delayed or prevented by mechanisms that obstruct or weakly violate ergodicity, such as integrability, many-body localization, Hilbert-space fragmentation, many-body scars, or prethermalization
\cite{Rigol_2008,Nandkishore_2015,Abanin_2019,Lukin_2017,Turner_2018,Mori_2018,Moudgalya_2022,Serbyn_2021}.
 Note that $C_{ij}(t)$ [Eq.~(\ref{eq:app_obs_C})] is quadratic in the expectation values $n_i(t)$, and time averaging does not commute with taking products. Hence, the time average $\overline{C_{ij}}(t_1,t_0)$ does not in general converge to the diagonal-ensemble value $C^{\mathrm{diag}}_{ij}=A^{\mathrm{nn,diag}}_{ij}-n_i^{\mathrm{diag}}n_j^{\mathrm{diag}}$; the two differ by the temporal covariance $\overline{n_i n_j}-\overline{n_i}\,\overline{n_j}$. Instead, we use
\begin{equation}
C_{ij}^{\infty}(t_1,t_0) = \overline{A^{\mathrm{nn}}_{ij}}(t_1,t_0) - \overline{n_i}(t_1,t_0)\,\overline{n_j}(t_1,t_0),
\label{eq:app_obs_C_infty}
\end{equation}
which converges to $C^{\mathrm{diag}}_{ij}$.

The diagonal ensemble still requires knowledge of the full eigensystem of
$\hat{\mathcal H}$, which is generally unavailable for large many-body systems.
The eigenstate thermalization hypothesis provides a practical interpretation.
If the dynamics are sufficiently ergodic for ETH to apply, the diagonal matrix elements of a few-body observable are smooth functions of energy, up to finite-size fluctuations,
\begin{align}
A_{\alpha\alpha} = \langle E_\alpha |\hat{A}|E_{\alpha} \rangle
\simeq
\langle \hat A\rangle_{\mathrm{micro}}(E_{\alpha}),
\end{align}
with the microcanonical expectation value $\langle \hat A\rangle_{\mathrm{micro}}(E)$.
The off-diagonal matrix elements are suppressed by the many-body entropy \cite{Deutsch_1991,Srednicki_1994,Rigol_2008,DAlessio_2016}.
If, in addition, the weights $|c_\alpha|^2$ have support in a sufficiently narrow energy window around the initial energy $E_{\mathrm{init}} = \langle \psi(0)|\hat{\mathcal H}|\psi(0)\rangle $, then the diagonal-ensemble value can be approximated by the microcanonical expectation value at $E_{\mathrm{init}}$, $A^{\mathrm{diag}} \simeq \langle \hat A\rangle_{\mathrm{micro}}(E_{\mathrm{init}})$.
For sufficiently large systems, the microcanonical and canonical ensembles are equivalent for local observables.
We therefore use the canonical thermal approximation at an effective inverse temperature $\beta(E_{\mathrm{init}})$,
\begin{align}
A^{\mathrm{diag}} \simeq  \langle \hat A\rangle_{\mathrm{th}}
\left[\beta(E_{\mathrm{init}})\right],
\label{eq:app_A_diag_canonical}
\end{align}
where
\begin{align}
A^{\mathrm{th}} &=\langle \hat A\rangle_{\mathrm{th}}[\beta] =
\mathrm{tr}\left\{ \hat A e^{-\beta\hat{\mathcal H}}
\right\} \, \big/ Z[\beta],
& Z[\beta] &= \mathrm{tr}\left\{
e^{-\beta\hat{\mathcal H}} \right\}.
\label{eq:app_thermal_exp_value}
\end{align}
The inverse temperature is not a free parameter. It is fixed by energy conservation,
\begin{align}
\langle \hat{\mathcal H}\rangle_{\mathrm{th}}[\beta]
= E_{\mathrm{init}}, \qquad
E_{\mathrm{init}}
= \langle\psi(0)|\hat{\mathcal H}|\psi(0)\rangle .
\label{eq:app_energy_conservation_canonical}
\end{align}

This approximation is used in Sec.~\ref{subsec:Pulse-shaping} and App.~\ref{app:quench_parameter_selection} as a diagnostic tool for selecting quench parameters $(\Omega, \delta)$.
It is not assumed to hold throughout the full parameter space.
In particular, finite-size effects, graph symmetries, proximity to integrable limits, many-body localization, Hilbert-space fragmentation, many-body scars, or prethermalization can obstruct or delay convergence to the thermal value \cite{Nandkishore_2015,Abanin_2019,Lukin_2017,Turner_2018,
Mori_2018,Moudgalya_2022,Serbyn_2021}.

Since the Hilbert space is finite, the spectrum of $\hat{\mathcal H}$ is bounded.
Consequently, the solution of Eq.~(\ref{eq:app_energy_conservation_canonical}) may have either positive or negative inverse temperature \cite{Ramsey_1956,Braun_2013}.
A negative-temperature ensemble of $\hat{\mathcal H}$ can equivalently be written as a positive-temperature ensemble of the negated Hamiltonian,
\begin{align}
e^{-\beta \hat{\mathcal H}}
&= e^{-|\beta|(-\hat{\mathcal H})},
& \beta&<0.
\end{align}
Thus, for $\beta<0$, the effective thermal description reverses the sign of all terms in the Hamiltonian.
In particular, the sign of the Ising interaction is reversed in this effective thermal description: turning the repulsive AFM model into an attractive FM model.

Conveniently, the sign of $\beta$ can be easily determined, knowing the initial energy $E_{\mathrm{init}}$ (\ref{eq:app_energy_conservation_canonical}) and the infinite-temperature energy
\begin{align}
\bar E &= \langle \hat{\mathcal H}\rangle_{\mathrm{th}}[\beta=0]
= \mathrm{tr}\left\{\hat{\mathcal H}\right\}\,\big/
\dim(\hat{\mathcal H}) ,
\label{eq:app_E_average}
\end{align}
see also Eq.~(\ref{eq:E_bar}).
For many Hamiltonians both quantities can be evaluated analytically.
The thermal energy $\langle \hat{\mathcal H}\rangle_{\mathrm{th}}[\beta]$ is a monotonically decreasing function of $\beta$.
Indeed,
\begin{align}
\frac{\partial \langle \hat{\mathcal H}\rangle_{\mathrm{th}}
}{\partial \beta} &= -
\left[ \langle \hat{\mathcal H}^2\rangle_{\mathrm{th}}
- \langle \hat{\mathcal H}\rangle_{\mathrm{th}}^2 \right]
= - \left\langle
\left( \hat{\mathcal H}
- \langle \hat{\mathcal H}\rangle_{\mathrm{th}}
\right)^2
\right\rangle_{\mathrm{th}}
\leq 0.
\label{eq:app_dH_Dbeta}
\end{align}
The inequality is strict unless the thermal state has zero energy variance.
Clearly, an initial energy $E_{\mathrm{init}}=\bar{E}$ implies a thermal state at infinite temperature with $\beta=0$.
Given the monotonicity (\ref{eq:app_dH_Dbeta}), it also follows that $E_{\mathrm{init}}<\bar{E}$ implies a positive solution $\beta$, whereas $E_{\mathrm{init}}>\bar{E}$ yields a negative solution $\beta$:
\begin{align}
E_{\mathrm{init}} < \bar E
&\quad\Rightarrow\quad
\beta(E_{\mathrm{init}})>0,
\nonumber\\
E_{\mathrm{init}} = \bar E
&\quad\Rightarrow\quad
\beta(E_{\mathrm{init}})=0,
\nonumber\\
E_{\mathrm{init}} > \bar E
&\quad\Rightarrow\quad
\beta(E_{\mathrm{init}})<0.
\label{eq:app_initial_energy_consequences}
\end{align}
Applying this criterion to the Rydberg Hamiltonian (\ref{eq:Rydberg_Ham}) where $\bar{E}$ evaluates to Eq.~(\ref{eq:E_bar}), and $E_\mathrm{init}=0$, one obtains the classification shown in App.~\ref{app:quench_parameter_selection}, Eq.~(\ref{eq:detuning_choices}).

\section{Quench-parameter selection}
\label{app:quench_parameter_selection}

\begin{figure*}[!t]
\includegraphics[width=\textwidth]{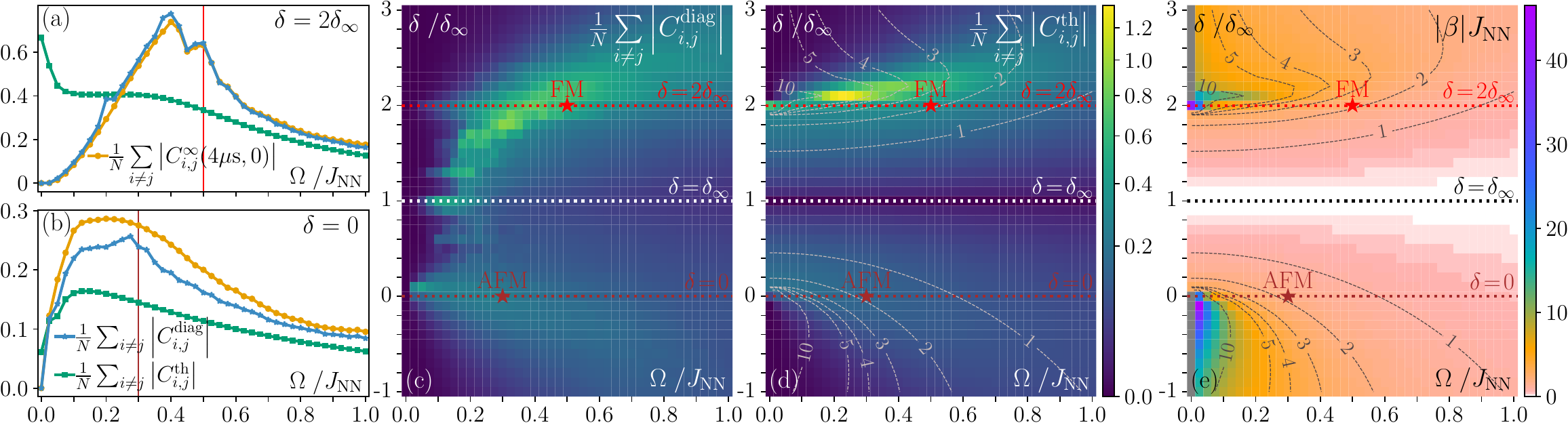}
\caption{
Exploration of the $(\Omega,\delta)$ parameter space for the same molecular graph as in Fig.~\ref{fig:time_evolution}, initialized in $|\psi(0)\rangle=|\boldsymbol{0}\rangle$.
The plotted quantity is the off-diagonal correlation weight per node,
$\mathcal{C}_{\mathrm{off}}=\sum_{i\neq j}|C_{i,j}|/N$, used as a proxy for the amount of correlation information contained in the feature matrix.
Panels (a)--(b) show the proxy as a function of $\Omega$ at the two relevant detunings: $\delta=2\delta_{\infty}$ for the FM setting and $\delta=0$ for the AFM setting, cf. Eq.~(\ref{eq:H_AFM_FM}) or (\ref{eq:app_H_AFM_FM}), with the selected $\Omega$ values indicated by vertical lines.
Panels (c)--(d) compare the diagonal-ensemble [Eq.~(\ref{eq:app_A_diag})] and thermal estimates [Eq.~(\ref{eq:app_thermal_exp_value})] over the full parameter window.
Panel (e) shows the absolute value of the effective inverse temperature $|\beta|$ obtained from Eq.~(\ref{eq:app_energy_conservation_canonical}).
The proxy for the time average
based on $C^\infty_{i,j}(t_1,t_0)$ [Eq.~(\ref{eq:app_obs_C_infty})]
is only shown in panels (a)--(b).
The line $\delta=\delta_{\infty}$ separates the positive-temperature AFM-like regime from the negative-temperature FM-like regime.
For this graph, $\delta_\infty = 0.53\,J_\mathrm{NN}/\hbar$ [cf. Eq.~(\ref{eq:detuning_choices})].
Gray clipping in panel (e) indicates $|\beta|=\infty$.
}
\label{fig:phase_diagram}
\end{figure*}

In this appendix, we motivate the quench-parameter selection that leads to the AFM-like and FM-like
Hamiltonians, $\hat{\mathcal{H}}_{\mathrm{AFM}}^{\mathcal{G}}$ and $\hat{\mathcal{H}}_{\mathrm{FM}}^{\mathcal{G}}$, used in this work, and defined in Eq.~(\ref{eq:H_AFM_FM}) in Sec.~\ref{subsec:Pulse-shaping}.
The quench settings $(\Omega,\delta)$ are chosen to systematically yield pronounced connected density correlations $C_{i,j}$ [Eq.~(\ref{eq:obs_C})].
Hereby, the selection strategy is guided by thermalization arguments which are detailed in App.~\ref{app:thermalization}.
In particular, we exploit that the diagonal-ensemble value $A^{\mathrm{diag}}$ (\ref{eq:app_A_diag}), experimentally approximated by the time average $\overline{A}(t_1,t_0)$ (\ref{eq:app_A_bar}), can be related to the canonical thermal expectation value $A^{\mathrm{th}}$ (\ref{eq:app_thermal_exp_value}) under the assumption of thermalization.
The main idea is illustrated on a small molecular graph in Fig.~\ref{fig:time_evolution} and Fig.~\ref{fig:phase_diagram}.

First, we apply the criterion (\ref{eq:app_initial_energy_consequences}) which determines the sign of the effective temperature on the Rydberg Hamiltonian [Eq.~(\ref{eq:Rydberg_Ham}) or (\ref{eq:Ising_Ham})] with an initial all-zero product state $\left|\psi\left(0\right)\right\rangle =\left|\boldsymbol{0}\right\rangle$ where $\left|\boldsymbol{0}\right\rangle \equiv\left|0\right\rangle ^{\otimes N}$.
The initial energy is $E_{\mathrm{init}} = \langle\boldsymbol{0}| \hat{\mathcal{H}} |\boldsymbol{0}\rangle =0$, and the infinite-temperature energy $\bar{E}$ is evaluated in Eq.~(\ref{eq:E_bar}).
Both energies equal each other, i.e. $E_{\mathrm{init}} = \bar E$, for the detuning value
\begin{align}
\delta_{\infty} = \bar{J} / (4\hbar) =\sum_{i\neq j}J_{ij} / (4N\hbar)
\label{eq:delta_infty},
\end{align}
such that the criterion (\ref{eq:app_initial_energy_consequences}) becomes
\begin{align}
\delta &< \delta_{\infty}
&&\Rightarrow
&\beta&>0
&&\leftrightarrow\;
\mathrm{AFM\text{-}like},
\nonumber \\
\delta &= \delta_{\infty}
&&\Rightarrow
&\beta&=0,&&
\nonumber \\
\delta &> \delta_{\infty}
&&\Rightarrow
&\beta&<0
&&\leftrightarrow\;
\mathrm{FM\text{-}like}.
\label{eq:detuning_choices}
\end{align}
Thus, increasing the detuning moves the thermal interpretation from an AFM-like positive-temperature regime through infinite temperature into an FM-like negative-temperature regime.
This classification does not imply that the system thermalizes for all values of $(\Omega,\delta)$; rather, it provides a useful organizing principle in the parameter regimes relevant for our feature-generation protocol for identifying regimes with robust AFM- or FM-like correlations.

For convenience, we repeat the two introduced Hamiltonians from Eq.~(\ref{eq:H_AFM_FM}) in Sec.~\ref{subsec:Pulse-shaping}
\begin{align}
\hat{\mathcal{H}}_{\mathrm{AFM}}^{\mathcal{G}} & =\hat{\mathcal{H}}\left(\mathbb{J}^{\mathcal{G}},\,\Omega=0.3\,J_{\mathrm{NN}}/\hbar,\,\delta=0\right),
&
\hat{\mathcal{H}}_{\mathrm{FM}}^{\mathcal{G}} & =\hat{\mathcal{H}}\left(\mathbb{J}^{\mathcal{G}},\,\Omega=0.5\,J_{\mathrm{NN}}/\hbar,\,\delta=2\delta_\infty^{\mathcal{G}}\right),
\label{eq:app_H_AFM_FM}
\end{align}
see Eqs.~(\ref{eq:Rydberg_Ham}),~(\ref{eq:Ising_Ham}),~(\ref{eq:graph_dependent_J}) for definitions.
To emphasize the graph dependence of $\delta_{\infty}=\delta_{\infty}\left(\mathbb{J}^{\mathcal{G}} \right)$ (\ref{eq:delta_infty}) we added the superscript $\mathcal{G}$ in Eq.~(\ref{eq:H_AFM_FM}) and (\ref{eq:app_H_AFM_FM}).

The choice $\delta=0$ places the AFM setting below $\delta_{\infty}$ and hence in the positive-temperature regime of Eq.~(\ref{eq:detuning_choices}).
The choice $\delta=2\delta_{\infty}$ places the FM setting in the negative-temperature regime and it has an additional useful interpretation: it cancels the spatially averaged longitudinal field.
Indeed, using $h_i^z=\delta/2-\bar J_i/(4\hbar)$ from Eq.~(\ref{eq:Ising_Ham}), one finds $\sum_i h_i^z=0$ is solved by $\delta=2\delta_{\infty}$.
Thus, the FM pulse setting approximately realizes a field-balanced transverse-field Ising regime, while the AFM setting probes the opposite side of the infinite-temperature line.

Figure~\ref{fig:time_evolution} in the main text confirms the expected qualitative behavior on a small molecular graph.
In the FM setting, neighboring atoms develop positive off-diagonal correlations, while in the AFM setting the corresponding correlations are negative.
The molecule is small enough for exact diagonalization, which allows us to compute both the diagonal-ensemble values $C_{i,j}^{\mathrm{diag}}$ [Eq.~(\ref{eq:app_A_diag})] and the thermal estimates $C_{i,j}^{\mathrm{th}}$ [Eq.~(\ref{eq:app_thermal_exp_value})].
The finite-time averages
$C^\infty_{i,j}(t_1,t_0)$ [Eq.~(\ref{eq:app_obs_C_infty})]
agree quantitatively well with the diagonal-ensemble values, indicating that the chosen measurement window captures the dominant long-time correlation signal.
The agreement with the thermal values is qualitative rather than quantitative, which is not surprising for such a small system, where finite-size effects, and the proximity to the $\Omega=0$ integrable limit can prevent fully ergodic dynamics.
Nevertheless, the thermal picture correctly predicts the AFM/FM character of the correlations and is therefore useful as a parameter-selection guide.

We next show that the chosen pulse settings not only produce the desired AFM- and FM-like correlation signs, but also lie in regimes with pronounced correlation magnitude.
To quantify this, we scan the parameter space $\Omega /(J_{\mathrm{NN}}/\hbar) \in [0, 1 ]$ and $\delta /\delta_\infty \in [-1, 3 ]$, and compute the off-diagonal correlation weight per node, $\mathcal{C}_{\mathrm{off}} = \sum_{i\neq j}\left| C_{i,j} \right|/N $.
This quantity serves as a simple proxy for the amount of nonlocal correlation information available in the correlation matrix.
We evaluate this proxy using the finite-time average (\ref{eq:app_A_bar}), the diagonal-ensemble value (\ref{eq:app_A_diag}), and the thermal expectation value (\ref{eq:app_thermal_exp_value}).
The results are shown in Fig.~\ref{fig:phase_diagram}.
We omit the full phase diagram of the finite-time average, since it is qualitatively very similar to the diagonal-ensemble result in Fig.~\ref{fig:phase_diagram}(c).

We first interpret the thermal phase diagram in Fig.~\ref{fig:phase_diagram}(d), together with the corresponding effective inverse temperature shown in Fig.~\ref{fig:phase_diagram}(e).
As predicted by Eq.~(\ref{eq:detuning_choices}), the line $\delta=\delta_{\infty}$ separates the AFM-like regime at positive temperature from the FM-like regime at negative temperature.
This separation is largely reproduced by the diagonal-ensemble estimate in Fig.~\ref{fig:phase_diagram}(c), except at very small $\Omega$, where the dynamics are more strongly affected by constrained subspaces and thermalization becomes less reliable.

On the FM side, the strongest correlations occur close to the regime where the average longitudinal field vanishes.
In this regime, the model approaches a field-balanced transverse-field Ising model with effectively ferromagnetic thermal correlations.
For the graph considered in Fig.~\ref{fig:phase_diagram}, the thermal correlation maximum appears near $\Omega/(J_{\mathrm{NN}}/\hbar)\approx0.23$ and $\delta/\delta_{\infty}\approx2.1$, which we interpret as a finite-size thermal crossover.
The existence of an enhanced-correlation regime is not tied to this particular molecular geometry: for a broad class of embedded graphs, one expects a finite-size thermal crossover at a graph-dependent effective temperature, provided the thermal description remains applicable.
We therefore choose $\delta=2\delta_{\infty}$ to place each graph near this crossover regime.
The thermal maximum, however, occurs at relatively small $\Omega$, where the finite-time dynamics and the diagonal ensemble show stronger deviations from the thermal estimate.
We therefore select the slightly larger value $\Omega=0.5\,J_{\mathrm{NN}}/\hbar$ [cf. Eq.~(\ref{eq:H_AFM_FM}) or (\ref{eq:app_H_AFM_FM})], which remains in a regime of enhanced correlations while producing more robust dynamical relaxation.

The crossover scale depends on the graph geometry through $\mathbb{J}^{\mathcal{G}}$ (\ref{eq:graph_dependent_J}).
For orientation, the effective crossover temperature observed in Fig.~\ref{fig:phase_diagram}(e), $|\beta_{\mathrm{cross}}|J_{\mathrm{NN}}\approx 7$, lies between familiar classical Ising limits.
In the nearest-neighbor field-free Ising model, the one-dimensional chain has no finite-temperature transition ($\beta_{\mathrm{crit}}=\infty$), whereas the square and triangular lattices have
$\beta_{\mathrm{crit}}J_{\mathrm{NN}}\approx1.76$ and $\beta_{\mathrm{crit}}J_{\mathrm{NN}}\approx1.10$, respectively, in the present convention \cite{Onsager_1944,Wannier_1950}.
These values include the factor-of-four conversion between the Rydberg interaction coefficient $J_{\mathrm{NN}}$ and the conventional Ising coupling in Eq.~(\ref{eq:Ising_Ham}).
The presence of the transverse field further shifts the crossover scale.
These values should be understood only as reference scales, since the molecular graph is finite and irregular and the present model includes a transverse field.

The AFM setting is chosen more empirically.
We fix $\delta=0$, which places the system on the positive-temperature side at a detuning distance from $\delta_{\infty}$ comparable to the FM choice $\delta=2\delta_{\infty}$.
We then choose $\Omega=0.3\,J_{\mathrm{NN}}/\hbar$, which yields a similar effective temperature scale and lies in a region with pronounced negative off-diagonal correlations, as shown in Fig.~\ref{fig:phase_diagram}(c).
Together, the two settings in Eq.~(\ref{eq:H_AFM_FM}) or (\ref{eq:app_H_AFM_FM}) provide complementary quantum feature maps: one dominated by FM-like positive correlations and one dominated by AFM-like negative correlations.

\paragraph{Sensitivity to the AFM working point and to the measurement window.}
The AFM point was selected without task-level tuning.
The scans of Fig.~\ref{fig:phase_diagram} show that $\Omega=0.3\,J_{\mathrm{NN}}/\hbar$ lies inside a broad AFM-like enhanced-correlation region rather than at an isolated optimum: for $\hbar\Omega/J_{\mathrm{NN}}\lesssim0.4$, this region extends over a finite neighborhood of $\delta=0$ on the positive-temperature side of the separatrix $\delta=\delta_\infty$.
Within it, the diagonal-ensemble correlations vary smoothly with $(\Omega,\delta)$ and remain AFM-like, with predominantly short-range negative correlations.
Small changes of $(\Omega,\delta)$ thus change the correlation magnitudes but not the regime, which is set by the sign of the effective interaction [Eq.~(\ref{eq:detuning_choices})].
The AFM/FM complementarity is not specific to molecules: the working points are computed per graph from $\delta_\infty^{\mathcal{G}}$, and chemistry enters only through which graphs admit a unit-disk embedding.
For denser graphs, $\delta_\infty^{\mathcal{G}}$ grows and the FM point $\delta=2\delta_\infty^{\mathcal{G}}$ moves to larger detuning, which rescales the working points without changing the physical picture; the ladder graphs of Sec.~\ref{sec:controls} are a non-molecular example.
These scans do not establish that every intrinsic or downstream score is stable across $\Omega$; this would require regenerating the features and retraining all models.
The exact sampling times are not expected to be critical either.
For a thermalizing system, the information in $A(t)$ is mainly carried by its diagonal-ensemble value $A^{\mathrm{diag}}$ [Eq.~(\ref{eq:app_A_diag})], which the window $[0.1,4]\,\mu$s approximates well (Fig.~\ref{fig:time_evolution}).
A suitable window must extend beyond the relaxation time, so that the time average approaches $A^{\mathrm{diag}}$, and must contain enough samples to average out the oscillatory off-diagonal contributions; consistently, a static component dominates the PCA of App.~\ref{app:pca-analysis}, and the four-statistic summaries retain most of the downstream performance.

\section{Emulation and QPU execution}
\label{app:emulation}

This appendix describes the settings used for emulation, neutral-atom hardware execution, and provides analyses of QGFs produced by our pipeline.

\subsection{Emulation budget and measurement-time window}
\label{app:emu-budget}

The emulator (Sec.~\ref{subsec:pipeline}) uses a state-vector backend for graphs with $N<25$ atoms (numerically exact expectation values from the full $2^N$-dimensional state) and the matrix-product-state backend for larger graphs (controlled bond-dimension truncation). Both interface with a common library so emulated and hardware pulse sequences follow the same protocol. Total cost: ${\sim}800$ GPU-days on an NVIDIA A100 for \ZINC (two pulses) and ${\sim}14$ GPU-days for MUTAG, which is roughly $60\times$ smaller.

We record observables at $N_t=20$ equally spaced times $t_\nu$ with $\nu\in \{1,\dots,N_t\}$ where $t_1=100\,\mathrm{ns}$ and $t_{N_t}=4\, \mu \mathrm{s}$. The lower bound is set by the onset of non-trivial many-body dynamics after the initial transverse-field-dominated regime; the upper bound is chosen long enough that the time-averaged correlations are well approximated.

\subsection{PCA analysis of quantum correlation features}
\label{app:pca-analysis}

\begin{figure*}[!htbp]
\includegraphics[width=\textwidth]{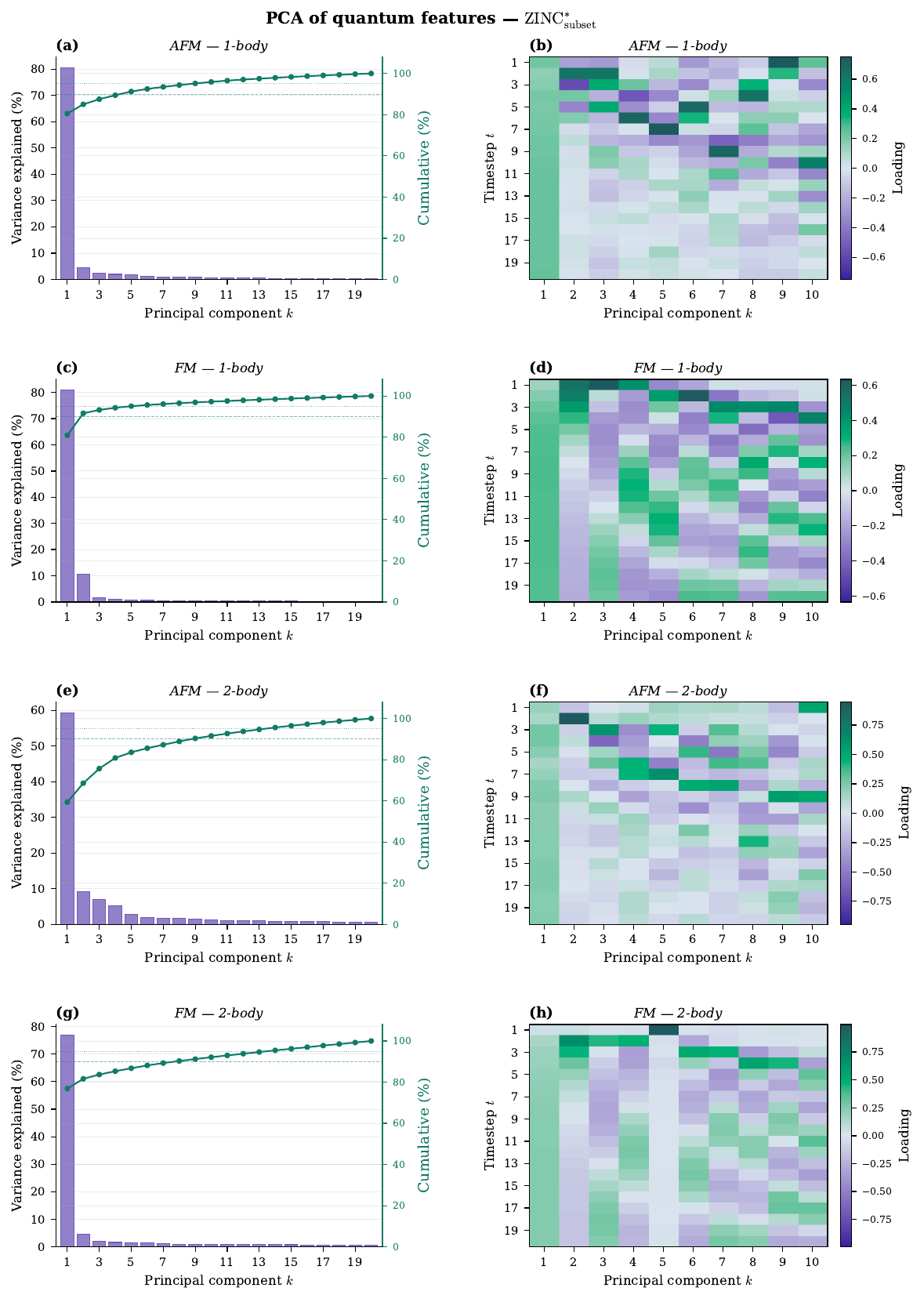}
\caption{
\textbf{PCA of quantum correlation features on $\mathrm{ZINC}_{\mathrm{subset}}^{*}$.} Each row shows a scree plot (left) and a loading heatmap (right) for a given pulse type and feature level. \textbf{(a,b)} AFM 1-body correlations $\boldsymbol{\mathcal{C}}^{(1)}_{i}$ [Eq.~(\ref{eq:quantum_descriptors})]; \textbf{(c,d)} FM 1-body correlations; \textbf{(e,f)} AFM 2-body correlations $\boldsymbol{\mathcal{C}}^{(2)}_{ij}$ (edge-level); \textbf{(g,h)} FM 2-body correlations. In every case the first principal component captures the bulk of the variance (80\%+ for 1-body, 60\%+ for 2-body), indicating that time-averaged activation dominates the representation. The loading heatmaps reveal a structured temporal pattern: PC1 loads uniformly across all timesteps (static component), while PC2 and PC3 exhibit oscillatory sign alternations (dynamic components), consistently across both pulse types and datasets.
}
\label{fig:zinc_pca}
\end{figure*}

To better understand the structure of the quantum feature space, we perform a principal component analysis (PCA) on the normalised AFM and FM correlation features. Fig.~\ref{fig:zinc_pca} presents, for each pulse type and feature type diagonal and off-diagonal, the collection features across all graphs in the dataset, yielding matrices of shape $[N_\text{total}, N_t]$ for diagonal (one row per node) and $[E_\text{total}, N_t]$ for off-diagonal (one row per directed edge), where $N_t = 20$ is the number of simulation time steps.

\paragraph{One-body features $\boldsymbol{\mathcal{C}}^{(1)}_{i}$ [Eq.~(\ref{eq:quantum_descriptors})].}
For both AFM and FM pulses, and on both datasets, the first principal component captures a dominant share of the total variance: approximately $80$--$83\%$ for AFM and a comparable fraction for FM.
This component corresponds to the average correlation level across timesteps, reflecting the time-averaged structural signature of each node.
Reaching $90\%$ explained variance requires 4--5 components, and $95\%$ requires 9--10 components.

\paragraph{Two-body features $\boldsymbol{\mathcal{C}}^{(2)}_{ij}$ [Eq.~(\ref{eq:quantum_descriptors})].}
The variance structure of two-body (pairwise) features is substantially more distributed.
The first principal component accounts for $59$--$65\%$ of the variance, with the second and third components contributing $7$--$9\%$ and $5$--$7\%$ respectively.
Reaching $90\%$ explained variance requires 9--10 components, and $95\%$ requires 14 components.

\paragraph{Interpretation.}
These results suggest that the quantum correlation features carry two distinct types of information:
\begin{enumerate}[topsep=2pt, itemsep=1pt, parsep=0pt]
  \item A dominant \emph{static} component (PC1), capturing the time-averaged correlation strength of each node or edge. This component is well approximated by the mean over timesteps and explains why the stat reduction ($\mu, \sigma, \min, \max$) retains most of the performance in the PE benchmark.
  \item A richer \emph{dynamic} component (PC2 onwards), encoding temporal modulations of the correlations. This component is more prominent in two-body features than in one-body features, consistent with the fact that pairwise correlations encode distance-dependent oscillatory behaviour that varies across the time evolution.
\end{enumerate}

Notably, this variance structure is nearly identical across the two datasets (\ZINC and MUTAG), suggesting that the informational structure of the quantum features is intrinsic to the Hamiltonian simulation rather than dataset-dependent.
This consistency supports the potential transferability of quantum PE across different molecular graph tasks.

\subsection{Details on hardware runs}
\label{annex:shot_number}

 \paragraph{Readout-error correction.}
The readout-error associated with the QPU runs is corrected in a post-processing step.
Hereby, one exploits the pre-calibrated values of false Rydberg detection $\epsilon = P_r\left(\mathrm{observed}=\left|1\right\rangle\,\big|\,\mathrm{true}=\left|0\right\rangle\right)$ and false ground-state detection $\epsilon^\prime = P_r\left(\mathrm{observed}=\left|0\right\rangle\,\big|\,\mathrm{true}=\left|1\right\rangle\right)$; for the employed QPU the values are $\epsilon = 0.025$ and $\epsilon^\prime = 0.1$.
The observed and true qubit states are related through the 1-qubit confusion matrix $A_{1q}$, via
\begin{align}
\left(\begin{array}{c}
P\left(\mathrm{true}=\left|0\right\rangle \right)\\
P\left(\mathrm{true}=\left|1\right\rangle \right)
\end{array}\right) & =A_{1q}^{-1}\;\left(\begin{array}{c}
P\left(\mathrm{obs}=\left|0\right\rangle \right)\\
P\left(\mathrm{obs}=\left|1\right\rangle \right)
\end{array}\right), & A_{1q} & =\left(\begin{array}{cc}
1-\epsilon & \epsilon^{\prime}\\
\epsilon & 1-\epsilon^{\prime}
\end{array}\right),\label{eq:app_confusion_matrix}
\end{align}
where $P\left(\mathrm{obs}=\left|b\right\rangle \right)$ denotes the probability for observing the qubit in state $\left|b\right\rangle$.
The true---readout-error corrected---occupation $\langle\hat{n}_i\rangle^{\mathrm{true}}=P\left(\mathrm{true}=\left|1_i\right\rangle \right)$ [cf. Eq.~(\ref{eq:app_obs_C})] follows from the observed occupation $\langle\hat{n}_i\rangle^{\mathrm{obs}}=P\left(\mathrm{obs}=\left|1_i\right\rangle \right)$ by the simple relation
$\langle\hat{n}_i\rangle^{\mathrm{true}} = \sum_{k=1}^2 \left(A_{1q}^{-1}\right)_{2,k}\,\left(v^{\mathrm{n,obs}}_i\right)_k$
where $\boldsymbol{v}^{\mathrm{n,obs}}_i=\left(1-\langle\hat{n}_i\rangle^{\mathrm{obs}},\,\langle\hat{n}_i\rangle^{\mathrm{obs}} \right)^T$.
The two-qubit correlation observable $\langle\hat{n}_i\hat{n}_j\rangle^{\mathrm{true}}=P\left(\mathrm{true}=\left|1_i1_j\right\rangle \right) $ [cf. Eq.~(\ref{eq:app_obs_C})] is treated analogously.
In extension to Eq.~(\ref{eq:app_confusion_matrix}) we introduce the two-qubit confusion matrix $A_{2q} = A_{1q} \otimes A_{1q}$.
The extended observation vector becomes
\begin{align}
\boldsymbol{v}^{\mathrm{nn,obs}}_{i,j}=\left(
\left\langle\left(1-\hat{n}_i\right)\left(1-\hat{n}_j\right)\right\rangle^{\mathrm{obs}},\,
\left\langle\left(1-\hat{n}_i\right)\hat{n}_j\right\rangle^{\mathrm{obs}},\,
\left\langle \hat{n}_i\left(1-\hat{n}_j\right)\right\rangle^{\mathrm{obs}},\,
\left\langle\hat{n}_i\hat{n}_j\right\rangle^{\mathrm{obs}}
\right)^T,
\label{eq:app_v_nn}
\end{align}
associated with observing the qubits in the states $\left(\left|0_i 0_j\right\rangle,\,
\left|0_i 1_j\right\rangle,\,
\left|1_i 0_j\right\rangle,\,
\left|1_i 1_j\right\rangle  \right)$.
Finally, the readout-error corrected correlation follows from the relation
$\langle\hat{n}_i\hat{n}_j\rangle^{\mathrm{true}} = \sum_{k=1}^4 \left(A_{2q}^{-1}\right)_{4,k}\,\left(v^{\mathrm{nn,obs}}_{i,j}\right)_k$.

\paragraph{Shot budget estimate for MUTAG connected correlations.}
Using the Central Limit Theorem (CLT) for the simultaneous estimation of all one-body and two-body connected correlations at global confidence level $1-\delta_\epsilon=95\%$, we count
$M=N(N+1) /2 $
that is, $N$ one-body observables together with $M-N=N(N-1)/2$ distinct off-diagonal correlations.
Here $\Phi$ denotes the cumulative distribution function (CDF) of a standard normal random variable $Z\sim\mathcal{N}(0,1)$, i.e., $\Phi(x)=\mathbb{P}(Z\le x)$, and $\Phi^{-1}$ is its quantile function. The corresponding Gaussian quantile is
\begin{align}
    k_N = \Phi^{-1}\!\left[1-\delta_\epsilon/(2M)\right],
\end{align}
and we use $\delta_\epsilon=0.05$.
Fixing an absolute error tolerance $\varepsilon$ and taking $v_{\max}=\sigma_{\max}^2$ as the maximal estimator variance (empirically estimated from the emulations performed), the required shot count per graph and per measurement time is
\begin{align}
    N_{\mathrm{shots}}^{\mathrm{CLT}}(N)
=
\left\lceil
\frac{v_{\max}}{\varepsilon^2}
\,k_N^2
\right\rceil.
\end{align}
With $v_{\max}=0.06$ and $\varepsilon=0.03$, this becomes $N_{\mathrm{shots}}^{\mathrm{CLT}}(N)
=
\left\lceil
66.67\,k_N^2
\right\rceil$.
The choice $\varepsilon=0.03$ is consistent with the natural scale of connected occupation correlations, which satisfy $|\boldsymbol{\mathcal{C}}|\leq 1/4$; it therefore corresponds to an absolute resolution of about $12\%$ of the full correlation range, sufficient to resolve medium-to-large local antiferromagnetic correlations.

For a standard MUTAG graph, taking $N\simeq 18$ gives $M=171$ and $k_{18}\approx 3.62$ such
that the shot count becomes
$N_{\mathrm{shots}}^{\mathrm{CLT}}(18)
=\left\lceil
\frac{0.06}{0.03^2}\times k_{18}^2
\right\rceil
=874$.

\paragraph{QPU cost scaling.}
The acquisition time scales as $T_{\mathrm{QPU}}\simeq G\,R\,N_t\,\bar N_{\mathrm{shots}}/f_{\mathrm{rep}}$, with $G$ graphs, $R$ quench regimes, $N_t$ measurement times, $\bar N_{\mathrm{shots}}$ the mean shot count per graph and time, and $f_{\mathrm{rep}}$ the effective repetition rate, including register loading, rearrangement and readout.
The shot count grows only logarithmically with graph size ($N_{\mathrm{shots}}^{\mathrm{CLT}}=734$ at $N=10$, $901$ at $N=20$, $983$ at $N=28$), so the cost is dominated by $f_{\mathrm{rep}}$.
For the MUTAG run ($G=188$, $R=1$, $N_t=4$, $\approx6.5\times10^5$ shots in total), the $\sim7.5$ QPU-days correspond to $f_{\mathrm{rep}}\approx1$\,Hz.
At this rate, a two-regime \ZINC run ($G=11{,}368$, $R=2$, $N_t=4$) would take $\sim950$ days, against $\sim95$ days at $10$\,Hz and $\sim10$ days at $100$\,Hz; using all $N_t=20$ times multiplies these figures by five.
Three-dimensional registers would additionally improve the embedding quality, with fewer unwanted interactions between non-adjacent atoms.

\subsection{Emulator--QPU concordance at \texorpdfstring{$t_4$}{t4}}
\label{app:emu-qpu-concordance}

We assess the agreement between idealized emulation and end-to-end QPU
execution on the MUTAG dataset at the AFM operating point
(Sec.~\ref{subsec:Pulse-shaping}), at the sampled measurement time
\texorpdfstring{$t_4$}{t4}. The reported times,
$t_{\mathrm{QPU}} = 940$~ns and $t_{\mathrm{emu}} = 715$~ns are different because, as the available max amplitude was lower in the QPU than in emulation, sampling times have to be rescaled accordingly (along with $r_{NN}$). All $188$ MUTAG graphs
admitting a 2D unit-disk embedding (Sec.~\ref{subsec:Graph_Embedding})
were measured under the shot budget of
App.~\ref{annex:shot_number} ($N_{\mathrm{shots}} = 874$).

For each graph and each correlation type we evaluate the
Spearman rank correlation $\rho_s$ between QPU and emulation entries,
and aggregate the result over graphs (per-graph distribution) and over
(graph, entry) pairs (pooled scatter).

\begin{figure}[t]
  \centering
  \includegraphics[width=\linewidth]{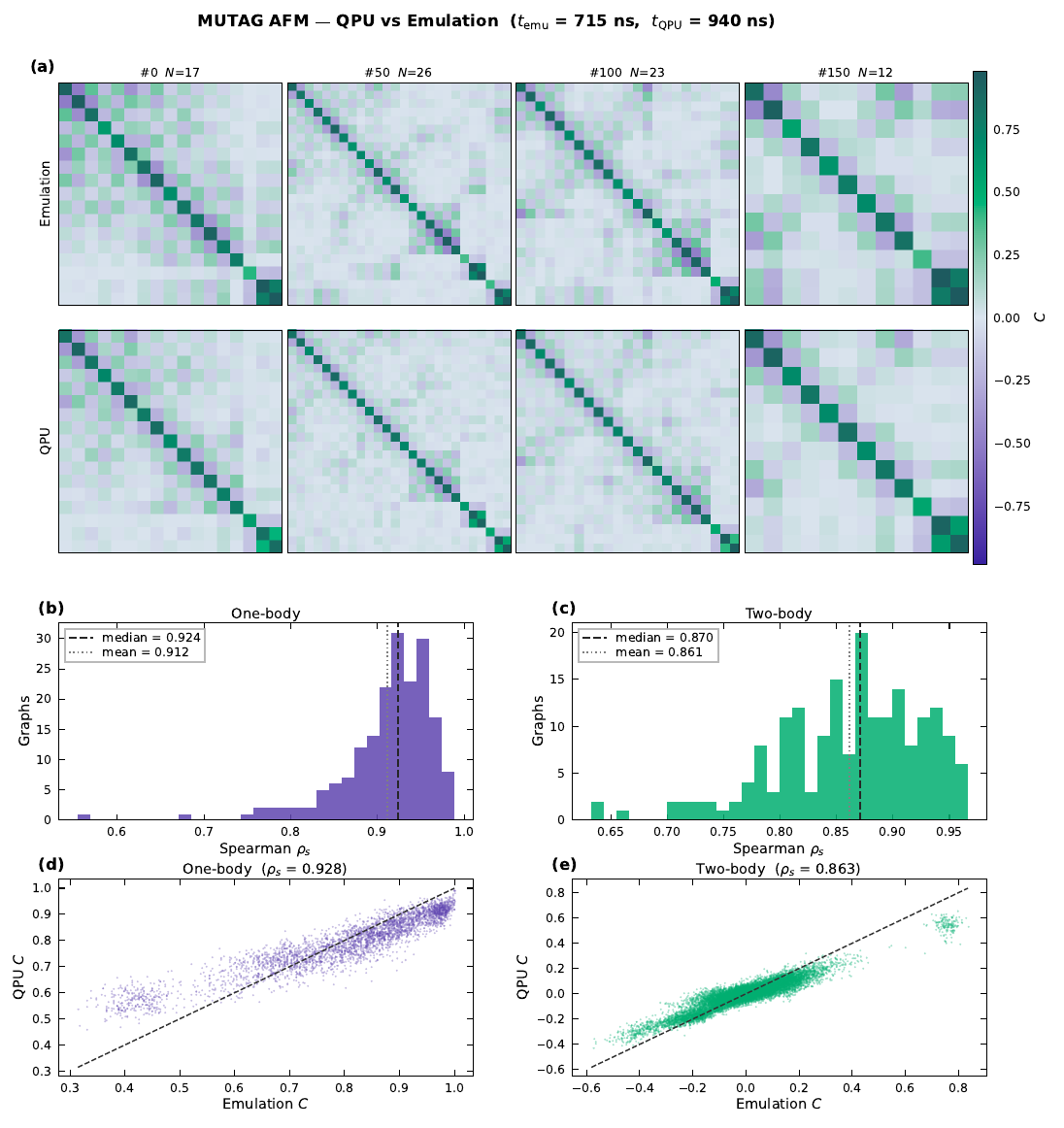}
  \caption{%
    \textbf{Emulator--QPU concordance on MUTAG at \texorpdfstring{$t_4$}{t4}}
    ($t_{\mathrm{emu}} = 715$~ns, $t_{\mathrm{QPU}} = 940$~ns; AFM
    operating point).
    \textbf{(a)}~Normalized connected correlation matrices $C_{i,j}(t)$ [Eq.~(\ref{eq:obs_C})]
    for the four representative graphs as functions of $i$ and $j$;
    top row: emulation, bottom row: QPU.
    The presented correlations are normalized across the dataset, showing what is actually supplied to the models.
    \textbf{(b,c)}~Per-graph distribution of the Spearman rank
    correlation $\rho_s$ between QPU and emulation entries, over the
    $188$ graphs for the
    (b) one-body [$C_{i,i}(t)$] and (c) two-body [$C_{i,j\neq i}(t)$] channels; dashed and dotted lines
    mark the median and mean.
    \textbf{(d,e)}~Pooled scatter of every QPU entry against its
    emulation counterpart, for one-body (d, $\rho_s = 0.928$) and
    two-body (e, $\rho_s = 0.863$); the dashed line marks $y = x$.
    Hardware preserves the rank ordering of correlation entries within
    each graph; the residual contraction towards zero at large $|C_{i,j}(t)|$
    is consistent with finite-time decoherence and
    finite-shot SPAM.}
  \label{fig:emu-qpu-concordance}
\end{figure}

Figure~\ref{fig:emu-qpu-concordance} summarizes the comparison.
Panel~(a) shows one timestep of the raw \C matrices for four
representative graphs of increasing size; the diagonal and the dominant
off-diagonal patterns are preserved on hardware. Panels~(b,c) report the
per-graph distribution of $\rho_s$: the one-body channel concentrates
above $\rho_s = 0.9$ (median $0.924$, mean $0.912$). The two-body
channel concentrates lower (median $0.870$, mean $0.861$), as expected
since two-body entries have lower magnitudes and are more sensitive to
shot noise and SPAM. Another explanation is the expected low magnitude of long-range correlations, which are concentrated around 0 and induce a lower rank-correlation as they are in the noise. Panels~(d,e) pool every correlation entry across the $188$ graphs into a single scatter: the one-body channel
($\rho_s = 0.928$) lies tightly around the diagonal with a mild
contraction towards zero on the high-magnitude side, consistent with
finite-time decoherence; the off-diagonal correlations ($\rho_s = 0.863$) show a
broader scatter and a small negative bias for entries with
$|\boldsymbol{\mathcal{C}}^{(2)}_{ij}| \gtrsim 0.5$, again compatible with depolarizing noise
on the most strongly correlated pairs.

Two points support the use of QPU features as drop-in replacements for
emulation features in Sec.~\ref{sec:gnn-downstream}. First, the rank
ordering of correlation entries is preserved within each graph at both
the one- and two-body level (panels~b,c), so GNN backbones using these
features as node or edge attributes see the same relative structure on
hardware as in simulation. Second, the worst-case graphs in panels~(b,c)
are systematically among the smallest of MUTAG, where the absolute
number of distinct entries is so small that a few miscalibrated shots
can shift the rank substantially; this regime does not extend to the
larger graphs of the dataset.

We emphasize that this concordance study is restricted to a single
operating point (AFM, $t_4$) on a single dataset (MUTAG); a full QPU
run on \ZINC remains beyond the budget of this study and
is left for future work.

\section{GNN Methods}
\label{app:methods}

In this appendix, we provide various methodological details about the GNN benchmark methodology and model details.

\subsection{Features notations}
\label{app:features}

Table~\ref{tab:feature-notation} summarizes all feature sets, their origin, and dimensionality.
\begin{table}[t]
\centering
\caption{%
Feature sets used throughout the paper.
Here $N_t=20$ denotes the number of readout times in the full time-resolved quantum feature trajectories.
}\label{tab:feature-notation}
\small
\begin{tabular}{l l c}
\toprule
\textbf{Feature set} & \textbf{Description} & $d$ \\
\midrule
RRWP & $\mathbf{M}^k,\; k{=}1{\dots}20$ & $20$ \\
Laplacian & $\phi_2{\dots}\phi_5$ of $\mathbf{L}$ & $4$ \\
AFM / FM & \Cone, \Ctwo at $N_t$ times & $20$ \\
AFM+FM & concat.\ AFM \& FM & $40$ \\
AFM$_{\mathrm{stat}}$ / FM$_{\mathrm{stat}}$ & $\operatorname{stat}(\cdot)$ over $N_t$ & $4$ \\
RRWP $+$ X & concat.\ RRWP \& X & $20{+}d_X$ \\
HW-AFM & QPU measurements & $4$ \\
HW-AFM$_{\mathrm{stat}}$ & $\operatorname{stat}(\cdot)$ of HW-AFM & $4$ \\
\bottomrule
\end{tabular}
\end{table}

\subsection{GNN details}
\label{app:GNN}

This appendix complements Sec.~\ref{subsec:architectures} with the per-backbone descriptions, the GT block structure, and the training protocol omitted from the main text.

\paragraph{Feature injection.} In every backbone the original dataset attributes (atom and bond types) are retained; structural descriptors are concatenated to the matching input channel. In the MPNN backbones, one-body features $\boldsymbol{\mathcal{C}}^{(1)}_{i}$ [Eq.~(\ref{eq:quantum_descriptors})] are concatenated to the node feature vector before the first layer. In GT, dataset attributes and one-body correlations are projected independently to the hidden dimension and summed (see below). Two-body features $\boldsymbol{\mathcal{C}}^{(2)}_{ij}$ [Eq.~(\ref{eq:quantum_descriptors})] are injected through whichever pairwise interface the architecture exposes: edge weights for GCN, edge attributes for GAT/GINE/Transformer/GeneralConv, and dense per-head attention biases for GT. This convention ensures that every architecture receives the same raw feature content; the architectural interface is the only differentiator.

\paragraph{Message-passing backbones.} GCN~\citep{kipf2017semi} performs isotropic neighborhood aggregation through a symmetric degree-normalized adjacency operator. GINE~\citep{hu2020strategies} extends the Graph Isomorphism Network (GIN)~\citep{xu2019powerful} with explicit edge-feature injection into the message function, inheriting GIN's $1$-WL expressiveness while accepting edge attributes. GeneralConv~\citep{li2020deepergcn} provides a more flexible aggregation rule with learnable message transformations.

\paragraph{Attention-based backbones.} GAT~\citep{velickovic2018graph} computes learned attention coefficients over local neighborhoods. The Transformer layer~\citep{shi2021masked} extends self-attention to the full node set with edge-aware biases, serving as a standard global-attention reference distinct from our custom GT.

\paragraph{Training protocol.} The downstream benchmark compares \emph{feature sets}, not architecture--feature combinations. We fix one set of training hyperparameters per architecture--dataset pair (Table~\ref{tab:training-hparams}) and reuse it across every feature configuration; no per-configuration hyperparameter search is performed. All downstream models are trained with AdamW~\citep{loshchilov2019decoupled} using a linear warm-up phase followed by ReduceLROnPlateau (factor $0.5$); early stopping monitors the validation loss (MAE for ZINC, cross-entropy for MUTAG). Folds for MUTAG are stratified by graph size and class, and folds are consistent across all experiments.

\begin{table}[h]
\centering
\caption{%
Main GT hyperparameter configurations used for each dataset.
The corresponding parameter budgets are used to approximately match model capacity across backbones.
}
\label{tab:training-hparams}
\small
\begin{tabular}{lcc}
\toprule
\textbf{Parameter} & \textbf{ZINC\textsubscript{subset}$^{*}$} & \textbf{MUTAG} \\
\midrule
\multicolumn{3}{l}{\textit{Architecture}} \\
Hidden dimension $D$       & 256   & 32 \\
Number of heads $H$        & 16    & 4 \\
Number of layers $L$       & 4     & 3 \\
FFN dimension $D_\text{ffn}$ & 512 & 64 \\
Dropout                    & 0.1   & 0.3 \\
Pooling                    & \multicolumn{2}{c}{Mean} \\
\midrule
\multicolumn{3}{l}{\textit{Training}} \\
Optimiser                  & \multicolumn{2}{c}{AdamW} \\
Learning rate              & \multicolumn{2}{c}{$5 \times 10^{-4}$} \\
Weight decay               & $10^{-5}$ & $10^{-3}$ \\
Batch size                 & 128   & 32 \\
Max epochs                 & 1000  & 300 \\
Linear warmup epochs       & 50    & 20 \\
Early stopping patience    & 200   & 75 \\
LR scheduler               & \multicolumn{2}{c}{ReduceLROnPlateau (factor $0.5$)} \\
\midrule
\multicolumn{3}{l}{\textit{Evaluation}} \\
Loss                       & MAE   & Cross-entropy \\
Metric                     & MAE   & F$_1$ macro \\
Seeds                      & 10    & 5 per fold \\
Folds                      & --    & 10 \\
\bottomrule
\end{tabular}
\end{table}

\subsection{Graph Transformer architecture}
\label{app:model-details}

We describe the Graph Transformer (GT) architecture used in the GNN downstream benchmark experiments of Section~\ref{sec:gnn-downstream}. The key distinction from the message-passing backbones is that the GT can consume the dense off-diagonal correlation tensor \Ctwo [Eq.~(\ref{eq:quantum_descriptors})] for all node pairs, rather than only using pairwise information on molecular bonds.

\paragraph{Input representation.}

Each molecular graph is represented by a set of atoms $\mathcal{V}_\mathcal{G}$ and a set of bonds $\mathcal{E}_\mathcal{G}$. Each atom type is mapped to a learned node embedding
\begin{equation}
    \mathbf{h}_i^{(0)}
    \in
    \mathbb{R}^{D},
    \qquad
    i \in \mathcal{V}_\mathcal{G} ,
\end{equation}
where $D$ is the hidden dimension. Bond attributes are embedded separately and used to construct additive attention biases, as described below.

\paragraph{Diagonal correlation injection.}

The diagonal correlations \Cone are injected at the node level before the first Transformer block. When these correlations are used, they are mapped to the hidden dimension by a residual MLP encoder:
\begin{equation}
    \mathbf{p}_i
    =
    f_{\theta}(\boldsymbol{\mathcal{C}}^{(1)}_{i}),
    \qquad
    f_{\theta}(\mathbf{x})
    =
    \mathbf{W}_{\mathrm{skip}}\mathbf{x}
    +
    \mathbf{W}_{\mathrm{out}}
    \sigma
    \bigl(
        \mathbf{W}_{\mathrm{in}}\mathbf{x}
    \bigr),
    \label{eq:diagonal-correlation-encoder}
\end{equation}
where $\sigma$ is the ReLU nonlinearity. The encoded vector is added to the learned atom embedding:
\begin{equation}
    \mathbf{h}_i^{(0)}
    \leftarrow
    \mathbf{h}_i^{(0)}
    +
    \mathbf{p}_i .
\end{equation}
When diagonal correlations are not used, this encoder is omitted and the initial node representation is the learned atom embedding alone.

\paragraph{Off-diagonal correlation biases.}

The off-diagonal correlations \Ctwo are injected as additive attention biases. Unlike message-passing backbones, which can only use pairwise information on molecular edges, the GT can consume \Ctwo for all node pairs $(i,j)$ in the dense graph.

When off-diagonal correlations are used, they are mapped to one scalar bias per attention head by a two-layer MLP:
\begin{equation}
    b_{ij}^{\mathrm{corr},(h)}
    =
    \left[
        \mathbf{W}_{4}
        \,
        \sigma
        \bigl(
            \mathbf{W}_{3}
            \boldsymbol{\mathcal{C}}^{(2)}_{ij}
        \bigr)
    \right]_h,
    \qquad
    h=1,\dots,H ,
    \label{eq:off-diagonal-correlation-bias}
\end{equation}
where $H$ is the number of attention heads. When off-diagonal correlations are not used, this correlation-bias encoder is omitted and $b_{ij}^{\mathrm{corr},(h)}$ is absent.

\paragraph{Edge biases.}

Molecular bond attributes are incorporated through a separate additive edge bias. For each attention head $h$, the bond representation associated with a node pair $(i,j)$ is projected to a scalar
\begin{equation}
    b_{ij}^{\mathrm{edge},(h)} .
\end{equation}
This term encodes molecular bond information, whereas $b_{ij}^{\mathrm{corr},(h)}$ encodes the dense off-diagonal correlations. The two sources of pair information are kept separate and combined only at the level of the attention logits.

\paragraph{Multi-head attention with additive pair biases.}

At Transformer layer $\ell$, node representations are projected to queries, keys, and values. For each head $h$, let
\begin{equation}
    \mathbf{q}_i^{(h)},
    \mathbf{k}_j^{(h)},
    \mathbf{v}_j^{(h)}
    \in
    \mathbb{R}^{d_{\mathrm{head}}},
    \qquad
    d_{\mathrm{head}} = D/H .
\end{equation}
The attention logit from node $i$ to node $j$ is
\begin{equation}
    \alpha_{ij}^{(h)}
    =
    \frac{
        \mathbf{q}_i^{(h)}
        \cdot
        \mathbf{k}_j^{(h)}
    }{
        \sqrt{d_{\mathrm{head}}}
    }
    +
    b_{ij}^{\mathrm{edge},(h)}
    +
    b_{ij}^{\mathrm{corr},(h)}
    +
    m_{ij},
    \label{eq:gt-attention-logit}
\end{equation}
where $m_{ij}$ is the padding mask, equal to $0$ for valid node pairs and $-\infty$ when node $j$ is a padding node. The attention weights are obtained by normalising over the key index:
\begin{equation}
    a_{ij}^{(h)}
    =
    \frac{
        \exp
        \bigl(
            \alpha_{ij}^{(h)}
        \bigr)
    }{
        \sum_{k \in V_G}
        \exp
        \bigl(
            \alpha_{ik}^{(h)}
        \bigr)
    } .
\end{equation}
The output of head $h$ at node $i$ is
\begin{equation}
    \mathbf{z}_i^{(h)}
    =
    \sum_{j \in V_G}
    a_{ij}^{(h)}
    \mathbf{v}_j^{(h)} .
\end{equation}
The head outputs are concatenated and linearly projected back to $\mathbb{R}^{D}$.

\paragraph{Transformer block.}

Each of the $L$ Transformer blocks consists of biased multi-head attention followed by a feed-forward network, with residual connections, dropout, and layer normalisation:
\begin{align}
    \tilde{\mathbf{h}}^{(\ell)}
    &=
    \operatorname{LayerNorm}
    \bigl(
        \mathbf{h}^{(\ell)}
        +
        \operatorname{Dropout}
        \bigl(
            \operatorname{MHA}
            \bigl(
                \mathbf{h}^{(\ell)},
                \mathbf{b}^{\mathrm{edge}},
                \mathbf{b}^{\mathrm{corr}}
            \bigr)
        \bigr)
    \bigr),
    \\
    \mathbf{h}^{(\ell+1)}
    &=
    \operatorname{LayerNorm}
    \bigl(
        \tilde{\mathbf{h}}^{(\ell)}
        +
        \operatorname{Dropout}
        \bigl(
            \operatorname{FFN}
            \bigl(
                \tilde{\mathbf{h}}^{(\ell)}
            \bigr)
        \bigr)
    \bigr).
    \label{eq:gt-transformer-block}
\end{align}
Here $\operatorname{MHA}$ denotes multi-head attention with the additive edge and correlation biases defined above. The feed-forward network is
\begin{equation}
    \operatorname{FFN}(\mathbf{x})
    =
    \mathbf{W}_{6}
    \,
    \operatorname{Dropout}
    \bigl(
        \sigma
        \bigl(
            \mathbf{W}_{5}\mathbf{x}
        \bigr)
    \bigr),
    \label{eq:gt-ffn}
\end{equation}
with hidden dimension $D_{\mathrm{ffn}}$.

\paragraph{Batching and padding.}

Graphs of different sizes are batched by densifying the node representations with \texttt{to\_dense\_batch} from PyTorch Geometric~\citep{fey2019fast}. This produces node tensors of shape
\begin{equation}
    B
    \times
    N_{\max}
    \times
    D ,
\end{equation}
where $B$ is the batch size and $N_{\max}$ is the maximum number of nodes in the batch. Off-diagonal correlation tensors and attention-bias tensors are padded consistently to shape
\begin{equation}
    B
    \times
    H
    \times
    N_{\max}
    \times
    N_{\max}.
\end{equation}
Padding nodes are excluded from the attention softmax by adding $-\infty$ to the corresponding logits through the mask term $m_{ij}$ in Eq.~\eqref{eq:gt-attention-logit}.

\paragraph{Graph-level output.}

After the final Transformer block, node representations are pooled over valid, non-padding nodes:
\begin{equation}
    \mathbf{h}_\mathcal{G}
    =
    \frac{1}{|\mathcal{V}_\mathcal{G}|}
    \sum_{i \in \mathcal{V}_\mathcal{G}}
    \mathbf{h}_i^{(L)} .
\end{equation}
The graph representation $\mathbf{h}_\mathcal{G}$ is passed through a two-layer MLP with ReLU activation and dropout. The output is a scalar for regression tasks and a vector of logits for classification tasks.

This architecture therefore separates three sources of information: atom identities enter through learned node embeddings, diagonal correlations \Cone are added to the node representations, and off-diagonal correlations \Ctwo enter as dense additive biases in the attention logits. This is the mechanism that allows the GT to use pairwise correlations beyond the molecular edge set $\mathcal{E}_\mathcal{G}$.

\subsection{Ablation: diagonal and off-diagonal quantum features}
\label{app:gnn-ablation}

We assess the impact of diagonal $\boldsymbol{\mathcal{C}}^{(1)}_i$ and off-diagonal $\boldsymbol{\mathcal{C}}^{(2)}_{ij}$ [Eq.~(\ref{eq:quantum_descriptors})] quantum features on model performance. We use a subset ZINC$_{\mathrm{tiny}}$ of $1.5\times 10^3$ graphs, with property distributions matched to the ZINC$_{\mathrm{small}}^{*}$ subset of $\approx 12\times 10^3$ graphs, and evaluate two representative architectures: a message-passing GCN and an attention-based GAT.
All experiments use the same fixed backbone as in the benchmark experiments.
All models use the training and evaluation settings, including dropout and pooling, reported in Table~\ref{tab:training-hparams}.
The model sizes are given in Table~\ref{tab:model-sizes}. We compare classical ZINC features (atom and bond types) against their augmentation with one- and two-body quantum features. Results in Table~\ref{tab:quantum-ablation-compact} show that quantum features consistently improve over classical features when averaged across six model sizes and five seeds.
Tables~\ref{tab:gcn-zinc-per-size-all} and~\ref{tab:gat-zinc-per-size-all}
show that the large configuration has better performance overall, while larger models provide no systematic further improvement.
Hence, we select the large backbone configuration for the emulated benchmark and QPU validation (Sec.~\ref{sec:gnn-downstream}).

\begin{figure}[htbp]
\centering
\includegraphics[width=\linewidth]{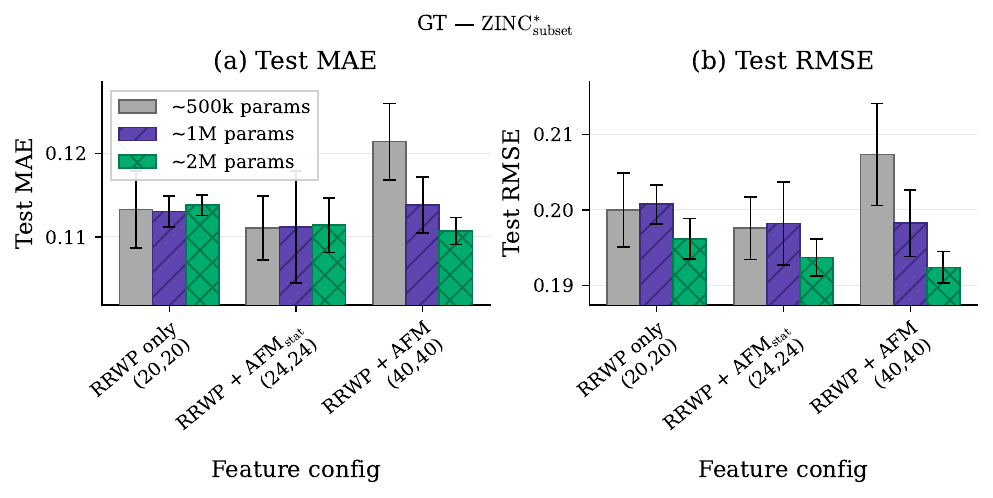}
\caption{
Results for model GT on \ZINC. Grouped bar charts show \textbf{(a)}~test MAE and \textbf{(b)}~test RMSE for three feature configurations: RRWP alone ($d_\mathrm{abs}=20,\;d_\mathrm{rel}=20$), RRWP + AFM$_\mathrm{stat}$ ($d_\mathrm{abs}=24,\;d_\mathrm{rel}=24$), and RRWP + AFM ($d_\mathrm{abs}=40,\;d_\mathrm{rel}=40$), each evaluated at three model capacities (Number of trainable parameters $\in \{500k,1M,2M\}$). Error bars report the standard deviation over 5 independent random seeds; in this study the reported epoch was selected on the test split. Increasing model capacity consistently reduces the gap between feature configurations, suggesting that the relative underperformance of RRWP + AFM at low capacity is a capacity bottleneck rather than a fundamental incompatibility between the two positional encodings. } \label{fig:capacity-study} \end{figure}

In Fig.~\ref{fig:capacity-study}, we show the error metrics used for selection of the graph transformer (GT) model with the best capacity for our benchmarking experiments.

\begin{table}
\caption{
  Model size configurations for GCN and GAT architectures.
  $N_l$: number of message-passing (backbone) or feedforward (prediction head) layers.
  $N_d$: hidden feature dimension per layer.
}
\label{tab:model-sizes}
\centering
\begin{tabular}{l c c c c}
\toprule
\multirow{2}{*}{Model} & \multicolumn{2}{c}{Backbone} & \multicolumn{2}{c}{Prediction head} \\
\cmidrule(lr){2-3} \cmidrule(lr){4-5}
 & $N_l$ & $N_d$ & $N_l$ & $N_d$ \\
\midrule
  Tiny\ \        & 2 & 16  & 2 & 16  \\
  Small\ \     & 2 & 32  & 2 & 32  \\
  Medium\ \    & 3 & 32  & 3 & 32  \\
  Large\ \    & 4 & 64  & 4 & 64  \\
  Huge\ \         & 4 & 128 & 4 & 8   \\
  V.~Huge\ \    & 5 & 256 & 3 & 256 \\
  Massive\ \ & 4 & 512 & 3 & 512 \\
  Giant\ \     & 5 & 512 & 3 & 512 \\
\bottomrule
\end{tabular}
\end{table}

\begin{table}
\small
\caption{%
Quantum feature ablation on ZINC$_{\text{tiny}}$ ($\sim$1.5k graphs, emulation).
All node and edge quantum features are used unless stated otherwise.
Mean MAE ($\downarrow$) $\pm$ standard deviation over 10 seeds is reported.
\mbox{$\best{\text{Best}}$} and \mbox{$\second{\text{2nd best}}$} indicate the best and runner-up feature sets in each column.
}
\label{tab:quantum-ablation-compact}
\centering
\resizebox{0.45\columnwidth}{!}{
\begin{tabular}{l c c}
\toprule
\multirow{2}{*}{Method} & \multicolumn{2}{c}{ZINC$_{\text{test}}^{*}$ (MAE $\downarrow$)} \\
\cmidrule(lr){2-3}
 & GCN & GAT \\
\midrule
\multicolumn{3}{l}{\textit{Baseline}} \\
  No PE & $0.520{_{\pm 0.050}}$ & $0.480{_{\pm 0.032}}$ \\
\midrule
\multicolumn{3}{l}{\textit{Quantum (FM)}} \\
  FM$_{1\text{b}}$ & $0.512{_{\pm 0.031}}$ & $0.473{_{\pm 0.045}}$ \\
  FM$_{2\text{b}}$ & $0.478{_{\pm 0.030}}$ & $0.456{_{\pm 0.031}}$ \\
  FM & $0.493{_{\pm 0.035}}$ & $0.482{_{\pm 0.055}}$ \\
\midrule
\multicolumn{3}{l}{\textit{Quantum (AFM)}} \\
  AFM$_{1\text{b}}$ & $\best{0.443{_{\pm 0.031}}}$ & $0.454{_{\pm 0.066}}$ \\
  AFM$_{2\text{b}}$ & $0.532{_{\pm 0.025}}$ & $\best{0.449{_{\pm 0.041}}}$ \\
  AFM & $\second{0.446}{_{\pm 0.020}}$ & $\second{0.450}{_{\pm 0.067}}$ \\
\bottomrule
\end{tabular}
}
\end{table}

\begin{table*}
\caption{%
Model-size ablation for GCN on ZINC$_{\text{tiny}}$ ($\sim$1.5k graphs, emulation).
Mean MAE ($\downarrow$) $\pm$ standard deviation over 10 seeds is reported.
\mbox{$\best{\text{Best}}$} and \mbox{$\second{\text{2nd best}}$} indicate the best and runner-up model sizes in each column.
}
\label{tab:gcn-zinc-per-size-all}
\centering
\resizebox{\textwidth}{!}{
\begin{tabular}{l c c c c c c c c}
\toprule
\multirow{2}{*}{Method} & \multicolumn{8}{c}{GCN on ZINC$_{\text{test}}^{*}$ (MAE $\downarrow$)} \\
\cmidrule(lr){2-9}
 & Tiny & Small & Medium & Large & Huge & V.~Huge & Massive & Giant \\
\midrule
\multicolumn{9}{l}{\textit{Baseline}} \\
  ZINC (Baseline) & $0.530{_{\pm 0.044}}$ & $0.529{_{\pm 0.025}}$ & $0.490{_{\pm 0.049}}$ & $0.513{_{\pm 0.062}}$ & $0.515{_{\pm 0.055}}$ & $0.527{_{\pm 0.048}}$ & $0.528{_{\pm 0.042}}$ & $0.546{_{\pm 0.042}}$ \\
\midrule
\multicolumn{9}{l}{\textit{Quantum (FM)}} \\
  FM$_{1\text{b}}$ & $0.518{_{\pm 0.031}}$ & $0.520{_{\pm 0.014}}$ & $0.514{_{\pm 0.033}}$ & $0.520{_{\pm 0.043}}$ & $0.503{_{\pm 0.028}}$ & $0.506{_{\pm 0.029}}$ & $0.508{_{\pm 0.021}}$ & $0.516{_{\pm 0.026}}$ \\
  FM$_{2\text{b}}$ & $0.478{_{\pm 0.026}}$ & $0.479{_{\pm 0.025}}$ & $0.461{_{\pm 0.018}}$ & $0.473{_{\pm 0.026}}$ & $0.468{_{\pm 0.028}}$ & $0.478{_{\pm 0.030}}$ & $0.494{_{\pm 0.028}}$ & $0.511{_{\pm 0.036}}$ \\
  FM & $0.524{_{\pm 0.013}}$ & $0.496{_{\pm 0.024}}$ & $0.488{_{\pm 0.011}}$ & $0.484{_{\pm 0.035}}$ & $0.481{_{\pm 0.042}}$ & $0.492{_{\pm 0.036}}$ & $0.496{_{\pm 0.033}}$ & $0.526{_{\pm 0.038}}$ \\
\midrule
\multicolumn{9}{l}{\textit{Quantum (AFM)}} \\
  AFM$_{1\text{b}}$ & $\second{0.469}{_{\pm 0.023}}$ & $\best{0.450{_{\pm 0.023}}}$ & $\best{0.441{_{\pm 0.036}}}$ & $\best{0.434{_{\pm 0.034}}}$ & $\best{0.434{_{\pm 0.031}}}$ & $\second{0.433}{_{\pm 0.026}}$ & $\second{0.451}{_{\pm 0.033}}$ & $\second{0.458}{_{\pm 0.021}}$ \\
  AFM$_{2\text{b}}$ & $0.554{_{\pm 0.016}}$ & $0.532{_{\pm 0.026}}$ & $0.538{_{\pm 0.005}}$ & $0.522{_{\pm 0.030}}$ & $0.534{_{\pm 0.023}}$ & $0.527{_{\pm 0.025}}$ & $0.533{_{\pm 0.024}}$ & $0.546{_{\pm 0.026}}$ \\
  AFM & $\best{0.464{_{\pm 0.009}}}$ & $\second{0.459}{_{\pm 0.010}}$ & $\second{0.455}{_{\pm 0.014}}$ & $\second{0.448}{_{\pm 0.021}}$ & $\second{0.438}{_{\pm 0.017}}$ & $\best{0.433{_{\pm 0.024}}}$ & $\best{0.442{_{\pm 0.016}}}$ & $\best{0.452{_{\pm 0.016}}}$ \\
\bottomrule
\end{tabular}
}
\end{table*}

\begin{table*}
\caption{%
Model-size ablation for GAT on ZINC$_{\text{tiny}}$ ($\sim$1.5k graphs, emulation).
Mean MAE ($\downarrow$) $\pm$ standard deviation over 10 seeds is reported.
\mbox{$\best{\text{Best}}$} and \mbox{$\second{\text{2nd best}}$} indicate the best and runner-up model sizes in each column.
}
\label{tab:gat-zinc-per-size-all}
\centering
\resizebox{\textwidth}{!}{
\begin{tabular}{l c c c c c c c c}
\toprule
\multirow{2}{*}{Method} & \multicolumn{8}{c}{GAT on ZINC$_{\text{test}}^{*}$ (MAE $\downarrow$)} \\
\cmidrule(lr){2-9}
 & Tiny & Small & Medium & Large & Huge & V.~Huge & Massive & Giant \\
\midrule
\multicolumn{9}{l}{\textit{Baseline}} \\
  ZINC (Baseline) & $0.471{_{\pm 0.024}}$ & $0.462{_{\pm 0.036}}$ & $0.458{_{\pm 0.019}}$ & $0.483{_{\pm 0.020}}$ & $0.474{_{\pm 0.024}}$ & $0.503{_{\pm 0.030}}$ & $0.505{_{\pm 0.041}}$ & $0.518{_{\pm 0.020}}$ \\
\midrule
\multicolumn{9}{l}{\textit{Quantum (FM)}} \\
  FM$_{1\text{b}}$ & $0.469{_{\pm 0.058}}$ & $0.486{_{\pm 0.059}}$ & $0.457{_{\pm 0.042}}$ & $0.459{_{\pm 0.039}}$ & $0.468{_{\pm 0.045}}$ & $0.488{_{\pm 0.029}}$ & $0.485{_{\pm 0.029}}$ & $0.511{_{\pm 0.018}}$ \\
  FM$_{2\text{b}}$ & $\best{0.447{_{\pm 0.025}}}$ & $\best{0.459{_{\pm 0.037}}}$ & $0.446{_{\pm 0.021}}$ & $0.465{_{\pm 0.040}}$ & $0.438{_{\pm 0.016}}$ & $0.457{_{\pm 0.021}}$ & $0.476{_{\pm 0.030}}$ & $0.497{_{\pm 0.015}}$ \\
  FM & $0.486{_{\pm 0.070}}$ & $0.495{_{\pm 0.069}}$ & $0.497{_{\pm 0.076}}$ & $0.454{_{\pm 0.041}}$ & $0.469{_{\pm 0.045}}$ & $0.484{_{\pm 0.027}}$ & $0.504{_{\pm 0.031}}$ & $0.510{_{\pm 0.029}}$ \\
\midrule
\multicolumn{9}{l}{\textit{Quantum (AFM)}} \\
  AFM$_{1\text{b}}$ & $0.501{_{\pm 0.175}}$ & $\second{0.460}{_{\pm 0.049}}$ & $\second{0.444}{_{\pm 0.039}}$ & $0.450{_{\pm 0.041}}$ & $0.438{_{\pm 0.032}}$ & $\best{0.444{_{\pm 0.022}}}$ & $0.459{_{\pm 0.025}}$ & $\second{0.465}{_{\pm 0.023}}$ \\
  AFM$_{2\text{b}}$ & $\second{0.450}{_{\pm 0.041}}$ & $0.468{_{\pm 0.058}}$ & $\best{0.443{_{\pm 0.031}}}$ & $\second{0.447}{_{\pm 0.046}}$ & $\best{0.430{_{\pm 0.020}}}$ & $\second{0.447}{_{\pm 0.032}}$ & $\second{0.453}{_{\pm 0.040}}$ & $0.487{_{\pm 0.032}}$ \\
  AFM & $0.504{_{\pm 0.155}}$ & $0.469{_{\pm 0.075}}$ & $0.457{_{\pm 0.064}}$ & $\best{0.421{_{\pm 0.026}}}$ & $\second{0.434}{_{\pm 0.023}}$ & $0.452{_{\pm 0.027}}$ & $\best{0.445{_{\pm 0.025}}}$ & $\best{0.461{_{\pm 0.029}}}$ \\
\bottomrule
\end{tabular}
}
\end{table*}

\subsection{\texorpdfstring{Hardware versus emulation per backbone}{Hardware versus emulation per backbone}}
\label{app:hw-deltas}

Table~\ref{tab:hw-deltas} lists, per backbone, the MUTAG F$_1$ of the RRWP hybrids with emulated and hardware AFM features.
For the compressed features, agreement holds on all six backbones, with a maximum absolute deviation of $0.009$.
For the full time-resolved features, five backbones are within $0.014$, while GCN deviates by $0.041$.

\begin{table}[h]
\centering
\caption{MUTAG, RRWP hybrids with emulated versus hardware (HW) AFM features, per backbone. Mean macro F$_1$ ($\uparrow$) over 10 stratified folds $\times$ 5 seeds (values of Table~\ref{tab:pe-comparison-mutag}); $\Delta$ = hardware $-$ emulation.}
\label{tab:hw-deltas}
\small
\setlength{\tabcolsep}{4pt}
\begin{tabular}{l c c c c c c}
\toprule
Backbone & RRWP+AFM & RRWP+HW-AFM & $\Delta$ (full) & RRWP+AFM$_{\text{stat}}$ & RRWP+HW-AFM$_{\text{stat}}$ & $\Delta$ (stat) \\
\midrule
GT & 0.855 & 0.854 & $-0.001$ & 0.854 & 0.849 & $-0.005$ \\
GCN & 0.878 & 0.837 & $-0.041$ & 0.852 & 0.852 & $+0.000$ \\
GAT & 0.862 & 0.852 & $-0.010$ & 0.860 & 0.853 & $-0.007$ \\
GINE & 0.848 & 0.841 & $-0.007$ & 0.846 & 0.852 & $+0.006$ \\
Transformer & 0.868 & 0.854 & $-0.014$ & 0.852 & 0.861 & $+0.009$ \\
General & 0.867 & 0.859 & $-0.008$ & 0.873 & 0.877 & $+0.004$ \\
\bottomrule
\end{tabular}
\end{table}

\section{Intrinsic benchmark}
\label{app:intrinsic-information-benchmark}

In this appendix, we present the targets, training details and full results for our intrinsic information benchmark.

\subsection{Intrinsic benchmark: targets and probe}
\label{app:benchmark-targets}

This appendix complements Sec.~\ref{sec:intrinsic-information-benchmark} with the training details and formal definition of each of the 15 probed targets and with the technical details of the probe omitted from the main text.

Feature sets, dimensionalities and the z-score normalisation with training-split statistics are defined once in (see App.~\ref{app:features}) and reused unchanged here.

\paragraph{Target definitions.}
All structural quantities are computed on the undirected simple graph $\mathcal{G} = (\mathcal{V}, \mathcal{E})$ underlying each molecule, i.e.\ after stripping bond multiplicities and atom-type information.

\subparagraph{Node-level (5).}
\begin{itemize}[topsep=0pt, itemsep=1pt, parsep=0pt]
  \item \textbf{Degree}: $d_i = |\{j : (i,j) \in \mathcal{E}\}|$.
  \item \textbf{Betweenness centrality}: $b_i = \sum_{s \neq t \neq i} \sigma_{st}(i)/\sigma_{st}$, where $\sigma_{st}$ is the number of shortest $s$-$t$ paths and $\sigma_{st}(i)$ the number of those passing through $i$.
  \item \textbf{Eccentricity}: $\varepsilon_i = \max_{j \in \mathcal{V}} d(i,j)$, the longest shortest-path distance from $i$.
  \item \textbf{Bridge-node detection} (binary): $1$ iff $i$ is incident to at least one bridge edge.
  \item \textbf{Smallest cycle through node}: $g_i = \min\{|C| : C \text{ simple cycle}, \, i \in C\}$, set to the graph girth when $i$ lies on no cycle.
\end{itemize}

\subparagraph{Edge-level (5).}
\begin{itemize}[topsep=0pt, itemsep=1pt, parsep=0pt]
  \item \textbf{Link prediction} (binary): $1$ iff $(i,j) \in \mathcal{E}$. Candidates are all pairs within two hops, so that negative examples lie in a realistic neighbourhood rather than being trivially far apart.
  \item \textbf{Edge betweenness}: $b_e = \sum_{s \neq t} \sigma_{st}(e)/\sigma_{st}$.
  \item \textbf{Bridge-edge detection} (binary): $1$ iff removing $e$ disconnects the graph.
  \item \textbf{Local edge connectivity}: $\kappa_{ij}$, the maximum number of edge-disjoint $i$-$j$ paths (Menger).
  \item \textbf{Smallest cycle through edge}: $g_{ij} = \min\{|C| : C \text{ simple cycle}, \, e \in C\}$.
\end{itemize}

\subparagraph{Graph-level (5).}
\begin{itemize}[topsep=0pt, itemsep=1pt, parsep=0pt]
  \item \textbf{Number of cycles}: circuit rank $|\mathcal{E}| - |\mathcal{V}| + c$ with $c$ connected components.
  \item \textbf{Diameter}: $\max_{i,j} d(i,j)$.
  \item \textbf{Density}: $2|\mathcal{E}| / \bigl(|\mathcal{V}|(|\mathcal{V}|-1)\bigr)$.
  \item \textbf{Wiener index}: $W = \tfrac{1}{2}\sum_{i,j \in \mathcal{V}} d(i,j)$.
  \item \textbf{Algebraic connectivity}: $\lambda_2(\mathbf{L})$, the second-smallest eigenvalue of the graph Laplacian.
\end{itemize}

\paragraph{Probe architecture.}
The shared encoder $\phi$ is a two-layer MLP mapping the input feature vector to $\mathbb{R}^{64}$ with batch normalisation and ReLU activations between layers.
The predictor $\rho$ has one hidden layer of width $64$ with batch normalisation and ReLU, and a linear output.
For graph-level probes, $\rho$ additionally applies a dropout layer with $p=0.1$ before the output; node- and edge-level probes use no dropout.
Neither $\phi$ nor $\rho$ has access to graph structure beyond what the input already encodes: there is no message passing, no attention, and no learned aggregation over neighbours, so any signal extracted reflects information already present in the encoding.

\paragraph{Entity-level input construction.}
\begin{itemize}[topsep=0pt, itemsep=1pt, parsep=0pt]
  \item \textbf{Node probes} use $\mathbf{x}_i =$\ \Cone [Eq.~(\ref{eq:quantum_descriptors})]
  \item \textbf{Edge probes} use $\mathbf{x}_{ij} =$\ \Ctwo for $(i,j) \in \mathcal{E}$. For link prediction we additionally sample, per graph, negative candidates from the two-hop complement $\{(i,j) \notin \mathcal{E} : d(i,j) \leq 2\}$ in a $1\!:\!1$ ratio with positive edges.
  \item \textbf{Graph probes} enrich each node with its node-averaged pairwise feature: $\mathbf{x}_i = [\,\boldsymbol{\mathcal{C}^{(1)}_{i}} \,\Vert\, \bar{\mathbf{c}}_i\,]$ with $\bar{\mathbf{c}}_i = \tfrac{1}{N}\sum_{j} \boldsymbol{\mathcal{C}^{(2)}_{ij}}$, yielding a $2d$-dimensional per-node vector before sum pooling.
\end{itemize}

\paragraph{Training.}

Probes are trained on \ZINC ($11{,}368$ graphs) and on MUTAG separately with $5$-fold cross-validation means and standard deviations are computed across the five folds.

All probes use AdamW (learning rate $10^{-3}$, default weight decay), MSE loss for regression, and binary cross-entropy with logits for classification.
5-fold cross-validation is performed at the \emph{graph} level: all nodes and edges of a given graph belong to the same fold, which prevents leakage between train and test when probing node- or edge-level quantities.
Folds are stratified by label for classification tasks.
Node- and edge-level probes run for at most $50$ epochs with patience $10$ and batch size $512$; graph-level probes for at most $100$ epochs with patience $20$ and batch size $32$.
The best checkpoint under validation loss is restored before computing the test metric, and the five test metrics are reported as mean $\pm$ standard deviation across folds.

\subsection{Intrinsic benchmark: full results}
\label{app:full-benchmark}

Table~\ref{tab:pe-benchmark} reports the complete ZINC$_{\text{subset}}^{*}$ intrinsic-benchmark results summarised in Table~\ref{tab:pe-benchmark-main} of Sec.~\ref{sec:intrinsic-information-benchmark}. It includes the random-feature control, the classical descriptors, the simulated FM and AFM descriptors, their compressed variants, the bare FM+AFM union, and the RRWP-based hybrid combinations. The corresponding MUTAG results are reported in Table~\ref{tab:pe-benchmark-mutag}. Compression effects are analysed in App.~\ref{app:intrinsic-compression}, and hardware AFM results in App.~\ref{app:intrinsic-hardware}.

\paragraph{Random-feature control and saturating tasks.}

The random-feature control is separated from the structural descriptors on the non-saturating regression tasks. On ZINC$_{\text{subset}}^{*}$, for instance, random features give $0.564$ MAE on degree prediction, whereas every non-random structural descriptor is substantially better. The same holds for cycle, betweenness, and graph-level distance targets, confirming that the probe is not solving the tasks from dataset-level shortcuts alone.

Link prediction, bridge-edge detection, and, to a lesser extent, bridge-node detection are close to saturation for many non-random feature sets. They are retained as sanity checks, but they should not be used to rank descriptor families.

\subsubsection{Compression variants}
\label{app:intrinsic-compression}

The columns denoted FM$_{\mathrm{stat}}$ and AFM$_{\mathrm{stat}}$ use the four temporal summary statistics defined in Sec.~\ref{sec:quantum-feature-extraction}: mean, variance, minimum, and maximum over the sampled evolution times. The columns denoted FM$_{\mathrm{mean}}$ and AFM$_{\mathrm{mean}}$ are a more aggressive compression in which only the temporal mean is retained.
 Note that our compression produces four descriptors per pair (mean, \dots), of which the mean $\overline{C_{ij}}$ [Eq.~(\ref{eq:app_A_bar})] is the leading one.
Alternatively, the data could be compressed into the asymptotically exact long-time correlation $C^{\infty}_{ij}$ [Eq.~(\ref{eq:app_obs_C_infty})], at the cost of retaining a single descriptor per pair.
In an ablation, adding $C^{\infty}_{ij}$ to the compressed features yielded no statistically significant change in performance.

The statistic-compressed descriptors preserve a substantial fraction of the signal despite reducing the temporal dimension from $N_t=20$ to four values. On ZINC$_{\text{subset}}^{*}$, AFM$_{\mathrm{stat}}$ remains competitive with, and often better than, Laplacian features on several non-spectral topological tasks, including degree, betweenness, edge betweenness, smallest cycle through a node or edge, number of cycles, and density. The degradation from AFM to AFM$_{\mathrm{stat}}$ is nevertheless systematic on informative regression tasks except density, typically increasing MAE by a factor of roughly $1.1$--$1.7$, and by $7.2$ on degree.

The mean-only variants lose more information. For example, AFM$_{\mathrm{mean}}$ gives $13.50 \times 10^{-2}$ MAE on degree prediction, compared with $1.26 \times 10^{-2}$ for full AFM, and FM$_{\mathrm{mean}}$ gives $2.36$ MAE on smallest-cycle-through-node prediction, compared with $1.42$ for full FM. Thus, the full descriptors do not owe their advantage only to average correlation amplitude; the temporal profile of the dynamics carries additional structural information.

\begin{table*}
\caption{%
Intrinsic information benchmark on ZINC$_{\text{subset}}^{*}$ graph tasks.
Mean $\pm$ standard deviation over 5-fold cross-validation is reported.
Metrics are MAE ($\downarrow$) for regression tasks and accuracy ($\uparrow$) for classification tasks.
\mbox{$\best{\text{Best}}$} and \mbox{$\second{\text{2nd best}}$} indicate the unique best and runner-up feature sets across the complete feature set, combining both halves of the table; entries marked \mbox{$\tied{\text{}}{}$} are tied with the best.
}
\label{tab:pe-benchmark}
\centering
  \resizebox{\textwidth}{!}{%
  \begin{tabular}{l c ccc ccc}
    \toprule
    & \multicolumn{1}{c}{Baseline} & \multicolumn{3}{c}{Classical} & \multicolumn{3}{c}{Quantum FM} \\
    \cmidrule(lr){2-2} \cmidrule(lr){3-5} \cmidrule(lr){6-8}
    Task
    & Random
    & RRWP
    & Laplacian
    & RRWP+Lap.
    & FM$_{\mathrm{mean}}$
    & FM$_{\mathrm{stat}}$
    & FM
    \\
    \midrule
    \multicolumn{8}{l}{\textit{Node Prediction}}\\
    Bridge Node (Acc $\uparrow$)
    & $0.58{_{\pm 0.00}}$
    & $0.93{_{\pm 0.01}}$ 
    & $0.76{_{\pm 0.00}}$
    & $0.95{_{\pm 0.00}}$
    & $0.68{_{\pm 0.00}}$
    & $0.74{_{\pm 0.00}}$ 
    & $0.83{_{\pm 0.00}}$ 
    \\
    \specialrule{0.4pt}{1pt}{1pt}
    Degree (MAE $\downarrow$) ($\times 10^{2}$)
    & $56.43{_{\pm 0.63}}$ 
    & $14.36{_{\pm 0.99}}$ 
    & $46.70{_{\pm 0.36}}$ 
    & $7.0{_{\pm 2.4}}$ 
    & $56.58{_{\pm 0.45}}$ 
    & $30.51{_{\pm 0.27}}$ 
    & $18.74{_{\pm 0.69}}$ 
    \\
    Betweenness (MAE $\downarrow$) ($\times 10^{2}$)
    & $16.48{_{\pm 0.05}}$ 
    & $7.77{_{\pm 0.72}}$ 
    & $12.46{_{\pm 0.11}}$ 
    & $4.76{_{\pm 0.29}}$ 
    & $16.17{_{\pm 0.07}}$ 
    & $9.10{_{\pm 0.13}}$ 
    & $7.17{_{\pm 0.05}}$ 
    \\
    Eccentricity (MAE $\downarrow$)
    & $2.14{_{\pm 0.01}}$ 
    & $1.88{_{\pm 0.02}}$ 
    & $1.51{_{\pm 0.02}}$ 
    & $\best{1.08}{_{\pm 0.04}}$ 
    & $2.10{_{\pm 0.01}}$ 
    & $1.90{_{\pm 0.01}}$ 
    & $1.65{_{\pm 0.01}}$ 
    \\
    Min Cycle (Node) (MAE $\downarrow$)
    & $2.66{_{\pm 0.01}}$
    & $0.73{_{\pm 0.02}}$ 
    & $1.68{_{\pm 0.02}}$ 
    & $0.60{_{\pm 0.03}}$ 
    & $2.36{_{\pm 0.02}}$ 
    & $1.80{_{\pm 0.01}}$ 
    & $1.42{_{\pm 0.04}}$ 
    \\
    \addlinespace[2pt]\midrule\addlinespace[1pt]
    \multicolumn{8}{l}{\textit{Edge Prediction}}\\
    Link Pred. (Acc $\uparrow$)
    & $0.91{_{\pm 0.00}}$
    & $\tied{1.00}{_{\pm 0.00}}$
    & $0.94{_{\pm 0.00}}$
    & $\tied{1.00}{_{\pm 0.00}}$
    & $0.97{_{\pm 0.00}}$ 
    & $0.99{_{\pm 0.00}}$ 
    & $0.99{_{\pm 0.00}}$
    \\
    Bridge Edge (Acc $\uparrow$)
    & $0.96{_{\pm 0.00}}$
    & $\tied{0.99}{_{\pm 0.00}}$
    & $0.98{_{\pm 0.00}}$
    & $\tied{0.99}{_{\pm 0.00}}$
    & $0.96{_{\pm 0.00}}$
    & $0.97{_{\pm 0.00}}$
    & $0.98{_{\pm 0.00}}$
    \\
    \specialrule{0.4pt}{1pt}{1pt}
    Edge Betw. (MAE $\downarrow$) ($\times 10^{3}$)
    & $35.20{_{\pm 0.14}}$ 
    & $11.6{_{\pm 1.7}}$ 
    & $20.57{_{\pm 0.35}}$ 
    & $\second{6.09}{_{\pm 0.95}}$ 
    & $16.01{_{\pm 0.14}}$ 
    & $13.04{_{\pm 0.28}}$ 
    & $9.89{_{\pm 0.08}}$ 
    \\
    Local Conn. (MAE $\downarrow$) ($\times 10^{2}$)
    & $26.96{_{\pm 0.13}}$ 
    & $10.01{_{\pm 0.25}}$ 
    & $13.92{_{\pm 0.16}}$ 
    & $7.39{_{\pm 0.38}}$ 
    & $15.24{_{\pm 0.11}}$ 
    & $13.89{_{\pm 0.16}}$ 
    & $12.60{_{\pm 0.12}}$ 
    \\
    Min Cycle (Edge) (MAE $\downarrow$) ($\times 10^{2}$)
    & $60.92{_{\pm 0.31}}$ 
    & $13.3{_{\pm 1.3}}$ 
    & $42.41{_{\pm 0.36}}$ 
    & $10.2{_{\pm 1.0}}$ 
    & $25.67{_{\pm 0.28}}$ 
    & $23.02{_{\pm 0.24}}$ 
    & $19.36{_{\pm 0.41}}$ 
    \\
    \addlinespace[2pt]\midrule\addlinespace[1pt]
    \multicolumn{8}{l}{\textit{Graph Prediction}}\\
    Num Cycles (MAE $\downarrow$) ($\times 10^{2}$)
    & $60.24{_{\pm 0.94}}$ 
    & $6.68{_{\pm 0.93}}$ 
    & $27.4{_{\pm 1.5}}$ 
    & $5.90{_{\pm 0.56}}$ 
    & $22.92{_{\pm 0.63}}$ 
    & $18.38{_{\pm 0.65}}$ 
    & $15.33{_{\pm 0.34}}$ 
    \\
    Diameter (MAE $\downarrow$)
    & $1.47{_{\pm 0.03}}$ 
    & $0.49{_{\pm 0.04}}$ 
    & $0.69{_{\pm 0.01}}$ 
    & $0.47{_{\pm 0.05}}$ 
    & $0.82{_{\pm 0.02}}$ 
    & $0.71{_{\pm 0.01}}$ 
    & $0.65{_{\pm 0.01}}$
    \\
    Density (MAE $\downarrow$) ($\times 10^{3}$)
    & $3.72{_{\pm 0.09}}$ 
    & $\tied{1.03}{_{\pm 0.22}}$ 
    & $2.48{_{\pm 0.22}}$ 
    & $\tied{1.14}{_{\pm 0.18}}$ 
    & $1.95{_{\pm 0.21}}$ 
    & $1.86{_{\pm 0.22}}$ 
    & $1.70{_{\pm 0.18}}$ 
    \\
    Wiener Index (MAE $\downarrow$) ($\times 10^{2}$)
    & $39.40{_{\pm 0.36}}$ 
    & $10.8{_{\pm 2.1}}$ 
    & $19.24{_{\pm 0.24}}$ 
    & $9.61{_{\pm 0.92}}$ 
    & $19.25{_{\pm 0.56}}$ 
    & $16.41{_{\pm 0.14}}$ 
    & $14.15{_{\pm 0.19}}$ 
    \\
    Alg. Connect. (MAE $\downarrow$) ($\times 10^{3}$)
    & $9.86{_{\pm 0.32}}$ 
    & $2.63{_{\pm 0.28}}$ 
    & $5.59{_{\pm 0.42}}$ 
    & $\best{2.22}{_{\pm 0.12}}$ 
    & $5.91{_{\pm 0.32}}$ 
    & $5.46{_{\pm 0.17}}$ 
    & $5.21{_{\pm 0.30}}$ 
    \\
    \bottomrule
  \end{tabular}
  }
\par\medskip
  \resizebox{\textwidth}{!}{%
  \begin{tabular}{l ccc cccc}
    \toprule
    & \multicolumn{3}{c}{Quantum AFM} & \multicolumn{4}{c}{Hybrid} \\
    \cmidrule(lr){2-4} \cmidrule(lr){5-8}
    Task
    & AFM$_{\mathrm{mean}}$
    & AFM$_{\mathrm{stat}}$
    & AFM
    & FM+AFM
    & RRWP+FM
    & RRWP+AFM
    & RRWP+FM+AFM
    \\
    \midrule
    \multicolumn{8}{l}{\textit{Node Prediction}}\\
    Bridge Node (Acc $\uparrow$)
    & $0.82{_{\pm 0.00}}$ 
    & $0.85{_{\pm 0.00}}$ 
    & $0.91{_{\pm 0.00}}$ 
    & $0.94{_{\pm 0.00}}$ 
    & $0.97{_{\pm 0.00}}$ 
    & $\tied{0.98}{_{\pm 0.00}}$ 
    & $\tied{0.98}{_{\pm 0.00}}$ 
    \\
    \specialrule{0.4pt}{1pt}{1pt}
    Degree (MAE $\downarrow$) ($\times 10^{2}$)
    & $13.50{_{\pm 0.21}}$ 
    & $9.07{_{\pm 0.09}}$ 
    & $1.26{_{\pm 0.16}}$ 
    & $\tied{1.09}{_{\pm 0.09}}$ 
    & $4.71{_{\pm 0.48}}$ 
    & $\tied{1.18}{_{\pm 0.34}}$ 
    & $1.25{_{\pm 0.39}}$ 
    \\
    Betweenness (MAE $\downarrow$) ($\times 10^{2}$)
    & $11.81{_{\pm 0.15}}$ 
    & $10.13{_{\pm 0.10}}$ 
    & $7.11{_{\pm 0.19}}$ 
    & $5.19{_{\pm 0.06}}$ 
    & $4.33{_{\pm 0.26}}$ 
    & $\second{3.90}{_{\pm 0.06}}$ 
    & $\best{3.31}{_{\pm 0.10}}$ 
    \\
    Eccentricity (MAE $\downarrow$)
    & $2.10{_{\pm 0.02}}$ 
    & $2.02{_{\pm 0.02}}$ 
    & $1.91{_{\pm 0.02}}$ 
    & $1.58{_{\pm 0.02}}$
    & $1.48{_{\pm 0.03}}$ 
    & $1.77{_{\pm 0.02}}$ 
    & $\second{1.47}{_{\pm 0.02}}$ 
    \\
    Min Cycle (Node) (MAE $\downarrow$)
    & $1.61{_{\pm 0.00}}$ 
    & $1.40{_{\pm 0.01}}$ 
    & $0.88{_{\pm 0.01}}$ 
    & $0.66{_{\pm 0.02}}$ 
    & $0.48{_{\pm 0.03}}$ 
    & $\second{0.36}{_{\pm 0.01}}$ 
    & $\best{0.28}{_{\pm 0.01}}$ 
    \\
    \addlinespace[2pt]\midrule\addlinespace[1pt]
    \multicolumn{8}{l}{\textit{Edge Prediction}}\\
    Link Pred. (Acc $\uparrow$)
    & $\tied{1.00}{_{\pm 0.00}}$ 
    & $\tied{1.00}{_{\pm 0.00}}$ 
    & $\tied{1.00}{_{\pm 0.00}}$ 
    & $\tied{1.00}{_{\pm 0.00}}$
    & $\tied{1.00}{_{\pm 0.00}}$
    & $\tied{1.00}{_{\pm 0.00}}$
    & $\tied{1.00}{_{\pm 0.00}}$
    \\
    Bridge Edge (Acc $\uparrow$)
    & $0.97{_{\pm 0.00}}$
    & $0.97{_{\pm 0.00}}$
    & $0.99{_{\pm 0.00}}$
    & $0.99{_{\pm 0.00}}$
    & $\tied{0.99}{_{\pm 0.00}}$
    & $\tied{0.99}{_{\pm 0.00}}$
    & $\tied{1.00}{_{\pm 0.00}}$ 
    \\
    \specialrule{0.4pt}{1pt}{1pt}
    Edge Betw. (MAE $\downarrow$) ($\times 10^{3}$)
    & $11.81{_{\pm 0.60}}$ 
    & $9.76{_{\pm 0.27}}$ 
    & $8.61{_{\pm 0.36}}$ 
    & $6.87{_{\pm 0.16}}$ 
    & $6.69{_{\pm 0.55}}$ 
    & $6.13{_{\pm 0.43}}$ 
    & $\best{5.36}{_{\pm 0.35}}$ 
    \\
    Local Conn. (MAE $\downarrow$) ($\times 10^{2}$)
    & $16.36{_{\pm 0.23}}$ 
    & $15.16{_{\pm 0.16}}$ 
    & $11.30{_{\pm 0.20}}$ 
    & $8.79{_{\pm 0.18}}$ 
    & $6.83{_{\pm 0.15}}$ 
    & $\second{6.07}{_{\pm 0.15}}$ 
    & $\best{5.14}{_{\pm 0.16}}$ 
    \\
    Min Cycle (Edge) (MAE $\downarrow$) ($\times 10^{2}$)
    & $26.15{_{\pm 0.42}}$ 
    & $23.07{_{\pm 0.50}}$ 
    & $13.21{_{\pm 0.49}}$ 
    & $11.75{_{\pm 0.63}}$ 
    & $10.03{_{\pm 0.92}}$ 
    & $\tied{7.62}{_{\pm 0.44}}$ 
    & $\tied{7.17}{_{\pm 0.73}}$ 
    \\
    \addlinespace[2pt]\midrule\addlinespace[1pt]
    \multicolumn{8}{l}{\textit{Graph Prediction}}\\
    Num Cycles (MAE $\downarrow$) ($\times 10^{2}$)
    & $21.41{_{\pm 0.57}}$ 
    & $16.36{_{\pm 0.50}}$ 
    & $9.85{_{\pm 0.85}}$ 
    & $7.20{_{\pm 0.29}}$ 
    & $\tied{4.83}{_{\pm 0.38}}$ 
    & $\tied{4.48}{_{\pm 0.71}}$ 
    & $\tied{4.67}{_{\pm 0.42}}$ 
    \\
    Diameter (MAE $\downarrow$)
    & $0.89{_{\pm 0.01}}$ 
    & $0.75{_{\pm 0.02}}$ 
    & $0.60{_{\pm 0.02}}$ 
    & $0.51{_{\pm 0.02}}$ 
    & $\second{0.36}{_{\pm 0.04}}$ 
    & $0.37{_{\pm 0.03}}$ 
    & $\best{0.32}{_{\pm 0.02}}$ 
    \\
    Density (MAE $\downarrow$) ($\times 10^{3}$)
    & $1.84{_{\pm 0.26}}$ 
    & $1.72{_{\pm 0.23}}$ 
    & $1.86{_{\pm 0.40}}$ 
    & $1.56{_{\pm 0.23}}$ 
    & $\tied{1.21}{_{\pm 0.33}}$ 
    & $\tied{1.10}{_{\pm 0.32}}$ 
    & $\tied{1.25}{_{\pm 0.26}}$ 
    \\
    Wiener Index (MAE $\downarrow$) ($\times 10^{2}$)
    & $23.22{_{\pm 0.29}}$ 
    & $19.46{_{\pm 0.32}}$ 
    & $14.99{_{\pm 0.84}}$ 
    & $11.7{_{\pm 1.3}}$ 
    & $\tied{7.13}{_{\pm 0.68}}$ 
    & $8.81{_{\pm 0.70}}$ 
    & $\tied{7.24}{_{\pm 0.71}}$ 
    \\
    Alg. Connect. (MAE $\downarrow$) ($\times 10^{3}$)
    & $7.34{_{\pm 0.23}}$ 
    & $6.30{_{\pm 0.35}}$ 
    & $5.32{_{\pm 0.27}}$ 
    & $4.29{_{\pm 0.56}}$ 
    & $2.70{_{\pm 0.15}}$ 
    & $\second{2.34}{_{\pm 0.22}}$ 
    & $2.38{_{\pm 0.20}}$ 
    \\
    \bottomrule
  \end{tabular}
  }
\end{table*}

\subsection{MUTAG corroborating results}
\label{app:mutag}

Table~\ref{tab:pe-benchmark-mutag} reports the same intrinsic-benchmark protocol on MUTAG. Because MUTAG is much smaller than ZINC$_{\text{subset}}^{*}$, fold-to-fold variability is larger and close numerical rankings should be interpreted cautiously. The table is therefore used as a qualitative robustness check rather than as a second precise ranking of all descriptors.

The main qualitative patterns are consistent with the ZINC$_{\text{subset}}^{*}$ results. AFM-based hybrids remain strong on local structural targets, while Laplacian and RRWP-based descriptors remain particularly effective on several spectral or distance-like quantities. As in the ZINC$_{\text{subset}}^{*}$ table, link prediction and bridge-edge detection are nearly saturated for the non-random feature sets and mainly serve as sanity checks.

\subsubsection{Hardware AFM analysis}
\label{app:intrinsic-hardware}

The MUTAG table also includes hardware AFM descriptors, denoted HW-AFM, and their RRWP hybrid. HW-AFM alone is generally weaker than the simulated AFM descriptor, which is expected given sampling noise, calibration effects, and the small size of the evaluation. However, RRWP+HW-AFM is consistently competitive with the simulated hybrids, is second on betweenness ($4.24\times10^{-2}$ MAE vs.\ $3.89\times10^{-2}$ for RRWP)
and ties the best on smallest cycle through a node (0.71 MAE). It also ties or
nearly ties the best feature sets on bridge-node detection, link prediction and
bridge-edge detection, but trails on diameter (0.62 vs.\ 0.51) and Wiener index
(0.17 vs.\ 0.11).

These hardware results should not be interpreted as a stable ranking against noiseless simulated descriptors, since several MUTAG differences are within fold-level variability. They nevertheless provide a useful consistency check: experimentally obtained AFM correlations still carry graph-theoretic information, and this information is most useful when combined with RRWP.

\begin{table*}
\caption{%
Intrinsic information benchmark on MUTAG graph tasks.
Mean $\pm$ standard deviation over 5-fold cross-validation is reported.
Metrics are MAE ($\downarrow$) for regression tasks and accuracy ($\uparrow$) for classification tasks.
\mbox{$\best{\text{Best}}$} and \mbox{$\second{\text{2nd best}}$} indicate the unique best and runner-up feature sets across the complete feature set, combining both halves of the table; entries marked \mbox{$\tied{\text{tied}}{}$} are tied with the best.
}
\label{tab:pe-benchmark-mutag}
\centering
  \resizebox{\textwidth}{!}{%
  \begin{tabular}{l c ccc ccc}
    \toprule
    & \multicolumn{1}{c}{Baseline} & \multicolumn{3}{c}{Classical} & \multicolumn{3}{c}{Quantum FM} \\
    \cmidrule(lr){2-2} \cmidrule(lr){3-5} \cmidrule(lr){6-8}
    Task
    & Random
    & RRWP
    & Laplacian
    & RRWP+Lap.
    & FM$_{\mathrm{mean}}$
    & FM$_{\mathrm{stat}}$
    & FM
    \\
    \midrule
    \multicolumn{8}{l}{\textit{Node Prediction}}\\
    Bridge Node (Acc $\uparrow$)
    & $0.58{_{\pm 0.01}}$ 
    & $0.96{_{\pm 0.01}}$ 
    & $0.84{_{\pm 0.03}}$ 
    & $0.97{_{\pm 0.01}}$
    & $0.72{_{\pm 0.02}}$ 
    & $0.76{_{\pm 0.03}}$ 
    & $0.83{_{\pm 0.02}}$ 
    \\
    \specialrule{0.4pt}{1pt}{1pt}
    Degree (MAE $\downarrow$) ($\times 10^{2}$)
    & $70.3{_{\pm 1.7}}$ 
    & $14.2{_{\pm 1.0}}$ 
    & $40.7{_{\pm 1.6}}$ 
    & $15.24{_{\pm 0.88}}$ 
    & $47.6{_{\pm 1.3}}$ 
    & $33.2{_{\pm 2.7}}$ 
    & $28.4{_{\pm 1.2}}$ 
    \\
    Betweenness (MAE $\downarrow$) ($\times 10^{2}$)
    & $12.79{_{\pm 0.57}}$ 
    & $\best{3.89}{_{\pm 0.28}}$ 
    & $8.21{_{\pm 0.24}}$ 
    & $4.53{_{\pm 0.40}}$ 
    & $9.95{_{\pm 0.94}}$ 
    & $6.65{_{\pm 0.44}}$ 
    & $6.34{_{\pm 0.18}}$ 
    \\
    Eccentricity (MAE $\downarrow$)
    & $1.82{_{\pm 0.05}}$ 
    & $\second{1.01}{_{\pm 0.10}}$ 
    & $1.09{_{\pm 0.07}}$ 
    & $\best{0.92}{_{\pm 0.08}}$ 
    & $1.35{_{\pm 0.08}}$ 
    & $1.16{_{\pm 0.07}}$ 
    & $1.12{_{\pm 0.04}}$ 
    \\
    Min Cycle (Node) (MAE $\downarrow$)
    & $3.09{_{\pm 0.09}}$ 
    & $0.73{_{\pm 0.02}}$ 
    & $1.61{_{\pm 0.06}}$ 
    & $0.81{_{\pm 0.08}}$ 
    & $1.70{_{\pm 0.12}}$ 
    & $1.62{_{\pm 0.10}}$ 
    & $1.62{_{\pm 0.07}}$ 
    \\
    \addlinespace[2pt]\midrule\addlinespace[1pt]
    \multicolumn{8}{l}{\textit{Edge Prediction}}\\
    Link Pred. (Acc $\uparrow$)
    & $0.88{_{\pm 0.00}}$
    & $\tied{1.00}{_{\pm 0.00}}$
    & $0.95{_{\pm 0.00}}$
    & $\tied{1.00}{_{\pm 0.00}}$
    & $0.94{_{\pm 0.01}}$ 
    & $0.98{_{\pm 0.00}}$ 
    & $0.99{_{\pm 0.00}}$ 
    \\
    Bridge Edge (Acc $\uparrow$)
    & $0.97{_{\pm 0.00}}$
    & $\tied{1.00}{_{\pm 0.00}}$
    & $0.99{_{\pm 0.00}}$
    & $\tied{1.00}{_{\pm 0.00}}$
    & $0.97{_{\pm 0.00}}$
    & $0.97{_{\pm 0.00}}$
    & $0.98{_{\pm 0.00}}$
    \\
    \specialrule{0.4pt}{1pt}{1pt}
    Edge Betw. (MAE $\downarrow$) ($\times 10^{3}$)
    & $45.0{_{\pm 2.0}}$ 
    & $\tied{9.00}{_{\pm 0.40}}$ 
    & $23.83{_{\pm 0.83}}$ 
    & $\tied{8.76}{_{\pm 0.43}}$ 
    & $20.0{_{\pm 1.9}}$ 
    & $17.1{_{\pm 1.6}}$ 
    & $13.34{_{\pm 0.66}}$ 
    \\
    Local Conn. (MAE $\downarrow$)
    & $0.53{_{\pm 0.00}}$
    & $0.27{_{\pm 0.02}}$ 
    & $0.33{_{\pm 0.01}}$
    & $0.21{_{\pm 0.01}}$ 
    & $0.33{_{\pm 0.01}}$
    & $0.30{_{\pm 0.01}}$ 
    & $0.26{_{\pm 0.00}}$ 
    \\
    Min Cycle (Edge) (MAE $\downarrow$)
    & $1.06{_{\pm 0.03}}$ 
    & $0.21{_{\pm 0.02}}$ 
    & $0.56{_{\pm 0.01}}$ 
    & $\best{0.17}{_{\pm 0.01}}$ 
    & $0.43{_{\pm 0.03}}$ 
    & $0.32{_{\pm 0.01}}$ 
    & $0.29{_{\pm 0.02}}$ 
    \\
    \addlinespace[2pt]\midrule\addlinespace[1pt]
    \multicolumn{8}{l}{\textit{Graph Prediction}}\\
    Num Cycles (MAE $\downarrow$)
    & $0.81{_{\pm 0.05}}$ 
    & $0.19{_{\pm 0.05}}$ 
    & $0.30{_{\pm 0.04}}$ 
    & $\best{0.14}{_{\pm 0.03}}$ 
    & $0.21{_{\pm 0.04}}$ 
    & $0.21{_{\pm 0.04}}$ 
    & $0.24{_{\pm 0.02}}$ 
    \\
    Diameter (MAE $\downarrow$)
    & $1.66{_{\pm 0.46}}$ 
    & $\tied{0.54}{_{\pm 0.04}}$ 
    & $0.84{_{\pm 0.10}}$ 
    & $\tied{0.51}{_{\pm 0.03}}$ 
    & $0.70{_{\pm 0.06}}$ 
    & $0.81{_{\pm 0.09}}$ 
    & $0.82{_{\pm 0.14}}$ 
    \\
    Density (MAE $\downarrow$) ($\times 10^{2}$)
    & $4.30{_{\pm 0.79}}$ 
    & $\tied{1.40}{_{\pm 0.49}}$ 
    & $4.5{_{\pm 2.1}}$ 
    & $2.8{_{\pm 1.1}}$ 
    & $\tied{1.89}{_{\pm 0.52}}$ 
    & $4.1{_{\pm 1.7}}$ 
    & $3.8{_{\pm 1.1}}$ 
    \\
    Wiener Index (MAE $\downarrow$)
    & $0.62{_{\pm 0.04}}$ 
    & $\best{0.11}{_{\pm 0.01}}$ 
    & $0.31{_{\pm 0.03}}$ 
    & $0.17{_{\pm 0.02}}$ 
    & $0.18{_{\pm 0.02}}$ 
    & $0.20{_{\pm 0.03}}$ 
    & $0.22{_{\pm 0.02}}$ 
    \\
    Alg. Connect. (MAE $\downarrow$) ($\times 10^{2}$)
    & $7.5{_{\pm 2.4}}$ 
    & $\best{1.71}{_{\pm 0.22}}$ 
    & $3.22{_{\pm 0.82}}$ 
    & $2.4{_{\pm 1.5}}$ 
    & $2.87{_{\pm 0.73}}$ 
    & $2.95{_{\pm 0.44}}$ 
    & $3.4{_{\pm 1.1}}$ 
    \\
    \bottomrule
  \end{tabular}
  }
\par\medskip
  \resizebox{\textwidth}{!}{%
  \begin{tabular}{l ccc cccc cc}
    \toprule
    & \multicolumn{3}{c}{Quantum AFM} & \multicolumn{4}{c}{Hybrid} & \multicolumn{2}{c}{Hardware} \\
    \cmidrule(lr){2-4} \cmidrule(lr){5-8} \cmidrule(lr){9-10}
    Task
    & AFM$_{\mathrm{mean}}$
    & AFM$_{\mathrm{stat}}$
    & AFM
    & FM+AFM
    & RRWP+FM
    & RRWP+AFM
    & RRWP+FM+AFM
    & HW-AFM
    & RRWP+HW-AFM
    \\
    \midrule
    \multicolumn{10}{l}{\textit{Node Prediction}}\\
    Bridge Node (Acc $\uparrow$)
    & $0.87{_{\pm 0.00}}$ 
    & $0.93{_{\pm 0.01}}$ 
    & $0.97{_{\pm 0.01}}$ 
    & $0.97{_{\pm 0.01}}$ 
    & $\tied{0.98}{_{\pm 0.01}}$ 
    & $\tied{0.98}{_{\pm 0.01}}$ 
    & $\tied{0.99}{_{\pm 0.01}}$ 
    & $0.92{_{\pm 0.01}}$ 
    & $\tied{0.98}{_{\pm 0.00}}$
    \\
    \specialrule{0.4pt}{1pt}{1pt}
    Degree (MAE $\downarrow$) ($\times 10^{2}$)
    & $5.59{_{\pm 0.62}}$ 
    & $5.72{_{\pm 0.45}}$ 
    & $\tied{4.13}{_{\pm 0.32}}$ 
    & $4.51{_{\pm 0.37}}$ 
    & $15.8{_{\pm 1.3}}$ 
    & $\tied{4.31}{_{\pm 0.64}}$ 
    & $4.83{_{\pm 0.39}}$ 
    & $7.79{_{\pm 0.26}}$ 
    & $7.91{_{\pm 0.79}}$ 
    \\
    Betweenness (MAE $\downarrow$) ($\times 10^{2}$)
    & $5.96{_{\pm 0.36}}$ 
    & $5.56{_{\pm 0.27}}$ 
    & $5.53{_{\pm 0.27}}$ 
    & $5.30{_{\pm 0.26}}$ 
    & $4.71{_{\pm 0.20}}$ 
    & $4.25{_{\pm 0.15}}$ 
    & $4.31{_{\pm 0.22}}$ 
    & $5.73{_{\pm 0.18}}$ 
    & $\second{4.24}{_{\pm 0.12}}$ 
    \\
    Eccentricity (MAE $\downarrow$)
    & $1.33{_{\pm 0.04}}$ 
    & $1.26{_{\pm 0.03}}$ 
    & $1.15{_{\pm 0.04}}$ 
    & $1.13{_{\pm 0.06}}$ 
    & $1.02{_{\pm 0.06}}$ 
    & $1.06{_{\pm 0.06}}$ 
    & $1.06{_{\pm 0.08}}$ 
    & $1.32{_{\pm 0.06}}$ 
    & $1.04{_{\pm 0.06}}$ 
    \\
    Min Cycle (Node) (MAE $\downarrow$)
    & $1.12{_{\pm 0.05}}$ 
    & $1.03{_{\pm 0.05}}$ 
    & $0.82{_{\pm 0.07}}$ 
    & $0.87{_{\pm 0.06}}$ 
    & $0.80{_{\pm 0.08}}$ 
    & $\tied{0.72}{_{\pm 0.04}}$ 
    & $0.79{_{\pm 0.06}}$ 
    & $0.92{_{\pm 0.06}}$ 
    & $\tied{0.71}{_{\pm 0.07}}$ 
    \\
    \addlinespace[2pt]\midrule\addlinespace[1pt]
    \multicolumn{10}{l}{\textit{Edge Prediction}}\\
    Link Pred. (Acc $\uparrow$)
    & $\tied{1.00}{_{\pm 0.00}}$ 
    & $\tied{1.00}{_{\pm 0.00}}$ 
    & $\tied{1.00}{_{\pm 0.00}}$ 
    & $\tied{1.00}{_{\pm 0.00}}$ 
    & $\tied{1.00}{_{\pm 0.00}}$
    & $\tied{1.00}{_{\pm 0.00}}$
    & $\tied{1.00}{_{\pm 0.00}}$
    & $\tied{1.00}{_{\pm 0.00}}$
    & $\tied{1.00}{_{\pm 0.00}}$
    \\
    Bridge Edge (Acc $\uparrow$)
    & $0.97{_{\pm 0.00}}$
    & $0.98{_{\pm 0.00}}$
    & $\tied{1.00}{_{\pm 0.00}}$ 
    & $\tied{1.00}{_{\pm 0.00}}$ 
    & $\tied{1.00}{_{\pm 0.00}}$
    & $\tied{1.00}{_{\pm 0.00}}$
    & $\tied{1.00}{_{\pm 0.00}}$
    & $0.99{_{\pm 0.00}}$
    & $\tied{1.00}{_{\pm 0.00}}$
    \\
    \specialrule{0.4pt}{1pt}{1pt}
    Edge Betw. (MAE $\downarrow$) ($\times 10^{3}$)
    & $9.65{_{\pm 0.87}}$ 
    & $11.0{_{\pm 1.0}}$ 
    & $\tied{9.46}{_{\pm 0.87}}$ 
    & $10.1{_{\pm 1.0}}$ 
    & $10.4{_{\pm 1.2}}$ 
    & $\tied{8.62}{_{\pm 0.98}}$ 
    & $\tied{8.76}{_{\pm 0.86}}$ 
    & $11.82{_{\pm 0.88}}$ 
    & $\tied{9.3}{_{\pm 1.2}}$ 
    \\
    Local Conn. (MAE $\downarrow$)
    & $0.43{_{\pm 0.00}}$ 
    & $0.40{_{\pm 0.00}}$ 
    & $0.34{_{\pm 0.01}}$ 
    & $0.22{_{\pm 0.01}}$
    & $\second{0.19}{_{\pm 0.01}}$
    & $0.22{_{\pm 0.01}}$
    & $\best{0.17}{_{\pm 0.01}}$ 
    & $0.40{_{\pm 0.01}}$ 
    & $0.25{_{\pm 0.01}}$ 
    \\
    Min Cycle (Edge) (MAE $\downarrow$)
    & $0.30{_{\pm 0.02}}$ 
    & $0.24{_{\pm 0.04}}$ 
    & $0.20{_{\pm 0.01}}$ 
    & $0.19{_{\pm 0.01}}$ 
    & $0.19{_{\pm 0.02}}$ 
    & $\tied{0.17}{_{\pm 0.02}}$ 
    & $\tied{0.18}{_{\pm 0.02}}$ 
    & $0.22{_{\pm 0.01}}$ 
    & $\tied{0.18}{_{\pm 0.02}}$ 
    \\
    \addlinespace[2pt]\midrule\addlinespace[1pt]
    \multicolumn{10}{l}{\textit{Graph Prediction}}\\
    Num Cycles (MAE $\downarrow$)
    & $0.34{_{\pm 0.01}}$ 
    & $0.29{_{\pm 0.02}}$ 
    & $0.40{_{\pm 0.06}}$ 
    & $0.22{_{\pm 0.01}}$ 
    & $\tied{0.15}{_{\pm 0.03}}$ 
    & $\tied{0.16}{_{\pm 0.02}}$ 
    & $\tied{0.14}{_{\pm 0.03}}$ 
    & $0.49{_{\pm 0.06}}$ 
    & $\tied{0.16}{_{\pm 0.04}}$ 
    \\
    Diameter (MAE $\downarrow$)
    & $0.99{_{\pm 0.10}}$ 
    & $0.79{_{\pm 0.04}}$ 
    & $1.28{_{\pm 0.14}}$ 
    & $0.82{_{\pm 0.15}}$ 
    & $0.64{_{\pm 0.05}}$ 
    & $0.55{_{\pm 0.02}}$ 
    & $0.67{_{\pm 0.11}}$ 
    & $1.34{_{\pm 0.17}}$ 
    & $0.62{_{\pm 0.08}}$ 
    \\
    Density (MAE $\downarrow$) ($\times 10^{2}$)
    & $2.24{_{\pm 0.57}}$ 
    & $2.29{_{\pm 0.85}}$ 
    & $4.6{_{\pm 1.5}}$ 
    & $3.7{_{\pm 1.7}}$ 
    & $2.5{_{\pm 1.2}}$ 
    & $2.4{_{\pm 1.1}}$ 
    & $2.8{_{\pm 1.2}}$ 
    & $7.2{_{\pm 2.1}}$ 
    & $2.4{_{\pm 1.1}}$ 
    \\
    Wiener Index (MAE $\downarrow$)
    & $0.41{_{\pm 0.07}}$ 
    & $0.23{_{\pm 0.01}}$ 
    & $0.51{_{\pm 0.10}}$ 
    & $0.27{_{\pm 0.06}}$ 
    & $0.16{_{\pm 0.03}}$ 
    & $0.20{_{\pm 0.04}}$ 
    & $\second{0.12}{_{\pm 0.02}}$ 
    & $0.53{_{\pm 0.07}}$ 
    & $0.17{_{\pm 0.04}}$ 
    \\
    Alg. Connect. (MAE $\downarrow$) ($\times 10^{2}$)
    & $3.2{_{\pm 1.5}}$ 
    & $2.23{_{\pm 0.19}}$ 
    & $4.7{_{\pm 1.7}}$ 
    & $4.1{_{\pm 1.6}}$ 
    & $\second{1.79}{_{\pm 0.12}}$ 
    & $2.28{_{\pm 0.66}}$ 
    & $2.67{_{\pm 0.60}}$ 
    & $5.7{_{\pm 2.3}}$ 
    & $2.32{_{\pm 0.63}}$ 
    \\
    \bottomrule
  \end{tabular}
  }
\end{table*}

\section{Statistical analysis of the GNN benchmark}
\label{app:stat-analysis}

The purpose of this appendix is to separate two questions that are easy to conflate: whether the hybrid champion is directionally favoured across benchmark cells, and whether the associated effect size is large enough to be practically resolved. The main text reports a one-sided sign test on the twelve paired (architecture, dataset) cell-level mean differences between the classical champion (RRWP) and the hybrid champion (RRWP$+$AFM$_{\mathrm{stat}}+$FM$_{\mathrm{stat}}$), giving $k=10$ wins out of $n=12$ and $p=0.019$ under a binomial null with $p_{0}=0.5$~\citep{demsar2006statistical}. The headline directional claim rests on this fixed-champion sign test. We complement them with three Bayesian decompositions of the same paired data, evaluated at increasingly fine sample-size resolutions, whose role is to characterise the \emph{amplitude} of the gap rather than its direction: a cell-level analysis on the twelve mean differences, a pooled signed-rank analysis on the $n=300$ trial-level MUTAG observations, and a hierarchical model that partial-pools the trial-level evidence across architectures. We do not apply a multiple-comparison correction because the analyses answer distinct questions; their results are reported jointly for transparency.

\paragraph{Bayesian sign test (cell-level).}
We run a Bayesian sign test~\citep{benavoli2017time} on the twelve paired cell-level mean differences using the \texttt{baycomp} library. To put MUTAG (F$_{1}$) and ZINC$_{\text{subset}}^{*}$ (MAE) on a common scale, each paired mean difference is normalised by the pooled standard deviation $\sqrt{\sigma_{\text{cl}}^{2}+\sigma_{\text{hy}}^{2}}$ of its cell, and the region of practical equivalence (ROPE) is set to $\pm 0.5$ pooled standard deviations, following the half-pooled-SD convention discussed by~\citet{kruschke2018rejecting} and corresponding to the medium-effect threshold of Cohen's $d$. Since differences are normalised by $\sqrt{\sigma_{\mathrm{cl}}^2+\sigma_{\mathrm{hy}}^2}$,
this corresponds to $d \approx 0.71$ in Cohen's convention. With the default Dirichlet-process prior ($s=0.5$, $z_{0}=0$), the posterior reads
\[
\bigl(P_{\text{hybrid}},\,P_{\text{rope}},\,P_{\text{classical}}\bigr) = \bigl(0.018,\;0.982,\;0.000\bigr),
\]
with 95\% bootstrap intervals $[0.000,\,0.389]$, $[0.611,\,1.000]$, and $[0.000,\,0.000]$ respectively, and where the first entry corresponds to the hybrid champion. The bulk of the posterior mass lies in the ROPE region, faithfully reflecting that per-cell mean gaps are individually below half the pooled noise; the decomposition is stable over the ROPE range $\pm 0.2$ to $\pm 0.8$ pooled standard deviations ($P_{\text{rope}}$ from $0.97$ at the tightest ROPE to $0.997$ at the widest), and the corresponding posterior over the population-level Cohen's $d$ has median $0.37$ with 95\% HDI $[-0.21,\,+0.95]$, consistent with anywhere from a small negative to a large positive normalised effect size. The small but nonzero $P_{\text{hybrid}}$ paired with a vanishing $P_{\text{classical}}$ is consistent with the directional sign-test conclusion of the main text without overstating the per-cell amplitude.

\paragraph{Bayesian signed-rank test (MUTAG, trial-level).}
We additionally run a Bayesian signed-rank test~\citep{benavoli2017time} on the trial-level MUTAG observations: $50$ paired (fold, seed) F$_{1}$ scores per architecture, pooled across the six architectures for a total of $n=300$ paired trials between the classical and hybrid champions. We focus on MUTAG at this resolution because it is the dataset where the quantum features appear to bring slightly more uplift across our sweep, and where the trial-level posterior is therefore most informative. The ROPE is set to $\pm 0.01$ F$_{1}$ points, the value recommended by~\citet{benavoli2017time} for classifier comparisons; the median within-cell trial-level standard deviation across the MUTAG champion cells is $0.077$ F$_{1}$, so the ROPE captures only differences smaller than one-eighth of the cross-trial noise. The posterior reads
\[
\bigl(P_{\text{hybrid}},\,P_{\text{rope}},\,P_{\text{classical}}\bigr) = \bigl(0.907,\;0.000,\;0.093\bigr),
\]
with 95\% bootstrap intervals $[0.281,\,1.000]$, $[0.000,\,0.000]$, and $[0.000,\,0.720]$ respectively. Augmenting the signed-rank decomposition with a Bayesian estimate of the pooled mean F$_{1}$ difference between the two champions yields a posterior median of $+0.005$ F$_{1}$ with 95\% HDI $[-0.003,\,+0.013]$, an interval that lies almost entirely on the positive side of zero but whose lower edge crosses zero; the signed-rank $91/9$ asymmetry should therefore be read as a directional rather than an amplitude statement. This pooled analysis treats the $n=300$ paired trials as exchangeable, which over-counts evidence because the $50$ trials within an architecture share architecture-level effects; the wide bootstrap interval on $P_{\text{classical}}$ reflects this fragility, and the hierarchical model below corrects for the clustering directly.

\paragraph{Hierarchical Bayesian test.}
We complement the pooled signed-rank with the hierarchical Bayesian test of~\citet{benavoli2017time}, which models a per-architecture mean difference drawn from a shared prior and partial-pools the trial-level evidence across architectures rather than treating all $n=300$ pairs as exchangeable. With six architectures, $50$ paired trials each, and the same $\pm 0.01$ F$_{1}$ ROPE, the posterior reads
\[
\bigl(P_{\text{hybrid}},\,P_{\text{rope}},\,P_{\text{classical}}\bigr) = \bigl(0.241,\;0.725,\;0.034\bigr),
\]
with 95\% bootstrap intervals $[0.000,\,0.457]$, $[0.057,\,1.000]$, and $[0.000,\,0.942]$ respectively. The bulk of the posterior returns to the practical-equivalence region, in agreement with the cell-level decomposition; the residual directional mass nevertheless leans toward the hybrid side, in agreement with the sign tests. The associated posterior over the global mean F$_{1}$ difference has median $+0.005$ with 95\% HDI $[-0.009,\,+0.019]$, the between-architecture standard deviation $\tau$ has median $0.006$ F$_{1}$ with 95\% HDI $[0.001,\,0.026]$, suggesting non-negligible between-architecture heterogeneity, and the per-architecture posteriors all assign 95\% HDIs to mean F$_{1}$ differences that include zero, with point medians from $+0.002$ (GeneralConv) to $+0.011$ (GCN). The qualitative decomposition is stable across the ROPE range $\pm 0.005$ to $\pm 0.02$ F$_{1}$: the bulk of the mass shifts from the directional region at the tightest ROPE ($P_{\text{hybrid}} = 0.56$) to overwhelmingly within the equivalence region at the widest ($P_{\text{rope}} = 0.97$), while the residual directional mass remains more concentrated on the hybrid side than on the classical side. We read the hierarchical posterior as the most defensible amplitude statement available: the per-architecture mean gap between the classical and hybrid champions on MUTAG is small with non-trivial probability of being practically zero, no individual architecture's mean is statistically resolvable from zero on its own, and the directional signal lives in the consistency of the per-architecture point estimates and in the residual asymmetry between $P_{\text{hybrid}}$ and $P_{\text{classical}}$.

\paragraph{Independence of cells.}
The cell-level sign tests treat the twelve (architecture, dataset) cells as exchangeable observations. Because the six architectures of each dataset are evaluated on the same train/validation/test splits (the standard PyG ZINC$_{\text{subset}}^{*}$ partition and a fixed stratified 10-fold partition for MUTAG with random state $42$), the per-cell mean differences may be correlated within each dataset. We therefore treat the sign-test $p$-values as descriptive evidence of directional consistency rather than as exact type-I error guarantees; this is also why the Bayesian analyses above explicitly separate the cell-level, pooled trial-level, and architecture-hierarchical resolutions.

\section{Linear and nonlinear regression of quantum features}
\label{app:walk-ceiling-setup}

In this appendix, we explain how we assess the capacity of combinations of classical features (powers of adjacency matrix or RRWP) to represent QGFs through linear and nonlinear regression experiments discussed in Sec.~\ref{subsec:predictability_of_quantum_features}.

\paragraph{Per-graph target.}

Each molecular graph $\mathcal{G}_\alpha$ of the dataset, with $\alpha \in \{1,\dots,N_\mathrm{data}\}$, is associated with a number of atoms $N_\alpha$, a classical adjacency matrix $A_\alpha$ and a row-normalized adjacency matrix $M_\alpha = D_\alpha^{-1}A_\alpha$, as well as---after the quantum feature extraction pipeline of Sec.~\ref{sec:quantum-feature-extraction}---with a sampled correlation matrix $C_{i,j}^\alpha(t)$ [cf. Eq.~(\ref{eq:obs_C})].
We use the $M_\alpha^\mathrm{off}=M_\alpha^\mathrm{tot} - N_\alpha$ off-diagonal (and independent) correlation entries to construct the target vector $y_\alpha(t) \in \mathbb{R}^{M_\alpha^\mathrm{off}}$, where $M^\mathrm{tot}_\alpha = N_\alpha(N_\alpha+1)/2$.

\paragraph{Pooling.}
For each graph $\mathcal{G}_\alpha$ with node-degree matrix $D_{\alpha}$, we construct two pair-feature families indexed by $k=0,\dots,K$,
\begin{equation}
\phi_k^{\mathrm{walk}}(A_\alpha)_{ij} = (A_\alpha^k)_{ij},
\qquad
\phi_k^{\mathrm{rrwp}}(A_\alpha)_{ij} = ((D_{\alpha}^{-1}A_\alpha)^k)_{ij},
\label{eq:phi}
\end{equation}
i.e.\ the number of length-$k$ walks from $i$ to $j$~\citep{diestel2017}, and the $k$-step random-walk transition probability respectively (RRWP, as in GRIT~\citep{ma2023graph}).
Each pair $(i,j)$ contributes one row $[\phi_0,\dots,\phi_K]$ to a per-graph design $X_\alpha^{\mathrm{walk}}, X_\alpha^{\mathrm{rrwp}} \in \mathbb{R}^{M_\alpha^\mathrm{off}\times(K+1)}$, which we then concatenate row-wise across all $N$ graphs:
\begin{equation}
X = \begin{bmatrix} X_1\\ X_2\\ \vdots\\ X_N \end{bmatrix} \in \mathbb{R}^{P\times(K+1)},
\qquad
y(t) = \begin{bmatrix} y_1(t)\\ y_2(t)\\ \vdots\\ y_N(t) \end{bmatrix} \in \mathbb{R}^{P},
\qquad
P = \sum_{\alpha=1}^{N_\mathrm{data}} M_\alpha^\mathrm{off}.
\end{equation}
Rows from different graphs share column indices $k=0,\dots,K$, so a single coefficient vector $\beta \in \mathbb{R}^{K+1}$ is fit jointly to \emph{all} pairs of \emph{all} graphs. This is what we mean by ``pooled'': one regression, $P$ data points, no per-graph parameters.
The reported pooled $R^2$ is one number per $(K, t, \text{family})$ cell, computed against the global mean $\bar y(t) = P^{-1}\sum_p y_p(t)$ (MUTAG: $P = 30{,}505$ pairs over $N_\mathrm{data}=188$ graphs; ZINC: $P \approx 3{\times}10^6$ over $N_\mathrm{data}=11{,}368$).

\paragraph{Column normalization.}
Columns of $X$ are rescaled by their max absolute value, $X^{\mathrm{norm}}_{:,k} = X_{:,k}/\max(\max_p |X_{p,k}|,\,1)$, to avoid $O(\lambda_{\max}^k)$ overflow at high $K$ (without this, $(A^k)_{ij}$ overflows \texttt{float64} around $K\approx 30$ on dense graphs), where $\lambda_{\max}$ is the maximal eigenvalue.

\paragraph{Probes.}
The linear probe solves ordinary least squares (OLS) regression,
$\hat\beta = \arg\min_\beta \|X^{\mathrm{norm}}\beta - y\|_2^2$,
via the SVD-based Moore--Penrose pseudoinverse~\citep{golub2013} (\texttt{numpy.linalg.lstsq}, \texttt{rcond=None}) and reports $R^2 = 1 - \mathrm{SS}_{\mathrm{res}}/\mathrm{SS}_{\mathrm{tot}}$ against the pooled mean. Nonlinear probes use kernel ridge with an RBF kernel~\citep{scholkopf2002} (Nystr\"om-approximated~\cite{williams2001} with $n_{\mathrm{components}}=500$, $\alpha=10^{-3}$, bandwidth from the median pairwise-distance heuristic~\citep{garreau2017}; verified within $1$\,pp of exact KernelRidge) and random forest~\citep{breiman2001} ($100$ trees, \texttt{max\_depth}=20, \texttt{min\_samples\_leaf}=10), scored by 5-fold held-out $R^2$ (\texttt{scikit-learn} implementations~\citep{pedregosa2011}).

\paragraph{Laplacian augmentation.}
We additionally augment each design $X_{\alpha}$ with the first $K$ eigenpairs $\{u_k(i),\,u_k(j),\,\lambda_k\}_{k=1}^{{K}}$ (sorted by ascending eigenvalue $\lambda$, zero-padded when $n_\alpha < K$, eigenvector signs canonicalized by largest-magnitude entry) of the symmetric normalized Laplacian $L = I - D^{-1/2}AD^{-1/2}$~\citep{vonluxburg2007}, chosen for its bounded spectrum $\lambda_k \in [0,2]$ that is comparable across graphs of different size.

\subsection{Regression results}\label{app:walk-ceiling-results}

We run the probes of App.~\ref{app:walk-ceiling-setup} on
MUTAG ($188$ graphs, $n\!\in\![10,28]$) and
\ZINC ($11{,}368$, $n\!\in\![9,37]$).

\paragraph{Linear probe (OLS).}
At $t\le 30$\,ns we observe $R^2 = 1.000$ on both pools.
At measurement times the explained variance plateaus (Table~\ref{tab:walk-ceiling}, walk-power, AFM connected, time-averaged on $t\in\mathrm{linspace}(100,6000)$\,ns):

\begin{table}[h]
\centering
\small
\caption{Pooled OLS $R^2$ of \Ctwo on walk-power features.
$K$-saturation is sharp ($\le 1 \%$ from $K{=}7$ to $K{=}50$).}
\begin{tabular}{lcccc}
\toprule
dataset & $R^2_{K=1}$ & $R^2_{K=7}$ & $R^2_{K=21}$ & $R^2_{K=50}$ \\
\midrule
MUTAG & 0.59 & 0.78 & 0.83 & 0.83 \\
ZINC  & 0.59 & 0.76 & 0.78 & 0.78 \\
\bottomrule
\end{tabular}
\label{tab:walk-ceiling}
\end{table}

\paragraph{Per-graph ceiling.}
The plateau is also \emph{per-graph}, not a pooling artifact: at the Cayley--Hamilton bound $K = n_\alpha - 1$~\citep{horn2013}, beyond which $\{I, A_\alpha, \dots, A_\alpha^K\}$ stops gaining rank, the per-graph saturated $R^2$ on MUTAG has median $0.876$ (IQR $[0.855, 0.882]$) for AFM walk-power, with a worst-case graph (MUTAG \#16) capping at $R^2 \approx 0.76$; RRWP gives essentially identical numbers (AFM median $0.879$, worst case $0.75$ for MUTAG \#149). We additionally verify that pooled $R^2$ and the per-graph mean $R^2_{\mathrm{pg}} = N_\mathrm{data}^{-1}\sum_\alpha R^2_\alpha$ (regression refit independently per graph) agree to within a few percentage points, ruling out the possibility that the pooled headline number is dominated by between-graph correlator-magnitude variance.

\paragraph{Nonlinear probe (KR + RF).}
To show that the ceiling is not just the linearity of OLS,
we fit the two nonlinear regressors above on the \emph{same} column-normalized designs. MUTAG, $K=21$, mid-plateau $t=1536$\,ns; the $5$ folds are formed at the graph level, so that pairs of the same graph never appear in both training and test sets:

\begin{table}[h]
\centering
\small
\caption{Nonlinear probes close at most $7$--$8 \%$ of the linear
gap for walk powers (OLS at the same time; graph-level folds, fold standard deviations $\le0.008$ for AFM and $\le0.033$ for FM). Two distinct ceilings separate cleanly.}
\begin{tabular}{lcccc}
\toprule
target / family   & OLS pool & KR (RBF) & RF      & residual \\
\midrule
AFM connected, walk     & 0.83     & 0.895    & 0.906   & $\sim$9\%  \\
AFM connected, RRWP     & --     & 0.898    & 0.909   & $\sim$9\%  \\
FM  connected, walk     & 0.70     & 0.753    & 0.779   & $\sim$22\% \\
FM  connected, RRWP     & --     & 0.693    & 0.751   & $\sim$25\% \\
\bottomrule
\end{tabular}
\label{tab:nonlinear}
\end{table}

The OLS-to-KR/RF gap is the \emph{nonlinear} content of \Ctwo
that walk-powers carry --- what an MPNN with a nonlinear readout could absorb. The KR/RF-to-1.0 gap is content that is \textbf{not a function of any walk-power / RRWP feature}: it requires features outside the adjacency-power algebra (e.g.\ eigenvector geometry).
Walk-powers and RRWP give numerically indistinguishable nonlinear-probe ceilings (within $\le 1 \%$), confirming the residual is a property of the feature family rather than of either specific parameterization.

\paragraph{Laplacian augmentation.}
Augmenting the design with normalized-Laplacian eigenpairs (App.~\ref{app:walk-ceiling-setup}) localizes where the residual lives. MUTAG, RF $5$-fold $R^2$, $K=21$, $t=1536$\,ns, graph-level folds:
\begin{table}[h]
\centering
\caption{Adding Laplacian eigenpairs lifts FM-conn from $0.75$--$0.78$
to $0.84$--$0.85$ --- the FM ``beyond-walk'' residual is partly spectral. The
AFM residual is mostly walk-content under nonlinearity (Laplacian
augmentation adds $\le 1 \%$).}
\small
\begin{tabular}{lcccc}
\toprule
target          & walk  & walk+Lap        & RRWP  & RRWP+Lap        \\
\midrule
AFM connected   & 0.906 & 0.912           & 0.909 & \textbf{0.913}           \\
FM  connected   & 0.779 & \textbf{0.847}  & 0.751 & 0.840  \\
\bottomrule
\end{tabular}
\label{tab:laplacian}
\end{table}

\begin{figure}[htbp]
\centering
\includegraphics[width=\linewidth]{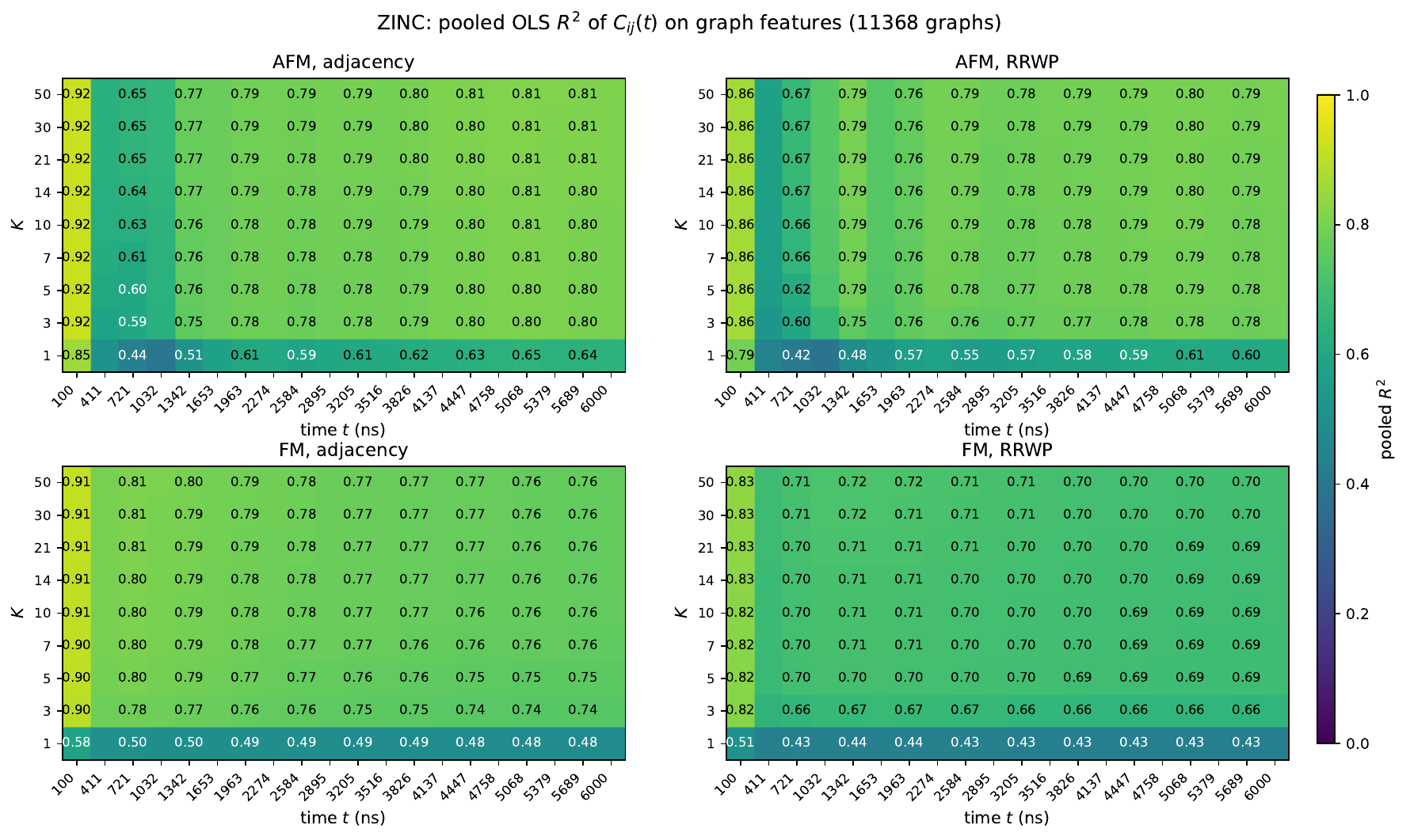}
\caption{
Pooled OLS results showing the $R^2$ score at different time steps for the AFM and FM regimes using $K$ classical features ($A^k$ for adjacency or $(D^{-1}A)^{k}$ for RRWP) on the \ZINC dataset.}
\label{fig:linear-zinc}
\end{figure}

\begin{figure}[htbp]
\centering
\includegraphics[width=\linewidth]{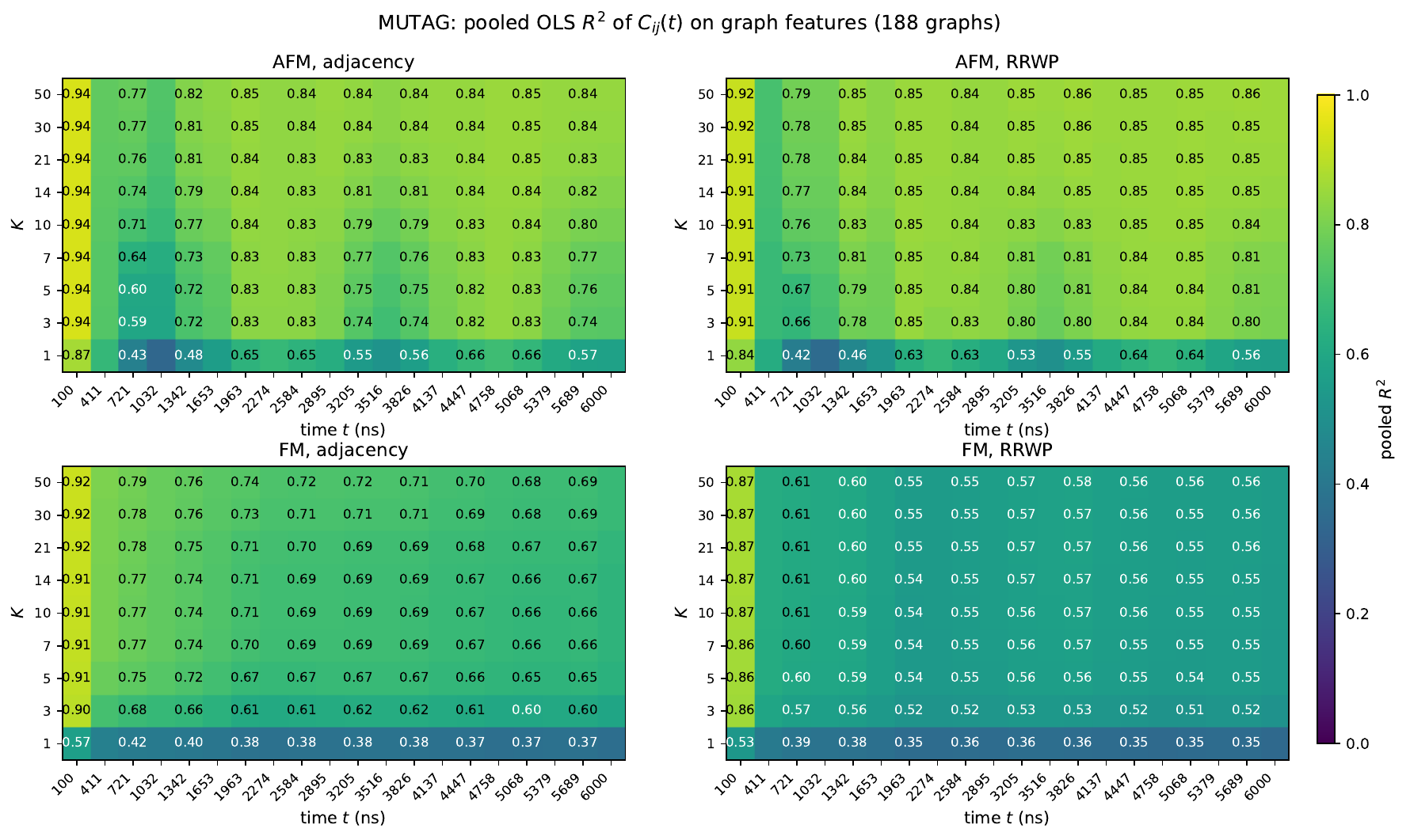}
\caption{
Pooled OLS results showing the $R^2$ score at different time steps for the AFM and FM regimes using $K$ classical features ($A^k$ for adjacency or $(D^{-1}A)^{k}$ for RRWP)  on the MUTAG dataset.
}
\label{fig:linear-mutag}
\end{figure}

\section{\texorpdfstring{Quantum graph features beyond classical structural encodings: details}{Quantum graph features beyond classical structural encodings: details}}
\label{app:controls}

This appendix details the controls summarized in Sec.~\ref{subsec:predictability_of_quantum_features} and Sec.~\ref{sec:controls}.

\subsection{\texorpdfstring{Expanded reconstruction basis and residual}{Expanded reconstruction basis and residual}}
\label{app:residual}

\begin{figure}[t]
  \centering
  \includegraphics[width=\linewidth]{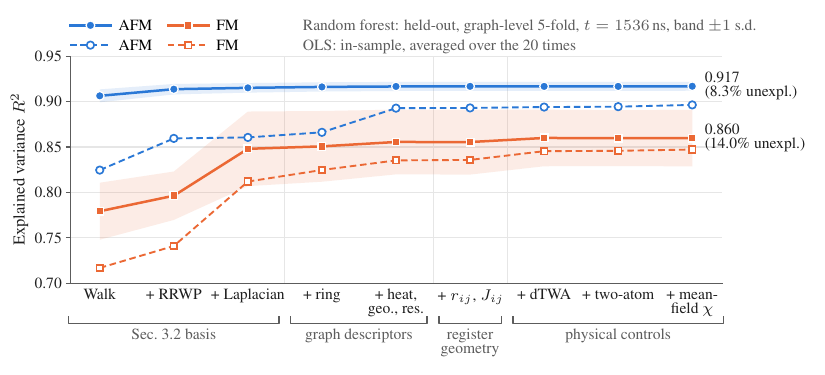}
  \caption{Explained variance $R^2$ of the off-diagonal correlations on MUTAG along
    the nine cumulative classical bases ($K=21$); brackets group the steps into the
    four rows of Table~\ref{tab:ceiling}, each row being the last step of its
    bracket. Solid: random forest, $R^2$ pooled over graph-level 5-fold held-out
    predictions at $t=1536$\,ns, band $\pm1$ standard deviation of the per-fold $R^2$;
    dashed: in-sample OLS, $R^2$ averaged over the 20 times. The band mainly reflects
    which graphs fall in each fold; the folds are identical at every step, so it is
    largely shared across steps and does not measure the uncertainty of the
    step-to-step changes. With the full basis the
    random forest leaves 8.3\% (AFM) and 14.0\% (FM) of the variance unexplained;
    register geometry adds nothing and the physical controls add at most 0.5
    percentage points (FM). $\chi$ is the static mean-field susceptibility
    $(1-\beta J)^{-1}$ of the register couplings.}
  \label{fig:recon-ceiling}
\end{figure}

Table~\ref{tab:ceiling} extends the reconstruction of App.~\ref{app:walk-ceiling-setup} with a cumulatively expanded basis.
Beyond walk powers, RRWP and Laplacian eigenpairs, it adds ring-membership indicators; heat-kernel, geodesic-distance and effective-resistance pair features; the register distances $r_{ij}$ and couplings $J_{ij}=C_6/r_{ij}^6$; and three restricted physical approximations of the same quench: classical-spin dynamics (discrete truncated Wigner approximation, dTWA), exact two-atom correlations under the same couplings, and mean-field correlations.
All nonlinear scores use graph-level folds.
With the full basis, the random forest leaves $8.3\%$ (AFM) and $14.0\%$ (FM) of the variance unexplained. The additive reconstruction process is shown in Fig.~\ref{fig:recon-ceiling}.
The register geometry adds nothing once the graph descriptors are present, and the restricted physical approximations add at most $0.5$ percentage points (FM), so the residual is explained neither by the layout nor by these approximations.

Table~\ref{tab:residual} characterizes the residual on \ZINC.
It is small in amplitude ($0.03$--$0.05$ at $d=1$, against $0.25$--$0.43$ for the initial correlations) but decays slowly with distance, so that it accounts for most of the long-range correlations (AFM: $d\gtrsim4$; FM: $d\gtrsim7$).
In the normalized-Laplacian eigenbasis, the FM residual is concentrated in global, low-$\lambda$ modes ($49\%$ of its energy below $\lambda=0.04$), whereas the AFM residual is spread across the spectrum.
Used as a standalone input to the intrinsic probe, the AFM$+$FM residual retains $97.4\%$ of the performance of the full features on graph-level tasks and $82.5\%$ on node-level tasks, against $42.4\%$ on edge-level tasks, which mostly depend on short-range structure.
No single graph invariant explains it: controlling for graph size, the mean absolute Spearman correlation between the residual and the benchmark targets is $0.46$ for graph-level, $0.33$ for edge-level and $0.17$ for node-level targets, with the strongest correlations for the number of cycles ($-0.57$), density ($-0.55$) and algebraic connectivity ($+0.46$).
A linear fit of the long-range residual ($d\ge4$) on classical-spin (dTWA) correlations gives $R^2<10^{-5}$ for AFM and $R^2=0.004$ for FM.

The residual is a low-amplitude, longer-ranged component, concentrated in low-$\lambda$ Laplacian modes for FM and spread across the spectrum for AFM; alone, it retains $97\%$ ($83\%$) of the full-feature probe performance on graph-level (node-level) tasks.

\begin{table}[h]
\centering
\caption{Reconstruction of the off-diagonal correlations on MUTAG from an expanded classical basis ($K=21$, reference time $t=1536$\,ns). Random forest, $R^2$ on graph-level 5-fold held-out data $\pm$ fold standard deviation; each row adds the indicated descriptors to all previous ones. Last row: unexplained variance $1-R^2$ of the full basis.}
\label{tab:ceiling}
\small
\begin{tabular}{l cc}
\toprule
Basis & AFM & FM \\
\midrule
Walk powers, RRWP, Laplacian eigenpairs (Sec.~\ref{subsec:predictability_of_quantum_features}) & $0.915\pm0.005$ & $0.848\pm0.041$ \\
$+$ ring, heat-kernel, geodesic and effective-resistance descriptors & $0.917\pm0.005$ & $0.856\pm0.036$ \\
$+$ register geometry ($r_{ij}$, $J_{ij}$) & $0.917\pm0.005$ & $0.855\pm0.036$ \\
$+$ physical controls (dTWA, exact two-atom, mean-field) & $0.917\pm0.005$ & $0.860\pm0.031$ \\
\midrule
Unexplained variance & $8.3\%$ & $14.0\%$ \\
\bottomrule
\end{tabular}
\end{table}

\begin{table}[h]
\centering
\caption{Structure of the unexplained residual of the off-diagonal correlations on \ZINC (time index 7), compared with the initial correlations. (a) Cumulative energy of the correlation matrix in the normalized-Laplacian eigenbasis up to eigenvalue $\lambda$. (b) Mean absolute value against shortest-path distance $d$ (ZZ convention, $4C_{i,j}$). (c) Retention of probe performance when the residual replaces the full features, $(\text{random}-\text{residual})/(\text{random}-\text{full})$, averaged per task category.}
\label{tab:residual}
\small
\setlength{\tabcolsep}{4pt}
\begin{tabular}{l cccc}
\toprule
\multicolumn{5}{l}{\textit{(a) Cumulative spectral energy}} \\
$\lambda$ & AFM initial & AFM residual & FM initial & FM residual \\
\midrule
$0.04$ & 0.050 & 0.018 & 0.737 & 0.493 \\
$0.54$ & 0.194 & 0.111 & 0.810 & 0.769 \\
$1.04$ & 0.289 & 0.351 & 0.848 & 0.928 \\
$1.54$ & 0.344 & 0.650 & 0.905 & 0.979 \\
$1.88$ & 0.527 & 0.888 & 0.964 & 0.997 \\
\midrule
\multicolumn{5}{l}{\textit{(b) Mean $|4C_{i,j}|$ versus distance}} \\
$d$ & AFM initial & AFM residual & FM initial & FM residual \\
\midrule
$1$ & 0.245 & 0.034 & 0.426 & 0.047 \\
$3$ & 0.049 & 0.031 & 0.153 & 0.048 \\
$5$ & 0.026 & 0.024 & 0.064 & 0.041 \\
$10$ & 0.021 & 0.021 & 0.036 & 0.033 \\
\midrule
\multicolumn{5}{l}{\textit{(c) Retention of probe performance}} \\
& Graph-level & Node-level & Edge-level & \\
\midrule
AFM$+$FM residual & $97.4\%$ & $82.5\%$ & $42.4\%$ & \\
\bottomrule
\end{tabular}
\end{table}

\subsection{\texorpdfstring{Geometry-augmented baselines}{Geometry-augmented baselines}}
\label{app:geometry}

The register geometry carries substantial information on its own (the couplings predict degree almost perfectly), so we compare against RRWP$+$Laplacian$+$Rings$+$Geometry (Sec.~\ref{subsec:classical_baselines}).
On the intrinsic benchmark on \ZINC, adding either the AFM$+$FM correlations or only their unexplained residual (Sec.~\ref{subsec:predictability_of_quantum_features}) to this baseline improves betweenness, bridge-node detection, edge betweenness and local connectivity beyond one standard error; eight tasks end in a statistical tie and the three cycle tasks are excluded, since the ring descriptors determine their targets almost by construction (Table ~\ref{tab:geometry-intrinsic}).

The geometry-augmented baseline (Base) adds ring descriptors and the register distances $r_{ij}$ and couplings $J_{ij}$ to RRWP$+$Laplacian.
For the GNN backbones, the geometry enters as five node-level and three pair-level register channels together with cycle-basis ring descriptors; for the GT, as the pair channels $(C_6/r_{ij}^6, r_{ij})$.
Table~\ref{tab:geometry-intrinsic} reports the intrinsic benchmark.
Geometry alone is highly informative (the couplings predict degree almost exactly), and Base already improves over RRWP$+$Laplacian on most tasks.
On top of it, both the AFM$+$FM correlations and their residuals improve betweenness, bridge-node detection, edge betweenness and local connectivity with disjoint standard-error bars, the residuals by a slightly larger margin.
Eight tasks end in a statistical tie (for the residuals, density and algebraic connectivity are marginally worse than Base, at the edge of the standard-error bars), and the three cycle tasks are excluded.
The geometry thus accounts for a substantial part of the signal, but not for all of it.

\begin{table}[t]
\centering
\caption{Intrinsic information benchmark on \ZINC with the geometry-augmented baseline.
Mean $\pm$ standard error over 5 folds; MAE ($\downarrow$) or accuracy ($\uparrow$).
Base: RRWP$+$Laplacian$+$Rings$+$Geometry ($r_{ij}$, $J_{ij}$); the last two columns add the AFM$+$FM correlations or only their unexplained residuals (Sec.~\ref{subsec:predictability_of_quantum_features}) to it.
Bold: improvement over Base with disjoint $\pm1$ standard-error bars.
Cycle tasks (--) are excluded for the ring-augmented columns, since the ring descriptors determine their targets almost by construction.}
\label{tab:geometry-intrinsic}
\small
\setlength{\tabcolsep}{4pt}
\begin{tabular}{l cccc}
\toprule
Task & RRWP$+$Laplacian & Base & Base$+$AFM$+$FM & Base$+$residuals \\
\midrule
Degree$^{\downarrow}$ ($\times 10^{2}$) & $6.98{\scriptstyle\pm 1.06}$ & $0.84{\scriptstyle\pm 0.25}$ & $0.86{\scriptstyle\pm 0.21}$ & $0.83{\scriptstyle\pm 0.28}$ \\
Betweenness$^{\downarrow}$ ($\times 10^{2}$) & $4.76{\scriptstyle\pm 0.13}$ & $2.01{\scriptstyle\pm 0.08}$ & $\mathbf{1.79}{\scriptstyle\pm 0.04}$ & $\mathbf{1.74}{\scriptstyle\pm 0.06}$ \\
Bridge Node$^{\uparrow}$ & $0.950{\scriptstyle\pm 0.002}$ & $0.992{\scriptstyle\pm 0.000}$ & $\mathbf{0.995}{\scriptstyle\pm 0.000}$ & $\mathbf{0.996}{\scriptstyle\pm 0.000}$ \\
Eccentricity$^{\downarrow}$ & $1.081{\scriptstyle\pm 0.018}$ & $0.632{\scriptstyle\pm 0.006}$ & $0.648{\scriptstyle\pm 0.015}$ & $0.631{\scriptstyle\pm 0.003}$ \\
Min Cycle (Node)$^{\downarrow}$ & $0.603{\scriptstyle\pm 0.012}$ & -- & -- & -- \\
Link Pred.$^{\uparrow}$ & $1.000{\scriptstyle\pm 0.000}$ & $1.000{\scriptstyle\pm 0.000}$ & $1.000{\scriptstyle\pm 0.000}$ & $1.000{\scriptstyle\pm 0.000}$ \\
Edge Betw.$^{\downarrow}$ ($\times 10^{3}$) & $6.09{\scriptstyle\pm 0.43}$ & $3.66{\scriptstyle\pm 0.08}$ & $\mathbf{2.98}{\scriptstyle\pm 0.11}$ & $\mathbf{2.67}{\scriptstyle\pm 0.09}$ \\
Min Cycle (Edge)$^{\downarrow}$ ($\times 10^{2}$) & $10.23{\scriptstyle\pm 0.47}$ & -- & -- & -- \\
Bridge Edge$^{\uparrow}$ & $0.994{\scriptstyle\pm 0.000}$ & $1.000{\scriptstyle\pm 0.000}$ & $1.000{\scriptstyle\pm 0.000}$ & $1.000{\scriptstyle\pm 0.000}$ \\
Local Conn.$^{\downarrow}$ ($\times 10^{2}$) & $7.39{\scriptstyle\pm 0.17}$ & $2.79{\scriptstyle\pm 0.04}$ & $\mathbf{2.43}{\scriptstyle\pm 0.08}$ & $\mathbf{2.34}{\scriptstyle\pm 0.09}$ \\
Num Cycles$^{\downarrow}$ ($\times 10^{2}$) & $5.90{\scriptstyle\pm 0.25}$ & -- & -- & -- \\
Diameter$^{\downarrow}$ & $0.473{\scriptstyle\pm 0.022}$ & $0.295{\scriptstyle\pm 0.010}$ & $0.290{\scriptstyle\pm 0.010}$ & $0.280{\scriptstyle\pm 0.010}$ \\
Density$^{\downarrow}$ ($\times 10^{3}$) & $1.14{\scriptstyle\pm 0.08}$ & $0.99{\scriptstyle\pm 0.05}$ & $1.12{\scriptstyle\pm 0.10}$ & $1.16{\scriptstyle\pm 0.12}$ \\
Wiener Index$^{\downarrow}$ ($\times 10^{2}$) & $9.61{\scriptstyle\pm 0.41}$ & $5.89{\scriptstyle\pm 0.06}$ & $6.24{\scriptstyle\pm 0.47}$ & $6.48{\scriptstyle\pm 0.70}$ \\
Alg. Connect.$^{\downarrow}$ ($\times 10^{3}$) & $2.22{\scriptstyle\pm 0.05}$ & $1.90{\scriptstyle\pm 0.12}$ & $1.84{\scriptstyle\pm 0.09}$ & $2.22{\scriptstyle\pm 0.19}$ \\
\bottomrule
\end{tabular}
\end{table}

\subsection{\texorpdfstring{Placement sensitivity}{Placement sensitivity}}
\label{app:placement}

We displace every atom of four \ZINC graphs by Gaussian noise of standard deviation $\sigma$ (relative to the minimal inter-atomic distance) and compare the AFM correlations with the reference ones (Table~\ref{tab:placement}).

\begin{figure}[t]
  \centering
  \includegraphics[width=\linewidth]{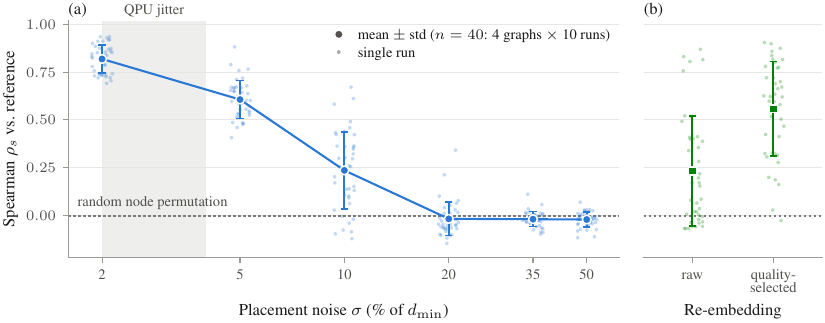}
  \caption{Sensitivity of the AFM correlations to atom placement, measured by the
    Spearman correlation $\rho_s$ between the off-diagonal correlations (all 20 times
    pooled) of a perturbed layout and of the reference one. (a)~Gaussian displacement
    of every atom with standard deviation $\sigma$, in units of the minimal
    inter-atomic distance $d_{\text{min}}\approx 5\,\mu$m (log scale); the perturbed
    register is then rescaled so that its minimal distance is $5\,\mu$m. The gray band
    marks hardware-level jitter ($\sigma\approx 2$--$4\%$). (b)~Ten alternative
    layouts per graph from re-running the force-based placement, without and with
    selection on embedding quality. Markers: mean $\pm$ standard deviation over
    $n=40$ cases (4 ZINC$^*$ subset graphs $\times$ 10 runs); dots: single runs;
    dotted line: random node permutation (Table~\ref{tab:placement}).}
  \label{fig:layout-noise}
\end{figure}

We distinguish small positional noise around a fixed layout from alternative valid layouts.
For the first, every atom $i$ with reference position $\boldsymbol{r}_i$ is displaced to $\boldsymbol{r}_i'=\boldsymbol{r}_i+\boldsymbol{\epsilon}_i$ with $\boldsymbol{\epsilon}_i\sim\mathcal{N}(0,\sigma^2 I_2)$, where the two Cartesian components are drawn independently and $\sigma$ is given in units of the minimal inter-atomic distance.
For the second, we rerun the force-based placement with different initializations and keep 10 valid layouts per graph.
In both cases, we re-emulate the AFM dynamics and compare the correlation matrices with the reference ones (Table~\ref{tab:placement}). Results are shown in Fig.~\ref{fig:layout-noise}.

At hardware-level noise ($\sigma=2\%$) the Spearman correlation is $0.82\pm0.07$, and $0.61\pm0.10$ at $\sigma=5\%$; above $\sigma=20\%$ it falls to the level of a random node permutation.
Ten alternative layouts obtained by re-running the force-based placement agree less ($0.56\pm0.25$ after selecting embeddings of good quality, $0.23\pm0.29$ without selection).
QGFs are thus robust to small positional noise around a fixed layout but not invariant under re-embedding: they describe a graph through a specified graph-to-register protocol.

Hardware-level jitter ($\sigma\approx2$--$4\%$, i.e., $0.1$--$0.2\,\mu$m) is already present in our QPU run, so the hardware results of Sec.~\ref{sec:gnn-downstream} partly validate the robustness to it.
The similarity decays smoothly with $\sigma$ and reaches the permutation floor near $\sigma=20\%$.
Re-embeddings change the long-range couplings more strongly and agree less, particularly without selection on embedding quality.
QGFs are therefore descriptors of a specified graph-to-register protocol; constructing canonical or placement-invariant quantum encodings is left for future work.

\begin{table}[h]
\centering
\caption{Placement sensitivity of the AFM features on four \ZINC graphs ($N=16,20,12,11$). Spearman correlation between the off-diagonal correlations (all 20 times pooled) of the perturbed and the reference layout; mean $\pm$ standard deviation over 4 graphs $\times$ 10 runs. Positional noise: Gaussian displacement of every coordinate with standard deviation $\sigma$, in units of the minimal inter-atomic distance; hardware-level jitter corresponds to $\sigma\approx2$--$4\%$.}
\label{tab:placement}
\small
\begin{tabular}{l c}
\toprule
Perturbation & Spearman $\rho_s$ \\
\midrule
$\sigma=2\%$ & $0.819\pm0.073$ \\
$\sigma=5\%$ & $0.607\pm0.099$ \\
$\sigma=10\%$ & $0.236\pm0.201$ \\
$\sigma=20\%$ & $-0.017\pm0.088$ \\
\midrule
10 re-embeddings, raw & $0.233\pm0.287$ \\
10 re-embeddings, quality-selected & $0.558\pm0.247$ \\
\midrule
Random node permutation & $-0.001$ \\
\bottomrule
\end{tabular}
\end{table}

\subsection{\texorpdfstring{Ladder graphs}{Ladder graphs}}
\label{app:ladder}

\begin{figure}[h]
\centering
\includegraphics[width=0.5\columnwidth]{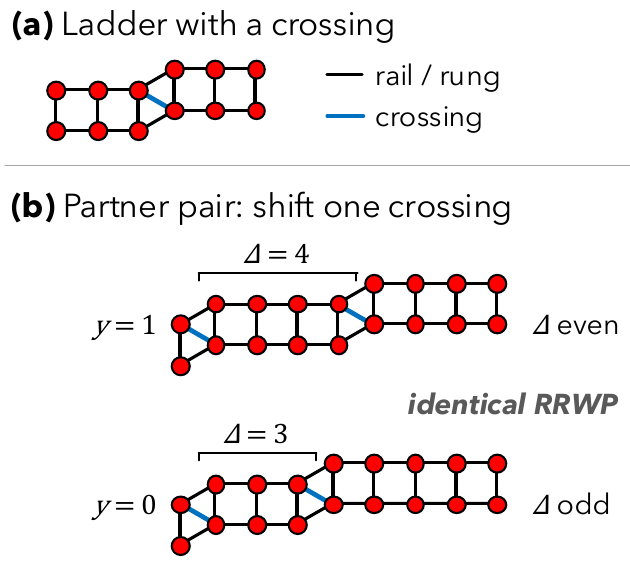}
\caption{
Ladder graphs (a) used to showcase the ability of QGFs to distinguish two classes $y$ (b) characterised by even/odd shifts $\Delta$ of the crossing.
}
\label{fig:ladders}
\end{figure}

In Fig.~\ref{fig:ladders}, a ladder with $m$ rungs $(T_i,B_i)$ has two alternating configurations of the blockade (maximum independent set) ground state, $T,B,T,B,\dots$ and $B,T,B,T,\dots$; a diagonal crossing added to a cell excludes one of them.
If two crossings are separated by an even number of cells $\Delta$, they exclude the same configuration and leave a unique ground state (pinned, $y=1$); if they are separated by an odd number of cells, they exclude opposite configurations and leave a degenerate ground-state manifold (frustrated, $y=0$).
Partners of opposite label thus differ only by moving one crossing by one cell; labels are computed exactly by a transfer-matrix recursion over the rungs.

To exclude short random walks as an explanation, we enumerate all ladders with $9$--$13$ rungs and group them by their pooled RRWP-2 signature (sorted entries of $M$ and $M^2$, $M=D^{-1}A$), keeping only groups that hold both labels, balanced per group.
Within a group, every pooled RRWP-2 and ring feature is identical, so these features are at chance by construction.
This yields 394 graphs in 98 groups, with $18$--$26$ atoms.
The registers use an exact unit-disk layout with square cells and rhombic crossing cells ($r_{\mathrm{NN}}=5.01\,\mu$m; all non-edges at distance $\ge\sqrt{2}\,r_{\mathrm{NN}}$).
The AFM and FM features are noiseless expectation values of the occupations and connected correlations at the 20 times of Sec.~\ref{subsec:pipeline} (matrix-product-state emulation, bond dimension 16), pooled over nodes, pairs and edges into permutation-invariant graph-level vectors; the classical families are pooled in the same way.
A standardized logistic regression is trained on 5 folds grouped by blind group, repeated 3 times.

Table~\ref{tab:ladder} shows that random-walk and geometric descriptors, alone or combined with spectral and ring descriptors, remain far below the quantum features, whereas AFM alone reaches $0.914$ ROC-AUC and AFM$+$FM $0.941$; adding all classical features to AFM$+$FM changes little.
The family is synthetic and its label is tied by construction to frustration and ground-state degeneracy; the result is therefore an existence statement, not evidence of a generic advantage on molecular benchmarks.

\begin{table}[h]
\centering
\caption{Ladder graphs: pinned versus frustrated ground state (394 graphs, 98 RRWP-2-blind groups, $18$--$26$ atoms). Logistic regression, 5 folds grouped by blind group $\times$ 3 repeats; mean $\pm$ standard deviation over the 15 test folds. Label-permutation test (100 shuffles): $p=0.010$, the minimum, for all feature sets except RRWP ($p=0.020$) and Geometry ($p=0.446$).}
\label{tab:ladder}
\small
\begin{tabular}{l c c c}
\toprule
Feature set & Accuracy & F$_1$ & ROC-AUC \\
\midrule
RRWP & $0.557\pm0.048$ & $0.559\pm0.047$ & $0.612\pm0.065$ \\
Geometry & $0.503\pm0.053$ & $0.495\pm0.080$ & $0.501\pm0.063$ \\
RRWP$+$Lap$+$Ring$+$Geom & $0.641\pm0.077$ & $0.643\pm0.091$ & $0.680\pm0.065$ \\
FM & $0.622\pm0.063$ & $0.620\pm0.071$ & $0.670\pm0.065$ \\
AFM & $0.847\pm0.045$ & $0.849\pm0.045$ & $0.914\pm0.026$ \\
AFM$+$FM & $0.875\pm0.043$ & $0.875\pm0.044$ & $\mathbf{0.941}\pm0.022$ \\
RRWP$+$Lap$+$Ring$+$Geom$+$AFM$+$FM & $\mathbf{0.882}\pm0.043$ & $\mathbf{0.884}\pm0.043$ & $\mathbf{0.941}\pm0.023$ \\
\bottomrule
\end{tabular}
\end{table}

\subsection{\texorpdfstring{Non-unit-disk couplings}{Non-unit-disk couplings}}
\label{app:non-unit-disk}
To probe couplings beyond what current two-dimensional registers implement, we compute exact state-vector dynamics for the 173 MUTAG graphs with $N\le24$, either with the register couplings $C_6/r_{ij}^6$ or with the ideal graph Hamiltonian $J_{\mathrm{NN}}A$ (cf. Eq.~(\ref{eq:graph_dependent_J})), i.e., without residual non-edge couplings, as accessible with three-dimensional registers or digital implementations.
Table~\ref{tab:mutag-exact-non-unit-disk} compares both, and the 1000-shot emulated features, on top of the geometry-augmented baseline for the five message-passing and attention backbones of Sec.~\ref{sec:gnn-downstream}.
The exact features slightly improve over the baseline ($+0.011$ and $+0.009$ mean F$_1$, $p\approx0.01$--$0.02$), and the ideal graph Hamiltonian performs on par with the register couplings.
This gain is not robust to the choice of backbones: including the two further backbones of the original run (GATv2 and GEN), none of the paired differences is significant ($p>0.08$).
The downstream signal therefore persists, but without a clear gain, when the couplings are replaced by those of the ideal graph Hamiltonian.

\begin{table}[h]
\caption{%
  MUTAG, the 173 graphs with $N \le 24$: geometry-augmented baseline (RRWP$+$Laplacian$+$Rings$+$Geometry) versus its extension by AFM$+$FM features from the 1000-shot emulation, from exact state-vector dynamics with the register couplings $C_6/r^6$, and from exact dynamics of the graph Hamiltonian $J_{\mathrm{NN}} A$ (not realizable by a unit-disk register). Mean macro F$_1$ ($\uparrow$) $\pm$ standard deviation over 10 stratified folds $\times$ 5 seeds; bold: best per backbone.
  Paired Wilcoxon tests over the 250 (backbone, fold, seed) trials, against the baseline: emulated $+0.004$ ($p=0.31$); exact $C_6/r^6$ $+0.011$ ($p=0.014$); exact $J_{\mathrm{NN}}A$ $+0.009$ ($p=0.021$).
}
\label{tab:mutag-exact-non-unit-disk}
\centering
\small
\setlength{\tabcolsep}{4pt}
\begin{tabular}{l c c c c}
\toprule
 & Baseline & \shortstack{${+}$AFM${+}$FM\\ \scriptsize emulated} & \shortstack{${+}$AFM${+}$FM\\ \scriptsize exact, $C_6/r^6$} & \shortstack{${+}$AFM${+}$FM\\ \scriptsize exact, $J_{\mathrm{NN}}A$} \\
\midrule
  GCN          & $0.853{_{\pm 0.071}}$ & $0.851{_{\pm 0.075}}$ & $0.856{_{\pm 0.087}}$ & $\mathbf{0.857}{_{\pm 0.085}}$ \\
  GAT          & $0.845{_{\pm 0.071}}$ & $\mathbf{0.856}{_{\pm 0.087}}$ & $0.852{_{\pm 0.076}}$ & $0.852{_{\pm 0.080}}$ \\
  GINE         & $0.845{_{\pm 0.083}}$ & $0.854{_{\pm 0.083}}$ & $\mathbf{0.860}{_{\pm 0.061}}$ & $0.856{_{\pm 0.076}}$ \\
  Transformer  & $0.854{_{\pm 0.072}}$ & $0.859{_{\pm 0.088}}$ & $\mathbf{0.872}{_{\pm 0.071}}$ & $0.864{_{\pm 0.071}}$ \\
  General      & $0.850{_{\pm 0.083}}$ & $0.845{_{\pm 0.099}}$ & $0.864{_{\pm 0.090}}$ & $\mathbf{0.865}{_{\pm 0.082}}$ \\
\midrule
  Mean & $0.850$ & $0.853$ & $\mathbf{0.861}$ & $0.859$ \\
\bottomrule
\end{tabular}
\end{table}

\clearpage

\newpage
\section*{NeurIPS Paper Checklist}

\begin{enumerate}

\item {\bf Claims}
    \item[] Question: Do the main claims made in the abstract and introduction accurately reflect the paper's contributions and scope?
    \item[] Answer: \answerYes{}
    \item[] Justification: First, 
using two physically motivated operating regimes of the graph-encoded Rydberg Hamiltonian, antiferromagnetic (AFM) and ferromagnetic (FM) [Sec.~\ref{subsec:Pulse-shaping}], we introduce an intrinsic benchmark of $15$ topological and spectral prediction tasks evaluated with permutation-invariant probes, showing that hybrid quantum--classical descriptors match or surpass purely classical baselines based on RRWP, RWSE and Laplacian eigenvectors, with AFM and FM correlations emphasizing different structural aspects of the graph [Sec.~\ref{subsec:intrinsic-method}].
Second, regression analyses indicate that QGFs are not accurately recoverable from the classical descriptors considered here [Sec.~\ref{subsec:predictability_of_quantum_features}], even after adding ring, distance, register-geometry and restricted physical descriptors. 
We also introduced geometry-augmented baselines, atom-placement perturbations and a synthetic family of ladder graphs to delimit where this residual information helps [App.~\ref{app:geometry}].
Third, we show that concatenating QGFs with molecular features yields small but consistent gains across six message-passing, attention-based, and graph-transformer backbones, including GCN~\citep{kipf2017semi}, GINE~\citep{hu2020strategies}, GeneralConv~\citep{li2020deepergcn}, GAT~\citep{velickovic2018graph}, Transformer~\citep{vaswani2017attention,shi2021masked}, and our customized graph transformer (GT) [Sec.~\ref{subsec:architectures}]; we further validate the trend on ESA~\citep{buterez2025end}, a graph architecture outside the standard transformer family [Sec.~\ref{subsec:esa}].
Finally, we execute the full feature-extraction pipeline on a neutral-atom QPU for all $188$ MUTAG graphs at four measurement time steps.

\item {\bf Limitations}
    \item[] Question: Does the paper discuss the limitations of the work performed by the authors?
    \item[] Answer: \answerYes{}
    \item[] Justification: The following limitations are stated and discussed: limitations in embeddability (restricted to 2D unit-disk graphs on current hardware), compute time on current QPUs, and small magnitude of observed improvements over classical baselines.

\item {\bf Theory assumptions and proofs}
    \item[] Question: For each theoretical result, does the paper provide the full set of assumptions and a complete (and correct) proof?
    \item[] Answer: \answerNA{}.
    \item[] Justification: The paper does not include theoretical results.

    \item {\bf Experimental result reproducibility}
    \item[] Question: Does the paper fully disclose all the information needed to reproduce the main experimental results of the paper to the extent that it affects the main claims and/or conclusions of the paper (regardless of whether the code and data are provided or not)?
    \item[] Answer: \answerYes{}
    \item[] Justification: Methodological details are given throughout the paper. Given the numerous experiments and calculations performed, the code and data is available for full reproducibility.

\item {\bf Open access to data and code}
    \item[] Question: Does the paper provide open access to the data and code, with sufficient instructions to faithfully reproduce the main experimental results, as described in supplemental material?
    \item[] Answer: \answerYes{}
    \item[] Justification: All code, datasets, features and raw QPU data are available, on github and zenodo cited at the end of main text.
    
    Code includes: QGF's emulation pipeline, intrinsic information benchmark pipeline, GNN downstream evaluation pipeline, predictability of quantum features from classical descriptors, statistical analysis.

    Everything is setup (seeds, dependencies, scripts) in a way that allows quick reproductability.

\item {\bf Experimental setting/details}
    \item[] Question: Does the paper specify all the training and test details (e.g., data splits, hyperparameters, how they were chosen, type of optimizer) necessary to understand the results?
    \item[] Answer: \answerYes{}
    \item[] Justification: Details about splits, hyperparameters and types of optimizers are given. Not all of them are given in an extensive manner, as they are numerous, but the information required to understand the work is present.

\item {\bf Experiment statistical significance}
    \item[] Question: Does the paper report error bars suitably and correctly defined or other appropriate information about the statistical significance of the experiments?
    \item[] Answer: \answerYes{}
    \item[] Justification: Standard deviations across folds and seeds are reported throughout the results tables. An exception is table 1, which (explicitly) omits standard deviations for clarity, which are instead reported in full in the annex.

\item {\bf Experiments compute resources}
    \item[] Question: For each experiment, does the paper provide sufficient information on the computer resources (type of compute workers, memory, time of execution) needed to reproduce the experiments?
    \item[] Answer: \answerYes{}
    \item[] Justification: The computational bottleneck of the paper is the calculation time of QGFs, either on a real QPU hardware or in emulation. Both are reported as resp. QPU and GPU days. Training time on all described tasks is moderate in comparison and therefore omitted.

\item {\bf Code of ethics}
    \item[] Question: Does the research conducted in the paper conform, in every respect, with the NeurIPS Code of Ethics \url{https://neurips.cc/public/EthicsGuidelines}?
    \item[] Answer: \answerYes{}
    \item[] Justification: The research respects fully the NeurIPS Code of Ethics.

\item {\bf Broader impacts}
    \item[] Question: Does the paper discuss both potential positive societal impacts and negative societal impacts of the work performed?
    \item[] Answer: \answerNA{}
    \item[] Justification: There is no societal impact of the work performed.
    
\item {\bf Safeguards}
    \item[] Question: Does the paper describe safeguards that have been put in place for responsible release of data or models that have a high risk for misuse (e.g., pre-trained language models, image generators, or scraped datasets)?
    \item[] Answer: \answerNA{}
    \item[] Justification: The paper poses no such risks.

\item {\bf Licenses for existing assets}
    \item[] Question: Are the creators or original owners of assets (e.g., code, data, models), used in the paper, properly credited and are the license and terms of use explicitly mentioned and properly respected?
    \item[] Answer: \answerYes{}
    \item[] Justification: Datasets and libraries used in the paper are open-licensed and cited.

\item {\bf New assets}
    \item[] Question: Are new assets introduced in the paper well documented and is the documentation provided alongside the assets?
    \item[] Answer: \answerYes{}
    \item[] Justification: All associated data is given in a Zenodo repository: 10.5281/zenodo.23099924

\item {\bf Crowdsourcing and research with human subjects}
    \item[] Question: For crowdsourcing experiments and research with human subjects, does the paper include the full text of instructions given to participants and screenshots, if applicable, as well as details about compensation (if any)? 
    \item[] Answer: \answerNA{}
    \item[] Justification: The paper does not involve crowdsourcing nor research with human subjects.

\item {\bf Institutional review board (IRB) approvals or equivalent for research with human subjects}
    \item[] Question: Does the paper describe potential risks incurred by study participants, whether such risks were disclosed to the subjects, and whether Institutional Review Board (IRB) approvals (or an equivalent approval/review based on the requirements of your country or institution) were obtained?
    \item[] Answer: \answerNA{}
    \item[] Justification: The paper does not involve crowdsourcing nor research with human subjects.

\item {\bf Declaration of LLM usage}
    \item[] Question: Does the paper describe the usage of LLMs if it is an important, original, or non-standard component of the core methods in this research? Note that if the LLM is used only for writing, editing, or formatting purposes and does \emph{not} impact the core methodology, scientific rigor, or originality of the research, declaration is not required.
    \item[] Answer: \answerNA{}
    \item[] Justification: No LLMs are used or required in the paper or the methods.

\end{enumerate}

\end{document}